\documentclass[aps,letterpaper,prb,twocolumn,showpacs,superscriptaddress]{revtex4}
\usepackage{bm,color,amsmath,amssymb,graphicx}
\usepackage{grffile} 
\usepackage{hyperref}
\usepackage{phonetic}

\providecommand{\range}[2][[]{#1 0~..~#2)}

\begin{document}

\title{Gauge-Fixed Wannier Wave-Functions for Fractional Topological Insulators}

\author{Yang-Le Wu}
\affiliation{Department of Physics, Princeton University, Princeton, NJ 08544}
\author{N. Regnault}
\affiliation{Department of Physics, Princeton University, Princeton, NJ 08544}
\affiliation{Laboratoire Pierre Aigrain, ENS and CNRS, 24 rue Lhomond, 75005 Paris, France}
\author{B. Andrei Bernevig}
\affiliation{Department of Physics, Princeton University, Princeton, NJ 08544}

\begin{abstract}
We propose an improved scheme to construct many-body trial wave functions for 
fractional Chern insulators (FCI), using one-dimensional localized Wannier basis.
The procedure borrows from the original scheme on a continuum cylinder, but is 
adapted to finite-size lattice systems with periodic boundaries.
It fixes several issues of the continuum description that made the overlap 
with the exact ground states insignificant.
The constructed lattice states are translationally invariant, and have the 
correct degeneracy as well as the correct relative and total momenta.
Our prescription preserves the (possible) inversion symmetry of the lattice 
model, and is isotropic in the limit of flat Berry curvature.
By relaxing the maximally localized hybrid Wannier orbital prescription,
we can form an orthonormal basis of states which, upon gauge fixing, can be 
used in lieu of the Landau orbitals.
We find that the exact ground states of several known FCI models at $\nu=1/3$ 
filling are well captured by the lattice states constructed from the Laughlin 
wave function.
The overlap is higher than $0.99$ in some models when the Hilbert space 
dimension is as large as $3\times 10^4$ in each total momentum sector.
\end{abstract}
\maketitle

\section{Introduction}

The Chern insulator~\cite{Haldane88:Honeycomb} is the first and simplest 
example of a topological insulator. It is defined by a non-zero Chern number 
of the occupied bands. It exhibits an integer Hall conductance similar to the 
integer quantum Hall effect but at zero overall magnetic field.
Recently, several groups added interactions to the one-body topological 
(Chern) insulator problem and 
reported~\cite{Sheng11:FCI,Neupert11:FCI,Regnault11:FCI,Venderbos12:t2g, 
Wang11:FCI-Boson,Wu12:Zoology}
strongly-correlated phases similar to the Fractional Quantum Hall (FQH) 
effect.
These fractional Chern insulators (FCI) have a partially filled band with 
non-zero Chern number and develop, similar to the FQH case, 
strongly correlated topological phases at specific values of $\nu$ of 
the filling factor.

At filling $\nu=1/q$ ($q=3,5$ for 
fermions~\cite{Sheng11:FCI,Neupert11:FCI,Regnault11:FCI,Venderbos12:t2g,
Wu12:Zoology}, $q=2,4$ for bosons~\cite{Wang11:FCI-Boson}), 
a phase similar to the Laughlin state~\cite{Laughlin83:Nobel} has been 
observed in numerical studies through several signatures.
First, the system has $q$-fold quasi-degenerate ground states
at filling $\nu=1/q$, separated by a finite energy gap from higher excitations. 
Second, twisted boundary conditions drive spectral flow within the ground 
state manifold, showing a Hall conductance $\sigma_{xy}=e^2/(qh)$. 
Third, the excitations from the topological ground state 
resemble the FQH quasiholes,~\cite{Regnault11:FCI,Wu12:Zoology}
identified by the counting~\cite{Bernevig12:Counting} of low-lying 
levels in the gapped energy and entanglement spectra.~\cite{Li08:ES}

Despite the similarity of the emergent features, the FCI and the FQH effects 
are hosted by substantially different systems.
The FQH effects are typically observed in two-dimensional electron gas.
This continuum system allows a holomorphic description, and the variational 
trial wave functions can be characterized by the asymptotic behavior when 
particles approach each other.~\cite{Bernevig08:Jack,Bernevig08:Jack2}
In sharp contrast, the Chern insulators are defined on a discrete lattice, and 
its continuum limit is different from the
FQH effect.~\cite{Kol93:FQH-Lattice,Moller09:FQH-Lattice}
This obstructs attempts to build model wave functions using the asymptotic 
behavior.

An open question is thus \emph{to what extent and in what form the FCI can be 
described by the FQH physics.}
In this paper we provide a generic scheme to compare FQH and FCI wave 
functions at finite size. In particular, we address the fermionic FCI phase at 
filling $\nu=1/3$ and clarify its relation to the Laughlin state.

The first step in this direction was made by Qi in 
Ref.~\onlinecite{Qi11:Wavefunction}.
The one-dimensional maximally localized Wannier states are plane waves 
localized in the direction perpendicular to propagation. They form an 
alternative basis in a topological flat band, and they resemble the Landau 
orbitals in the lowest Landau level (LLL) in the continuum.
The central, elegant idea of Ref.~\onlinecite{Qi11:Wavefunction} is to exploit this 
similarity and transcribe FQH wave functions written in the second-quantized 
form to the FCI system.
The two-dimensional (2D) fractional topological insulator (FTI) system was 
similarly analyzed using decoupled FQH states in the Wannier basis.

Upon closer inspection, the original proposal suffers from several issues that 
prevent direct application of this scheme to make contact with existing 
numerical studies.
The formalism in Ref.~\onlinecite{Qi11:Wavefunction} was built upon a 
smooth gauge in the continuum. This poses technical challenges for numerical 
implementation on a finite-size lattice,
and conceptually, the one-dimensional (1D) \emph{maximally} localized Wannier 
states are orthogonal to each other \emph{only in the continuum limit}. The 
non-orthogonality at finite size spoils the translational invariance of the 
constructed many-body states.
More importantly, the maximally localized Wannier orbital to LLL mapping in 
Ref.~\onlinecite{Qi11:Wavefunction} is a mapping between the basis 
states of two \emph{separate} systems. The two sets of basis states have 
\emph{independent gauge freedoms}. Merely introducing a 1D relabeling of the 
Wannier states is not enough to properly fix the mapping from a Chern 
insulator to the LLL.
A naive implementation of the formulas in Ref.~\onlinecite{Qi11:Wavefunction} 
results in variational states very low overlaps (lower than $0.04$ for a 
system of $8$ particles on a $6\times 4$ lattice) with the exact 
diagonalization ground states of the two-orbital model discussed therein.

In this paper, we build upon Ref.~\onlinecite{Qi11:Wavefunction} and provide 
an improved prescription for constructing FCI wave functions using gauge-fixed 
(non-maximally) localized Wannier states, with large overlap with the exact 
diagonalization ground states.
Our procedure is defined on a lattice of finite size with periodic boundary 
conditions in both directions.
We do not need gauge smoothing, and we trade maximal localization for the 
orthogonality between the 1D localized Wannier states at 
finite size. We explicitly fix the phase choice of the Wannier states so that 
it best matches that of the Landau orbitals.
Our prescription keeps both the center-of-mass translational symmetry and the 
inversion symmetry at the many-body level,
and it implements the folding rule~\cite{Bernevig12:Counting} that accounts 
for the total momenta of the degenerate FCI ground states on a torus.
We also show that, in our prescription, using the Wannier bases localized in 
either $x$ or $y$ directions produces the same set of many-body states in the 
limit of flat Berry curvature.

As a numerical test, we construct the Laughlin state for several FCI models at 
filling $\nu=1/3$.
In all cases, the lattice Laughlin states have very high overlap with the ground 
states, and their entanglement spectra share the same gapped structure.
The overlap for the best FCI models found so far~\cite{Wu12:Zoology} reaches 
$0.99$ for a system of $8$ particles on a $6\times 4$ lattice, of which
the Hilbert space dimension is as large as $3\times 10^4$ in each total 
momentum sector.
Our results provide the first \emph{direct, quantitative} evidence that the 
fractionalized phase in Chern insulators at $\nu=1/3$ can be well approximated 
by the Laughlin state.

The paper is organized as follows.
In Sec.~\ref{sec:FQH} we discuss the many-body translational 
symmetries~\cite{Haldane85:TorusBZ} of the torus FQH states in the continuum.
We construct a recombined set of degenerate FQH states that allow a direct 
mapping to the lattice.
Sec.~\ref{sec:wannier-finite-size} summarizes with the basic properties of the 
1D localized Wannier states on a lattice. Except for a few subtleties, this 
discussion is similar to the one in Ref.~\onlinecite{Qi11:Wavefunction}. 

In Sec.~\ref{sec:wannier-construction} describes the Wannier construction of the 
many-body FQH states on a lattice. We elaborate upon the \emph{phase} choice 
of the 1D localized Wannier states. We show that the resulting many-body 
states are translationally invariant in both directions. We also demonstrate 
that the counting of the $q$-fold states in each momentum sector agrees with 
the counting rule proposed in Ref.~\onlinecite{Bernevig12:Counting}.
Eq.~\eqref{eq:bloch-amplitude} provides an explicit and ready-to-use formula 
to rewrite \emph{any FQH wave function} in a given Chern insulator model.

Sec.~\ref{sec:inversion} shows that our prescription preserves inversion 
symmetry on the many-body level when such symmetry exists at the 
single-particle level.
Sec.~\ref{sec:results} presents the numerical verification that the ground 
states of various FCI models are well approximated by the Laughlin state 
constructed using our prescription.
We also examine how close our phase choice for the Wannier basis is from the 
absolute ideal by a brute-force optimization.
Sec.~\ref{sec:nematicity} examines the isotropy of the Wannier construction. 
We prove that the many-body lattice states constructed using the Wannier bases 
localized in either direction are the same in the limit of flat Berry 
curvature. The isotropy in the presence of curvature fluctuations is tested 
numerically.
Sec.~\ref{sec:conclusion} concludes the paper and discusses a few future 
directions.

\section{FQH Translational Symmetries}\label{sec:FQH}

It is instructive to review the many-body translational symmetries of the 
FQH system in the continuum. 
We begin by recapitulating the main results of the 
PRL~\cite{Haldane85:TorusBZ} by Haldane.

We study the problem of $N_e$ electrons moving on a \emph{twisted} torus 
with a perpendicular magnetic field. 
The twisted torus can be represented by a parallelogram with opposite edges 
identified.
We set up a Cartesian coordinate system $(\widetilde{x},\widetilde{y})$, with 
orthonormal basis vectors $(\hat{e}_x,\hat{e}_y)$.
The tilde over $\widetilde{x}$ and $\widetilde{y}$ emphasizes the continuous 
nature of these variables.
We consider the torus spanned by
$\mathbf{L}_1=L_1\hat{e}_v$ and $\mathbf{L}_2=L_2\hat{e}_y$ (see 
Fig.~\ref{fig:orbitals-y}).\footnote{It is always possible to line up 
$\mathbf{L}_2$ with the $\hat{e}_y$ axis for a torus of any aspect ratio.}
Here, $L_1$ and $L_2$ are the length of the two fundamental cycles of the 
torus, and the unit vector $\hat{e}_v$ is defined by
$\hat{e}_v=\sin\theta\,\hat{e}_x+\cos\theta\,\hat{e}_y$,
where the twist angle $\theta\in(0,\pi)$ is the angle between the two 
fundamental cycles. The rectangular torus corresponds to $\theta=\pi/2$.
We introduce the normal unit vector 
$\hat{e}_z\equiv \hat{e}_x\times\hat{e}_y$,
and define the reciprocal primitive vectors
\begin{align}\label{eq:reciprocal-G}
\mathbf{G}_1&=2\pi\hat{e}_x/(L_1\sin\theta),&
\mathbf{G}_2&=2\pi(\hat{e}_y-\cot\theta\,\hat{e}_x)/L_2.
\end{align}
They satisfy $\mathbf{G}_a\cdot\mathbf{L}_b=2\pi\delta_{ab}$ 
for $a,b\in\{1,2\}$.

The torus is pierced by a magnetic field in the $-\hat{e}_z$ 
direction, $\mathbf{B}=\nabla\times\mathbf{A}=B\hat{e}_z$ with $B<0$.
The Hamiltonian of the interacting electrons reads
\begin{equation}
H=\sum_i \frac{[-i\hbar\nabla_i-e\mathbf{A}(\widetilde{\mathbf{r}}_i)]^2}{2m}
+\sum_{i<j}V(\widetilde{\mathbf{r}_i}-\widetilde{\mathbf{r}}_j).
\end{equation}
We denote by $e<0$ the charge of the electron.
The magnetic length is $l_B=\sqrt{\hbar/(eB)}$. 
The total number of fluxes $N_\phi$ penetrating the 
torus has to be an integer, given by $L_1L_2\sin\theta=2\pi l_B^2N_\phi$.
We define $N=\mathrm{GCD}(N_e,N_\phi)$, where GCD stands for the Greatest 
Common Divisor. Then $p\equiv N_e/N$ and $q\equiv N_\phi/N$ are coprime.

The magnetic translation operator of a single particle is defined by
$T(\mathbf{a})=e^{-i\mathbf{a}\cdot\mathbf{K}/\hbar}$,
where 
$\mathbf{K}=-i\hbar\nabla-e\mathbf{A}(\widetilde{\mathbf{r}})
+e\mathbf{B}\times \widetilde{\mathbf{r}}$
is the guiding center momentum of the $i$-th particle. 
We pick Landau gauge 
$\mathbf{A}(\widetilde{x},\widetilde{y})=B\widetilde{x}\,\hat{e}_y$,
and impose periodic (not twisted) boundary conditions $T(\mathbf{L}_a)=1$, 
$a=1,2$.
The wave functions $\langle\widetilde{x},\widetilde{y}|j\rangle$ of
the $N_\phi$ single-particle states in the LLL are given by
\begin{widetext}
\begin{equation}\label{eq:LLL-wf}
\phi_j(\widetilde{x},\widetilde{y})=
\frac{1}{(\sqrt{\pi}L_2 l_B)^{1/2}}
\sum_n^{\mathbb{Z}}
\exp\left[
2\pi(j+nN_\phi)\frac{\widetilde{x}+i\widetilde{y}}{L_2}
-i\frac{\pi}{N_\phi}\frac{L_1e^{-i\theta}}{L_2}(j+nN_\phi)^2
\right]
e^{-\widetilde{x}^2/(2l_B^2)}.
\end{equation}
\end{widetext}
where the state index $j$ is an integer defined modulo $N_\phi$.

We can decompose the translation operator of the $i$-th 
electron, $T_i(\mathbf{a})$, into a relative part and a center-of-mass part,
$T_i(\mathbf{a})=T_{\mathrm{rel},i}(\mathbf{a})T_{\mathrm{cm}}(\mathbf{a}/N_e)$,
where
\begin{align}
T_{\mathrm{rel},i}(\mathbf{a})&=T_i(\mathbf{a})\prod_{j}^{N_e}
T_j({\textstyle -\frac{\mathbf{a}}{N_e}}),
&T_{\mathrm{cm}}(\mathbf{a})&=\prod_j^{N_e}T_j(\mathbf{a}).
\end{align}
Here and hereafter throughout the paper, we use shorthand notation for 
products and summations:
\begin{align}\label{eq:sum-prod-convention}
\prod_i^M&\equiv \prod_{i=0}^{M-1},&
\sum_i^M&\equiv \sum_{i=0}^{M-1}.
\end{align}
We are particularly interested in the following translation operators and we 
define the shorthand notations
\begin{equation}
\begin{aligned}
T_{\mathrm{rel},i}^x&=T_{\mathrm{rel},i}(p\mathbf{L}_1),&
T_{\mathrm{rel},i}^y&=T_{\mathrm{rel},i}(p\mathbf{L}_2),\\
T_\mathrm{cm}^x&=T_{\mathrm{cm}}(\mathbf{L}_1/N_\phi),&
T_\mathrm{cm}^y&=T_{\mathrm{cm}}(\mathbf{L}_2/N_\phi).
\end{aligned}
\end{equation}
Thanks to the periodic boundary condition,
we can drop the particle index $i$ from the relative translation operators,
since $T_{\mathrm{rel},i}^x=[(T_\mathrm{cm}^x)^q]^\dagger$ and 
$T_{\mathrm{rel},i}^y=[(T_\mathrm{cm}^y)^q]^\dagger$ do not depend on the 
particle index.

The operators $T_\mathrm{rel}^x$ and $T_\mathrm{cm}^y$, and of course 
$T_\mathrm{rel}^y$ being a power of $T_\mathrm{cm}^y$, commute with each other 
and the Hamiltonian $H$.~\cite{Haldane85:TorusBZ}
We use them to block-diagonalize the Hamiltonian into momentum sectors labeled 
by two-dimensional (2D) wave numbers $(\kappa_x,\kappa_y)$ in an 
$N\times N_\phi$ Brillouin zone,
defined by the eigenvalues\footnote{
Notice that the original treatment in Ref.~\onlinecite{Haldane85:TorusBZ} put 
an additional factor of $(-1)^{pq(N_e-1)}$ in the formula, but this is not 
necessary for our purposes: $(\kappa_x,\kappa_y)$ are 
\emph{defined} by Eq.~\eqref{eq:kappa-x-y}. Our choice can be regarded as an 
alternative labeling aimed to simplify the formulas in this paper, at the cost 
of the identity of $(\kappa_x,\kappa_y)$ as true momentum.}
\begin{equation}\label{eq:kappa-x-y}
\begin{aligned}
T_{\mathrm{rel}}^x&=e^{i2\pi\kappa_x/N},&
T_\mathrm{cm}^y&=e^{-i2\pi\kappa_y/N_\phi}.
\end{aligned}
\end{equation}

The operator $T_\mathrm{cm}^x$ commutes with $T_\mathrm{rel}^x$ and $H$, but 
not with $T_\mathrm{cm}^y$:
\begin{equation}\label{eq:Tcm-not-commute}
T_{\mathrm{cm}}^yT_{\mathrm{cm}}^x=T_{\mathrm{cm}}^xT_{\mathrm{cm}}^y
e^{-i2\pi p/q}.
\end{equation}
Therefore, the many-body energy eigenstates can be grouped into $q$-fold 
center-of-mass multiplets; the $q$ states in each multiplet can be transversed 
by successive applications of $T_\mathrm{cm}^x$ and they share the same energy 
and the value of $\kappa_x$ and $\kappa_y$ mod $N$.

We set the cyclotron energy to infinity and focus on the states in the LLL.
In this approximation, the many-body Hilbert space is spanned by the 
occupation-number basis states (Slater determinants) constructed from the LLL 
states $|j\rangle$,
and the projected interacting Hamiltonian is given in 
Appx.~\ref{sec:FQH-Hamiltonian}.
We use the curly braces $\{\cdot\}$ to denote a list of quantum numbers for 
the $N_e$ electrons. For succinctness, we always omit the electron indices 
from $\{\cdot\}$ and related expressions.

Now we are in a position to establish a concrete representation of the 
many-body translation algebra.
The basis states in the LLL are 
$|\{j\}\rangle\equiv|j_0,j_1,\ldots,j_{N_e-1}\rangle$ with an implied 
anti-symmetrization. The action of the many-body translation operators 
reads~\cite{Bernevig12:Counting}
\begin{equation}\label{eq:j-translation}
\begin{aligned}
T_\mathrm{rel}^x|\{j\}\rangle&\!=\!|\{j-q\}\rangle,&
\!T_\mathrm{rel}^y|\{j\}\rangle&\!=\!e^{i2\pi\!\sum j/N}|\{j\}\rangle,\\
T_\mathrm{cm}^x|\{j\}\rangle&\!=\!|\{j+1\}\rangle,&
\!T_\mathrm{cm}^y|\{j\}\rangle&\!=\!
e^{-i2\pi\!\sum j/N_\phi}|\{j\}\rangle.
\end{aligned}
\end{equation}
Here the sum $\sum j$ runs over all the particles, and the state $|\{j+l\}\rangle$ 
is obtained from $|\{j\}\rangle$ by shifting the $j$ quantum number of each 
particle by $l$.
As expected, the relative and the center-of-mass translations are related by 
the periodic boundary conditions,
$T_{\mathrm{rel}}^x=[(T_\mathrm{cm}^x)^q]^\dagger$,
$T_{\mathrm{rel}}^y=[(T_\mathrm{cm}^y)^q]^\dagger$.

\subsection{Recombination of the $q$-Fold States}\label{sec:recombination}

It has been suggested that the FCI with $|C|=1$ on an $N_x\times N_y$ lattice 
corresponds to a FQH system with flux
$N_\phi=N_xN_y$.~\cite{Qi11:Wavefunction,Bernevig12:Counting}
We now look for an alternative representation of the many-body translational 
symmetries with a Brillouin zone commensurate to the lattice system, in 
preparation for the analysis of the translational symmetry of the Wannier 
construction in Sec.~\ref{sec:translational-inv}.
We emphasize that we are still working in the continuum and we are merely 
providing another representation of the center-of-mass translation algebra. 
The integers $N_x$ and $N_y$ should be understood as a factorization of 
$N_\phi$ at this stage.

Following Ref.~\onlinecite{Bernevig12:Counting}, we define the integers 
$N_{0x}=\mathrm{GCD}(N_e,N_x)$, $N_{0y}=\mathrm{GCD}(N_e,N_y)$, and 
$p_x=N_e/N_{0x}$, $p_y=N_e/N_{0y}$, $q_x=N_x/N_{0x}$, $q_y=N_y/N_{0y}$. 
Obviously $p_x,q_x$ are coprime, so are $p_y,q_y$. Less obviously, $q_xq_y$ 
divides $q$,~\cite{Bernevig12:Counting}
and thus $p,q_x$ are coprime, so are $p,q_y$.
We define the translation operators\footnote{
We emphasize that the operators $S_x$ and $R_y$ are defined in the continuum. 
In contrast to the discussions in Ref.~\onlinecite{Bernevig12:Counting}, here 
we do not add a lattice pinning potential to the FQH setup.
Were we to do that, $S_x$ and $R_y$ would not be legitimate translation 
operators as they contain one-body translations that are a fraction of a 
unit cell size.
We bridge the FQH and the FCI sides through the properties of the 
second-quantized amplitudes, without directly migrating the continuum 
translation operators and their algebra to the lattice.}
\begin{align}\label{eq:Sx-Ry}
S_x&=(T_\mathrm{cm}^x)^{q/q_x},
&R_y&=(T_\mathrm{cm}^y)^{q_x}.
\end{align}
They commute with each other and the Hamiltonian $H$. 
Therefore, we can use $(S_x,R_y)$, instead of 
$(T_\mathrm{rel}^x,T_\mathrm{cm}^y)$, to block-diagonalize the 
Hamiltonian into momentum sectors. Since 
$(S_x)^{q_x}=(T_\mathrm{rel}^x)^\dagger$, the new set of 2D wave numbers 
defined by $(S_x,R_y)$ takes value from an $(Nq_x)\times (N_{0x}N_y)$ 
Brillouin zone [as opposed to the $N\times N_\phi$ Brillouin zone of 
Eq.~\eqref{eq:j-translation}].

Notice that $S_x$ operates within the $q$-fold multiplet. Successive 
applications of $S_x$ break the multiplet into non-overlapping orbits. The 
states in each orbit have the same $R_y$ eigenvalue due to $[S_x,R_y]=0$.
The orbit structure can be revealed by the eigenvalues of $T_\mathrm{cm}^y$, 
as they are distinct for the $q$ states.
Plugging Eq.~\eqref{eq:Sx-Ry} into Eq.~\eqref{eq:Tcm-not-commute}, we have
\begin{align}
\label{eq:Sx-not-commute}
T_\mathrm{cm}^y S_x&=S_x T_\mathrm{cm}^y e^{-i2\pi p/q_x},\\
\label{eq:Ry-not-commute}
T_\mathrm{cm}^x R_y&=R_y T_\mathrm{cm}^x e^{i2\pi p/(q/q_x)}.
\end{align}
From Eq.~\eqref{eq:Sx-not-commute}, every orbit of $S_x$ has the same length 
$q_x$. The total number of orbits is thus $q/q_x$. From
Eq.~\eqref{eq:Ry-not-commute}, for $r\in\range{q/q_x}$, $(T_\mathrm{cm}^x)^r$ 
brings any given state in the $q$-fold multiplet to $q/q_x$ distinct states 
discriminated by $R_y$ eigenvalues, and thus belonging to $q/q_x$ different 
orbits of $S_x$. This covers \emph{each} of the orbits exactly \emph{once}.

We can recombine the $q$-fold states to form simultaneous eigenstates of $S_x$ 
and $R_y$.
We start the construction from an arbitrary energy eigenstate $|\Psi\rangle$ in 
the $q$-fold multiplet diagonal in $(T_\mathrm{rel}^x,T_\mathrm{cm}^y)$, 
labeled by $\kappa_x\in\range{N}$ and $\kappa_y\in\range{N_y}$.
The usual, relative $y$-momentum which labels the eigenvalue of 
$T_\mathrm{rel}^y$ can be easily obtained as $\kappa_y$ mod $N$.
For the Laughlin case at filling $\nu=1/3$, $|\Psi\rangle$ is just one of the 
threefold degenerate states on the torus.
The $q$ states can be regrouped into the orbits of $S_x$, labeled by $r$:
\begin{equation}
\Big\{(S_x)^m(T_\mathrm{cm}^x)^r|\Psi\rangle\,\Big|\,m\in\range{q_x}\Big\},\quad
r\in\range{q/q_x}.
\end{equation}
We can recombine the $q_x$ states in each orbit into $q_x$ eigenstates of $S_x$.
Within the multiplet, ince 
$(S_x)^{q_x}=(T_\mathrm{rel}^x)^\dagger=e^{-i2\pi \kappa_x/N}$, we have
\begin{equation}
(S_x)^{q_x}=e^{-i2\pi q_x(\kappa_x+sN_e)/(Nq_x)}.
\end{equation}
Note that the values of $e^{i2\pi(\kappa_x+sN_e)/(Nq_x)}$ are distinct for 
$s\in\range{q_x}$, thanks to $p,q_x$ being coprime. We define the 
$q_x\times (q/q_x)$ states
\begin{equation}\label{eq:psi-s-r}
|\Psi;s,r\rangle=
\frac{1}{\sqrt{q_x}}\sum_m^{q_x}e^{i2\pi m(\kappa_x+sN_e)/(Nq_x)}
(S_x)^m(T_\mathrm{cm}^x)^r|\Psi\rangle,
\end{equation}
for $s\in\range{q_x}$ and $r\in\range{q/q_x}$.
These states are orthonormal, and they are simultaneous eigenstates of $S_x$ 
and $R_y$ in the $(Nq_x)\times (N_{0x}N_y)$ Brillouin zone,\footnote{
A notable special case is when $q$ divides $N_y$. Since
$\mathrm{GCD}(N_y,q)=\mathrm{GCD}(pN_y,q)=N_yN_{0x}/N=q/q_x$,
in this case we have $q_x=1$, and thus
$S_x=(T_\mathrm{rel}^x)^\dagger$ and $R_y=T_\mathrm{cm}^y$.
Hence, the original $q$-fold states diagonal in 
$(T_\mathrm{rel}^x,T_\mathrm{cm}^y)$ are already the simultaneous eigenstates 
of $S_x$ and $R_y$ without any recombination.
}
\begin{equation}
\begin{aligned}
S_x|\Psi;s,r\rangle&=e^{-i2\pi(\kappa_x+sN_e)/(Nq_x)}|\Psi;s,r\rangle,\\
R_y|\Psi;s,r\rangle&=e^{-i2\pi(\kappa_y+rN_e)/(N_{0x}N_y)}|\Psi;s,r\rangle.
\end{aligned}
\end{equation}

The $q_x\times(q/q_x)$ states are related by center-of-mass translations, namely
\begin{equation}\label{eq:T-psi-s-r}
\begin{aligned}
T_\mathrm{cm}^x|\Psi;s,r\rangle
&=|\Psi;s,r+1\rangle,\\
T_\mathrm{cm}^y|\Psi;s,r\rangle
&=e^{-i2\pi(\kappa_y+rN_e)/N_\phi}|\Psi;s-1,r\rangle.
\end{aligned}
\end{equation}
In terms of amplitudes, the second equation above can be written as [using 
Eq.~\eqref{eq:j-translation}]
\begin{multline}\label{eq:psi-s-shift}
e^{-i2\pi\sum j/N_\phi}\langle\{j\}|\Psi;s,r\rangle\\
=e^{-i2\pi(\kappa_y+rN_e)/N_\phi}\langle\{j\}|\Psi;s-1,r\rangle.
\end{multline}
Incidentally, we note that $|\Psi;s,r\rangle$ is periodic under $s\rightarrow s+q_x$, 
but it acquires a phase when $r\rightarrow r+q/q_x$ [since 
$(T_\mathrm{cm}^x)^{q/q_x}=S_x$ as defined in Eq.~\eqref{eq:Sx-Ry}]
\begin{equation}\label{eq:psi-r-quasiperiodic}
|\Psi;s,r+q/q_x\rangle=e^{-i2\pi(\kappa_x+sN_e)/(Nq_x)}|\Psi;s,r\rangle.
\end{equation}

\subsection{Translational Symmetries in Amplitudes}\label{sec:FQH-translation}

We now consider the manifestation of the translation symmetries in the 
amplitudes $\langle\{j\}|\Psi;s,r\rangle$.
The action of $S_x$ and $R_y$ on the basis states $|\{j\}\rangle$ reads
\begin{equation}
\begin{aligned}
S_x|\{j\}\rangle&=|\{j+q/q_x\}\rangle,\\
R_y|\{j\}\rangle&=e^{-i2\pi\sum j/(N_{0x}N_y)}|\{j\}\rangle.
\end{aligned}
\end{equation}
Therefore, from the first equation above we have
\begin{equation}
\label{eq:FQH-wf-sym-x}
e^{-i2\pi (\kappa_x+sN_e)/(Nq_x)}\langle\{j\}|\Psi;s,r\rangle
=\langle\{j-q/q_x\}|\Psi;s,r\rangle,
\end{equation}
while from the second, we find that the amplitude 
$\langle\{j\}|\Psi;s,r\rangle$ vanishes unless
\begin{equation}\label{eq:FQH-wf-sym-j}
\sum j=\kappa_y+rN_e\text{ mod } N_{0x}N_y.
\end{equation}
The last two equations above summarize the information that we need from the 
FQH side to establish the translational invariance of the FCI many-body wave 
functions to be constructed in Sec.~\ref{sec:wannier-construction} on an 
$N_x\times N_y$ lattice.

\section{Hybrid Localized Wannier States}\label{sec:wannier-finite-size}

We now proceed to establish the hybrid Wannier basis in a Chern band on a 
lattice.
The ultimate goal is to construct 1D localized Wannier states which are plane 
waves in the second direction (hybrid). Such states mimic
the Landau orbitals $|j\rangle$. We begin by reviewing the construction and 
the properties of the 1D maximally localized Wannier 
states.~\cite{Kivelson82:Wannier,Marzari97:MLWF,Yu11:Z2}

Consider a 2D band insulator with $N_b$ orbitals per unit cell, indexed by 
$\alpha$.
We assume lattice translational symmetry and periodic boundary conditions.
Denote the primitive translation vectors by $\mathbf{b}_1$ and $\mathbf{b}_2$, 
with $\hat{e}_z\cdot(\mathbf{b}_1\times\mathbf{b}_2)>0$.
Then the Bravais lattice is indexed by $x\mathbf{b}_1+y\mathbf{b}_2$, with 
$(x,y)\in\mathbb{Z}^2$ (be to differentiated from 
$(\widetilde{x},\widetilde{y})\in\mathbb{R}^2$ in the continuum.)
We pick the principal region to be 
$(x,y)\in\range{N_x}\times \range{N_y}$.
The momentum space is given by the 
reciprocal lattice $\mathbf{k}$. We label points in the momentum space 
by wave numbers $(k_x,k_y)\in\mathbb{Z}^2$, defined by
$k_x=\mathbf{k}\cdot\mathbf{b}_1$ and
$k_y=\mathbf{k}\cdot\mathbf{b}_2$.
The single-particle orbitals can be written as $|x,y,\alpha\rangle$, or 
$|k_x,k_y,\alpha\rangle$ in the momentum space. 

The energy eigenstates are Bloch waves. We focus on a single, isolated 
band $|k_x,k_y\rangle$, which will be fractionally filled in the many-body 
construction. For now we only focus on the one-band problem. Generalization to 
the 2D FTI problem is straightforward as that problem decouples in the Wannier 
basis.
The wave function of the Bloch band is
$u_\alpha(k_x,k_y)=\langle k_x,k_y,\alpha|k_x,k_y\rangle$.
We assume periodic boundary condition in \emph{both} $k_x$ and $k_y$ 
directions, $|k_x+l_xN_x,k_y+l_yN_y,n\rangle=|k_x,k_y,n\rangle$,
for $l_x,l_y\in\mathbb{Z}$. We set the first Brillouin zone to 
$\range{N_x}\times\range{N_y}$. 

We denote the position of an orbital $\alpha$ relative to its unit cell 
coordinate by $\epsilon^x_\alpha \mathbf{b}_1+\epsilon^y_\alpha\mathbf{b}_2$, 
where $(\epsilon^x_\alpha,\epsilon^y_\alpha)\in\mathbb{R}^2$ are the relative 
displacements of the orbitals within an unit cell.
Taking into account the periodic boundary, we can define the (exponentiated) 
position operators
\begin{equation}
\widehat{x}=\sum_{x,y}\sum_\alpha
|x,y,\alpha\rangle e^{-i2\pi(x+\epsilon^x_\alpha)/N_x}\langle x,y,\alpha|,
\end{equation}
and similarly $\widehat{y}$.
Here the integers $(x,y)$ are summed over the Bravais lattice points in the 
principal region of the torus, and $\alpha$ is summed over all $N_b$ orbitals.

Using the projector to the occupied band
\begin{equation}
P=\sum_{k_x,k_y}|k_x,k_y\rangle\langle k_x,k_y|,
\end{equation}
we can define the projected position operators $\widehat{\mathcal{X}}=P\widehat{x}P$ 
and $\widehat{\mathcal{Y}}=P\widehat{y}P$. In the momentum space, they take the form
\begin{equation}
\label{eq:non-unitary-XY}
\begin{aligned}
\widehat{\mathcal{X}}\!
&=\!\sum_{k_x,k_y}
|k_x,k_y\rangle\mathcal{A}_x(k_x,k_y)\langle k_x+1,k_y|,\\
\widehat{\mathcal{Y}}\!
&=\!\sum_{k_x,k_y}
|k_x,k_y\rangle\mathcal{A}_y(k_x,k_y)\langle k_x,k_y+1|,
\end{aligned}
\end{equation}
where $(k_x,k_y)$ is summed over the first Brillouin zone, and
$\mathcal{A}_{x,y}(k_x,k_y)$ are the (exponentiated) Berry connections
given by
\begin{equation}
\label{eq:discrete-A-nonunitary}
\begin{aligned}
\mathcal{A}_x(k_x,k_y)\!
&=\langle k_x,k_y|\widehat{x}|k_x+1,k_y\rangle\\
&=\!\sum_\alpha e^{-i2\pi\epsilon^x_\alpha/N_x}
u_{\alpha}^*(k_x,k_y)u_{\alpha}(k_x\!+\!1,k_y),\\
\mathcal{A}_y(k_x,k_y)\!
&=\langle k_x,k_y|\widehat{y}|k_x,k_y+1\rangle\\
&=\!\sum_\alpha e^{-i2\pi\epsilon^y_\alpha/N_y}
u_{\alpha}^*(k_x,k_y)u_{\alpha}(k_x,k_y\!+\!1).
\end{aligned}
\end{equation}
Notice that the Berry connections depend on the embedding of the lattice and 
its orbitals in real space, as does the Berry curvature.~\footnote{We thank 
F.D.M.~Haldane for stressing the importance of this point to us.}
The link to the usual continuum definition of the Berry connections is 
explicitly demonstrated in Appx.~\ref{sec:continuum-limit}.
We emphasize that the definition of the exponentiated discrete Berry 
connections does not require a smooth gauge.

\subsection{Localization and Orthogonality}\label{sec:orthogonality}
Thanks to the translational invariance, we can view the 2D system as $N_y$ 
decoupled 1D subsystems labeled by the momentum $k_y$. The position operator 
$\widehat{\mathcal{X}}$ operates within each subsystem and is diagonal in $k_y$.
At each value of $k_y$, the eigenstates of this operator are the closest 
analogue to the states $|x,k_y,\alpha\rangle$ that we can construct within the 
occupied bands; they are the maximally localized Wannier state for the 1D 
subsystem.~\cite{Kivelson82:Wannier,Marzari97:MLWF,Yu11:Z2}
The eigenvalue of $\widehat{\mathcal{X}}$ is the (exponentiated) Wannier 
center position in the $\mathbf{b}_1$ direction.

Ref.~\onlinecite{Qi11:Wavefunction} proposed to build FQH wave functions 
using the maximally localized Wannier orbitals as one body basis with a 
one-to-one mapping into the LLL orbitals. A major issue at finite 
size, however, is that the $N_x$ eigenstates of the projected position 
operator at each value of $k_y$ are \emph{not orthogonal} due to the 
\emph{non-unitarity} of the projected position operator at finite $N_x$. This 
point is demonstrated in Appx.~\ref{sec:orthogonality-appx}.

Gram-Schmidt orthogonalization is not a suitable solution, as it 
mixes the Wannier states localized in different unit cells, thereby 
spoiling the translational invariance.
The proper resolution is to keep only the \emph{phase} part of the exponentiated 
Berry connection and use the alternative, unitary projected position operators 
to define the 1D localized Wannier states.
Specifically, we define the unitary connections\footnote{
The non-unitary 
connection $\mathcal{A}_a(k_y,k_y)$ ($a=x,y$) may vanish on a very small 
lattice. This problem is most pronounced on an $N_x\times N_y=2\times 2$ 
lattice with inversion symmetry. In this case, all the four Bloch states are 
inversion eigenstates. The connection $\mathcal{A}_a$ between two Bloch states 
vanishes when the two states belong to different inversion eigenvalues.
We do not consider such pathological cases in our treatment.}
\begin{equation}
A_a(k_x,k_y)=\mathcal{A}_a(k_x,k_y)/|\mathcal{A}_a(k_x,k_y)|,
\quad a=x,y,
\end{equation}
and the unitary projected position operators
\begin{align}
\label{eq:unitary-X}
\widehat{X}&=\sum_{k_x,k_y}
|k_x,k_y\rangle A_x(k_x,k_y)\langle k_x+1,k_y|,\\
\label{eq:unitary-Y}
\widehat{Y}&=\sum_{k_x,k_y}
|k_x,k_y\rangle A_y(k_x,k_y)\langle k_x,k_y+1|.
\end{align}
Technically, the eigenstates of these operators are not \emph{maximally} 
localized (they are almost so), but this is a small price to pay for the 
orthogonality.

If there is more than one occupied band, such in the case of the quantum spin 
Hall (QSH) effect and 2D topological insulators,
the prescription for obtaining orthonormal orbitals is to perform a singular 
value decomposition (SVD) of the non-Abelian exponentiated Berry connection 
into $V\Sigma W^\dagger$, 
and use $VW^\dagger$ in place of the Berry connection to define 
the projected position operators.~\cite{Marzari97:MLWF}
Specifically, if the occupied bands are denoted by $|k_x,k_y,n\rangle$, we 
replace the matrix 
$\langle k_x,k_y,m|k_x+1,k_y,n\rangle$ (of indices $m,n$) by its SVD when 
defining the projected position operator $\widehat{X}$.~\cite{Yu11:Z2}

Thanks to the gauge periodicity in both $k_x$ and $k_y$, we can define 
the unitary Wilson loops
\begin{align}\nonumber
W_x(k_y)&=\prod_{k_x}^{N_x}A_x(k_x,k_y),&
\!\!W_y(k_x)&=\prod_{k_y}^{N_y}A_y(k_x,k_y).
\end{align}
For later convenience, we introduce $\lambda_x(k_y)$ and $\lambda_y(k_x)$ defined by
\begin{align}\label{eq:lambda-W}
\left[\lambda_x(k_y)\right]^{N_x}&=W_x(k_y),&
\left[\lambda_y(k_x)\right]^{N_y}&=W_y(k_x).
\end{align}
We pick $\lambda_x(k_y)$ to be the $N_x$-th root with argument angle in 
$(-2\pi/N_x,0]$,
and $\lambda_y(k_x)$ to be the $N_y$-th root with argument angle in 
$(-2\pi/N_y,0]$.
We can interpret $\lambda_y(k_x)$ and $\lambda_y(k_x)$ as the ``average'' 
phase (connection) of the Wilson loop evenly distributed to each bond along the 
loop.
In general, for more than one occupied band, $\lambda_x(k_y)$ and 
$\lambda_y(k_x)$ are obtained from the sets of eigenvalues of the 
corresponding Wilson loop matrices.

The eigenvalues and the eigenstates of $\widehat{X}$ can be found easily, due 
to its decoupling into $N_y$ blocks labeled by $k_y$.
Each block $\widehat{X}_{k_y}$ is an $N_x\times N_x$ 
matrix in the Bloch basis. The $N_x$ eigenvalues of $\widehat{X}_{k_y}$ are 
given by the $N_x$-th roots of $W_x(k_y)$,~\cite{Yu11:Z2}
\begin{equation}
\widehat{X}_{k_y}|X,k_y\rangle=\Lambda_x(X,k_y)|X,k_y\rangle,
\end{equation}
where $\Lambda_x(X,k_y)=e^{-i2\pi X/N_x}\lambda_x(k_y)$,
indexed by $X\in\range{N_x}$.
The eigenstate belonging to the eigenvalue $\Lambda_x(X,k_y)$ can be written
\begin{multline}\label{eq:wannier-eigenstate}
|X,k_y\rangle=\frac{e^{i\Phi_y(X,k_y)}}{\sqrt{N_x}}\\
\times\sum_{k_x}^{N_x}
e^{-i2\pi k_xX/N_x}
\left\{\frac{\left[\lambda_x(k_y)\right]^{k_x}}
{\prod_{\kappa}^{k_x}A_x(\kappa,k_y)}\right\}
|k_x,k_y\rangle.
\end{multline}
Recall from Eq.~\eqref{eq:sum-prod-convention} that we use the shortand
\begin{equation}
\prod_\kappa^{k_x}A_x(\kappa,k_y)\equiv\prod_{\kappa=0}^{k_x-1}A_x(\kappa,k_y).
\end{equation}
Since the curly-braced prefactor in Eq.~\eqref{eq:wannier-eigenstate} is
unitary, these states inherit at finite size the orthonormality of 
$|k_x,k_y\rangle$ through the Fourier transform.
These orthogonal Wannier orbitals are not maximally localized for the 
finite-size lattice, but quickly become so as the number of sites $N_x$ 
increases.
They exhibit a gauge freedom, $\Phi_y(X,k_y)\in\mathbb{R}$, which is a phase 
to be specified in Sec.~\ref{sec:phase-fixing}.

The center of the 1D localized Wannier state is given by
\begin{equation}
\langle X,k_y|\widehat{x}|X,k_y\rangle=\Lambda_x(X,k_y)
\left[\frac{1}{N_x}\sum_{k_x}^{N_x}\Big|\mathcal{A}_x(k_x,k_y)\Big|\right],
\end{equation}
where the bracketed factor is a positive real number. Therefore, we can 
interpret
\begin{equation}\label{eq:wannier-center}
\chi^{X,k_y}=-\frac{N_x}{2\pi}\;\mathrm{arg}[\Lambda_x(X,k_y)]
=X - \frac{1}{2\pi}\;\mathrm{arg}[W_x(k_y)]
\end{equation}
as the center of the maximally localized Wannier function, where we pick the 
branch of the argument function with $\mathrm{arg}(z)\in(-2\pi,0]$ to make 
$\chi^{X,k_y}\in[X,X+1)$. We place the $X$-th unit cell over $[X,X+1)$.
As expected, in each unit cell there are $N_y$ Wannier centers, corresponding 
to the $N_y$ values of $k_y$.

For a flat-band Hamiltonian with no band dispersion, such as the 
single-particle part of the FCI Hamiltonians in 
Ref.~\onlinecite{Wu12:Zoology}, the Wannier states are actually 
single-particle energy eigenstates of the lowest, fractionally filled band. 
This qualifies the set of all the 1D 
localized Wannier states $\{|X,k_y\rangle\}$ as a single-particle basis for 
constructing the many-body trial states for the FCI, which lie entirely in the 
lowest band, in the same spirit as the LLL projected FQH wave functions.

\subsection{Gauge Freedom}\label{sec:gauge-freedom}

In the language of Ref.~\onlinecite{Marzari97:MLWF}, the construction of 
eigenstates in Eq.~\eqref{eq:wannier-eigenstate}
can be interpreted as transforming the Bloch states with an arbitrary phase 
factor to the ``parallel transport'' (pt) gauge, in which
\begin{equation}\label{eq:pt-gauge}
|k_x,k_y\rangle_\mathrm{pt}=
\frac{\left[\lambda_x(k_y)\right]^{k_x}}
{\prod_{\kappa}^{k_x}A_x(\kappa,k_y)}\;
|k_x,k_y\rangle,
\end{equation}
and the 1D localized Wannier states are just Fourier transform of 
the Bloch states $|k_x,k_y\rangle_\mathrm{pt}$.
Despite the name, technically the phase of $|k_x,k_y\rangle_\mathrm{pt}$
contains not only the parallel transport $\prod_\kappa^{k_y}A_x(\kappa,k_y)$ 
from $|0,k_y\rangle$, it also 
contains a rotation to accommodate the Wilson loop $W_x(k_y)$.

Under the gauge transform 
\begin{equation}
|k_x,k_y\rangle\rightarrow e^{i\eta(k_x,k_y)}|k_x,k_y\rangle,
\end{equation}
the Wilson line transforms by 
\begin{equation}
\prod_\kappa^{k_x}A_x(\kappa,k_y)\rightarrow
e^{-i\eta(0,k_y)}\prod_\kappa^{k_x}A_x(\kappa,k_y)\,
e^{i\eta(k_x,k_y)},
\end{equation}
and thus the Wannier state transforms by
\begin{equation}\label{eq:wannier-gauge-transform}
|X,k_y\rangle\rightarrow e^{i\eta(0,k_y)}|X,k_y\rangle.
\end{equation}
The reason for the appearance of $e^{i\eta(k_x,k_y)}$ with only $k_x=0$ 
is obvious: the parallel transport gauge [Eq.~\eqref{eq:pt-gauge}] depends 
only on the initial condition at $k_x=0$ and the gauge-invariant Wilson loop 
$W_x(k_y)$. Had we chosen to start the parallel transport from $k_x^0$, the 
Wannier state would acquire a phase $e^{i\eta(k_x^0,k_y)}$ upon the gauge 
transform. The full gauge freedom of the Wannier states is kept in the phase 
$e^{i\Phi_y(X,k_y)}$. Due to the $X$-dependency, this is much larger than 
the gauge freedom in $e^{i\eta(0,k_y)}$.

In Ref.~\onlinecite{Qi11:Wavefunction}, a similar gauge transformation formula 
was derived in a different, smooth gauge in the continuum limit that is 
\emph{not} periodic in the $k_y$ direction. However, as detailed in 
Appx.~\ref{sec:wannier-continuum}, we find in that case a gauge transformation 
formula that is fundamentally different from the one given in 
Ref.~\onlinecite{Qi11:Wavefunction}:
due to the absence of periodic boundary condition in the $k_y$ direction in 
the setup of Ref.~\onlinecite{Qi11:Wavefunction},
a gauge transformation with a non-trivial winding may shift the 1D localized 
Wannier state to another unit cell, changing the $X$ index.

Incidentally, we point out that any gauge fixing in the $k_y$ direction at 
$k_x\neq 0$ is futile, as the ``parallel transport'' construction would 
effectively override any gauge choice at $k_x\neq 0$. This is reflected by
the independence of Eq.~\eqref{eq:wannier-gauge-transform} from 
$e^{i\eta(k_x,k_y)}$ with $k_x\neq 0$.

We emphasize that the phase factor $e^{i\Phi_y(X,k_y)}$, or the 
single-particle gauge for $|k_x,k_y\rangle$, do \emph{not} affect the 
localization of the Wannier state, but they do affect essentially the 
similarity with the LLL. We will discuss the choice of 
$e^{i\Phi_y(X,k_y)}$ in Sec.~\ref{sec:phase-fixing}.

\subsection{Connection to the Lowest Landau Level}\label{sec:wannier-j}

Ultimately, we want to bridge the physics of a partially filled band with a 
Chern number $C=\pm 1$ and that of the spinless FQH effects in the LLL on a 
continuum torus.
(Unless noted otherwise, in this paper we specialize to Chern number $|C|=1$.) 

\begin{figure}[tb]
\centering
\includegraphics[width=3.2in]{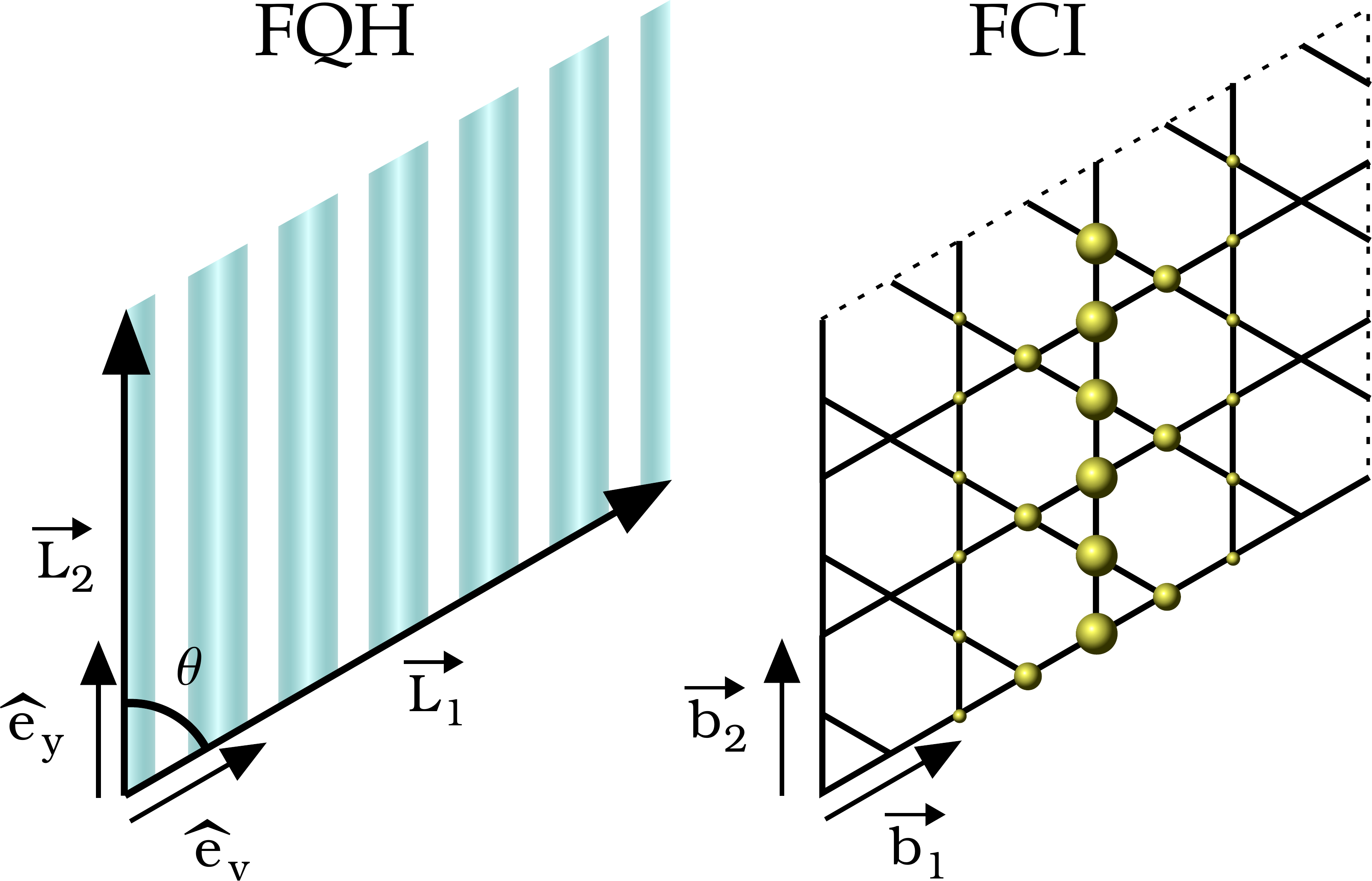}%
\caption{\label{fig:orbitals-y}
Single-particle orbitals in the lowest Landau level (LLL) and the Chern band.
In the left panel we show the Landau orbitals in the LLL on a torus. The two 
fundamental cycles are marked by $\mathbf{L}_1$ and $\mathbf{L}_2$, and the 
twist angle is labeled by $\theta$.
The Landau orbitals are plane waves in the $\hat{e}_y$ direction that are 
localized in the direction perpendicular to propagation. 
In the right panel we show a 1D Wannier state localized in the $\mathbf{b}_1$ 
direction and in the lowest band of the Kagome lattice model. The size of the 
spheres depicts the weights of the Wannier state on each lattice site.
}
\end{figure}

As illustrated in Fig.~\ref{fig:orbitals-y},
at the single-particle level, both the Wannier states $|X,k_y\rangle$ in a 
Chern band and the Landau-gauge orbitals $|j\rangle$ in the LLL are plane 
waves localized in the direction perpendicular to wave propagation.
Building upon the correspondence between the center lines of the localized 
states, Ref.~\onlinecite{Qi11:Wavefunction} proposed a linear relabeling 
of the Wannier states. The resulting 1D label is analogous to the LLL state 
index $j$.
We now examine this relabeling of the Wannier states in more details.
As we will see, the original prescription lacks a few ingredients that are 
essential for a concrete numerical implementation of the scheme.

First, we need to specify the torus geometry in the continuum.
The number of fluxes in the corresponding LLL problem is given 
by $N_\phi=N_xN_y$.~\cite{Qi11:Wavefunction,Bernevig12:Counting}. 
We want the continuum torus to take the same shape as the lattice system with 
periodic boundaries.
The aspect ratio $L_1/L_2$ is set to\footnote{
Arguably we could also use the aspect ratio as a variational parameter. We 
leave this for future work.}
\begin{equation}
\frac{L_1}{L_2}=\frac{N_x|\mathbf{b}_1|}{N_y|\mathbf{b}_2|}.
\end{equation}
The only remaining parameter is the twist angle $\theta$ between $\mathbf{L}_1$ 
and $\mathbf{L}_2$ (defined in Sec.~\ref{sec:FQH}). For brevity,
we want to map both $C=1$ and $C=-1$ Chern bands to the LLL with a magnetic 
field in the $-\hat{e}_z$ direction as studied in Sec.~\ref{sec:FQH}.
For $C=1$, we put $\mathbf{L}_1$ and $\mathbf{L}_2$ in the direction of 
$\mathbf{b}_1$ and $\mathbf{b}_2$, respectively.
For $C=-1$, we need to ``flip'' the fundamental parallelogram. This 
corresponds to choosing the twist angle 
$\theta=\pi-\langle\mathbf{b}_1,\mathbf{b}_2\rangle$, 
where $\langle\mathbf{b}_1,\mathbf{b}_2\rangle\in(0,\pi)$ is the angle between 
$\mathbf{b}_1$ and $\mathbf{b}_2$.
This choice of $\theta$ is justified numerically in Sec.~\ref{sec:delta-y-theta}.

We now make a detailed comparison between the LLL orbitals $|j\rangle$ and the 
Wannier orbitals $|X,k_y\rangle$.
In the LLL, the center coordinate $L_1 j/N_\phi$ of orbital $|j\rangle$ in the 
$\mathbf{b}_1$ direction is a monotonically increasing function of $j$, and 
the $j=0$ orbital is centered along the line $x=0$. 
In the Chern band, the unitary Wilson loop $W_x(k_y)$ is 
a pure phase; when $k_y$ changes, it winds around the unit circle at the 
origin of the complex plane. In the continuum limit, $W_x(k_y)$ has winding 
number $C=\pm 1$ when $k_y$ goes around a single Brillouin zone. Assume for the 
moment that the winding motion of $W_x(k_y)$ on the unit circle is 
\emph{unidirectional} (clockwise when $C=1$ and counterclockwise when $C=-1$) 
when $k_y$ increase from $0$ to $N_y$, i.e. there is no ``zigzag'' pattern of 
going back-and-forth anywhere.
In this case, when the $k_y$ Brillouin zone boundary is 
\emph{properly} chosen, $\mathrm{arg}[W_x(k_y)]\in(-2\pi,0]$ depends 
\emph{monotonically} on $k_y$.

\begin{figure}[tb]
\centering
\includegraphics[width=3.2in]{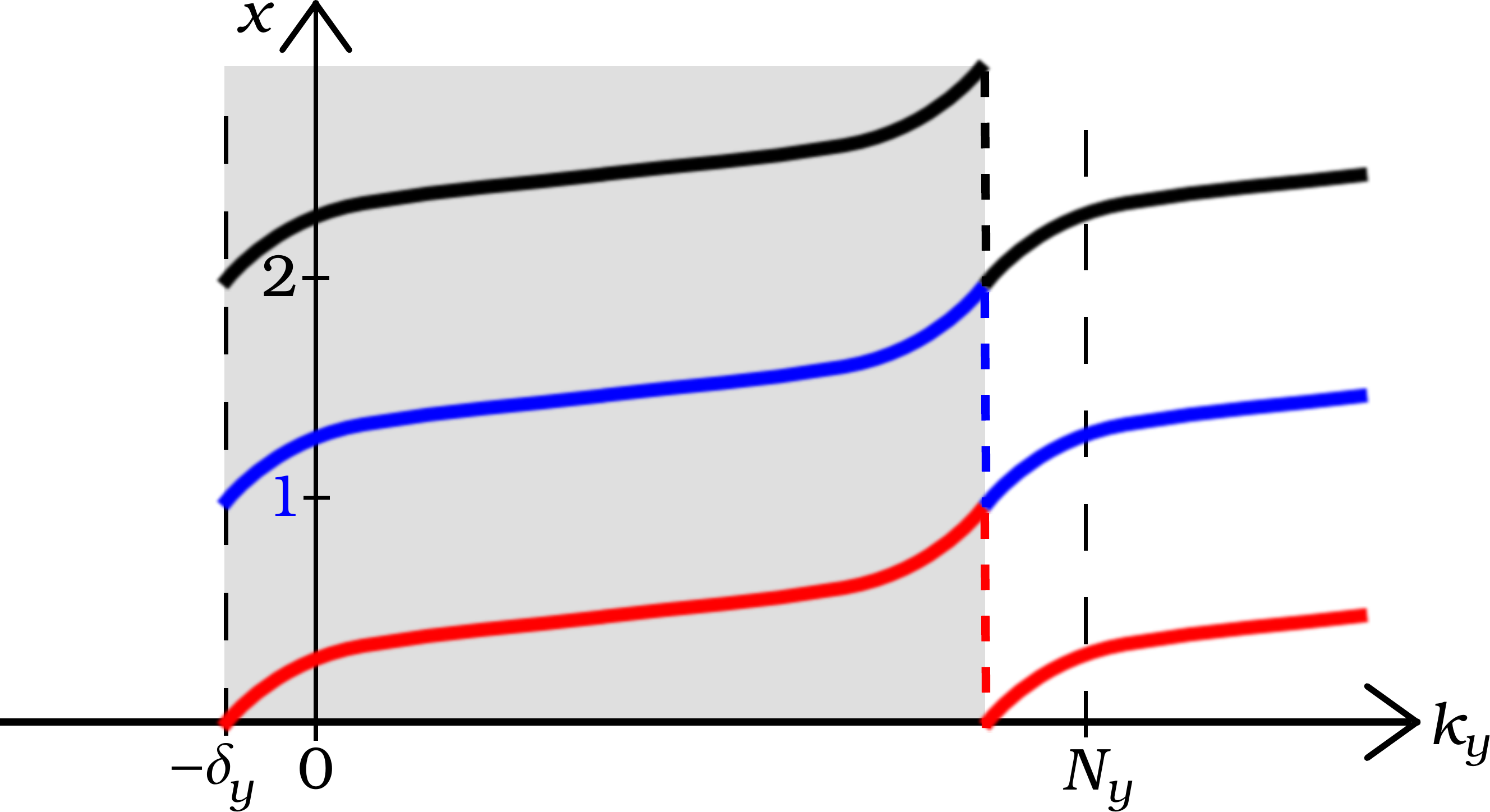}%
\caption{\label{fig:delta-y}
The principal Brillouin zone for $C=+1$. Plotted are the flows of the Wannier 
center position $\chi^{X,k_y}=X-\mathrm{arg}[W_x(k_y)]/(2\pi)$ 
[Eq.~\eqref{eq:wannier-center}] as a function of $k_y$, color-coded according 
to the unit cell index $X$.
The movement of the Wannier center in each unit cell is monotonic in the 
principal Brillouin zone $Ck_y+\delta_y\in\range{N_y}$ (marked by the gray 
shade), but not monotonic in the Brillouin zone $k_y\in\range{N_y}$.
}
\end{figure}

The proper choice of the Brillouin zone is illustrated in 
Fig.~\ref{fig:delta-y}. 
We start the $k_y$ Brillouin zone from the point where 
$W_x(k_y)$ is closest to $1$ in the lower half of the complex plane.
The specific prescription is the following.
We pick the branch of $\mathrm{arg}[W_x(k_y)]$ that takes value in 
$(-2\pi,0]$. 
We define the shift $\delta_y\in\range{N_y}$ as the cardinality of the set
\begin{equation}
\Big\{\,k_y\in\range{N_y}\;\;\Big|\;\;
\mathrm{arg}[W_x(k_y)]>\mathrm{arg}[W_x(0)]\,\Big\}.
\end{equation}
Then, for $Ck_y+\delta_y\in\range{N_y}$, $\mathrm{arg}[W_x(k_y)]$ depends 
\emph{monotonically} on $k_y$; $Ck_y+\delta_y$ labels the Wannier centers in an 
unit cell sequentially in ascending order of the center position $\chi^{X,k_y}$.
We refer to the set of $k_y$ given by
\begin{equation}\label{eq:pBZ}
Ck_y+\delta_y\in\range{N_y}
\end{equation}
as the ``principal Brillouin zone'' (pBZ).
We introduce the 1D label of Wannier states
\begin{equation}\label{eq:j-Xky}
j^{X,k_y}=XN_y+Ck_y+\delta_y
\end{equation}
with $Ck_y+\delta_y\in\range{N_y}$.
We need to shift $k_y$ back to the pBZ \emph{before} 
performing the mapping.
Thanks to the explicit choice of the domain for $k_y$, the 
mapping $(X,k_y)\rightarrow j$ is invertible.
In the presence of ``zigzag'' patterns in $W_x(k_y)$, we stick to the above
$j^{X,k_y}$ formula, and take $\delta_y$ as a variational parameter. This simple 
dependence on $k_y$ is necessary for the translational invariance of the 
FQH-analog wave functions, as we will see in 
Sec.~\ref{sec:translational-inv}.
In the rest of this paper we will focus on the case without ``zigzag'' patterns.

The phase factor $e^{i\Phi_y(X,k_y)}$ is subject to periodicity constraints.
Since in the LLL $|j\rangle=|j+N_\phi\rangle$, the Wannier states 
$|X,k_y\rangle$ and $|X+N_x,k_y\rangle$ are mapped to the same LLL state. This 
mandates the gauge periodicity in $X$, i.e. $|X,k_y\rangle=|X+N_x,k_y\rangle$.
Eq.~\eqref{eq:wannier-eigenstate} then leads to
\begin{equation}\label{eq:Phi-periodicity-x}
e^{i\Phi_y(X,k_y)}=e^{i\Phi_y(X+N_x,k_y)}.
\end{equation}
The periodic gauge of the Bloch state in the $k_y$ direction, i.e. 
$|k_x,k_y\rangle=|k_x,k_y+N_y\rangle$, requires that
\begin{equation}\label{eq:Phi-periodicity-y}
e^{i\Phi_y(X,k_y)}=e^{i\Phi_y(X,k_y+N_y)}.
\end{equation}

It should be noted that the analogy between the Wannier states and the 
LLL states is not exact even in the continuum limit, due to the inevitable
fluctuations in the Berry curvature. For example, if the 
Berry curvature is not uniform in the $y$ direction, the Wannier centers 
$\chi^{X,k_y}$ are not distributed evenly over each unit cell, in contrast to 
the uniform distribution of the LLL orbitals. This is in line with the 
comparison between the LLL and the Chern band made in
Refs.~\onlinecite{Parameswaran11:W-inf,Goerbig12}.

\section{Wannier Construction of FQH States on a Lattice}
\label{sec:wannier-construction}

We now turn to the construction of model wave functions in a fractional Chern 
insulator defined on an $N_x\times N_y$ lattice.
We consider finding model wave functions that approximate the ground states of 
generic density-density interaction with translational invariance, namely
\begin{equation}
V_\mathrm{lat}=\sum_{(\mu,\nu,\boldsymbol{\delta})}\sum_{\mathbf{r}}
\psi_{\mathbf{r}+\boldsymbol{\delta},\mu}^\dagger
\psi_{\mathbf{r},\nu}^\dagger
\psi_{\mathbf{r},\nu}^{\phantom{\dagger}}
\psi_{\mathbf{r}+\boldsymbol{\delta},\mu}^{\phantom{\dagger}},
\end{equation}
Here $\boldsymbol{\delta}$ is the unit cell displacements between neighboring 
orbitals $\mu$ and $\nu$. The sum $(\mu,\nu,\boldsymbol{\delta})$ is over all 
distinct (up to lattice translation) orbital pairs within some interaction 
range.
The Bravais lattice coordinate $\mathbf{r}=(x,y)\in\mathbb{Z}^2$ is summed 
over the principal region of the torus.
We refer the reader to Ref.~\onlinecite{Wu12:Zoology} for a series of
examples with concrete choices of $(\mu,\nu,\boldsymbol{\delta})$.
Longer range interactions can also be implemented, but our experience is that 
they diminish the strength of the FQH-like state.

In the flat-band limit of the single band 
approximation~\cite{Regnault11:FCI,Wu12:Zoology}, the interacting Hamiltonian 
is just the projected density-density interaction $PV_\mathrm{lat}P$:
\begin{multline}\label{eq:Hlat-generic}
H_\mathrm{lat}=
\frac{1}{N_\phi}\sum_{\{\mu,\nu,\boldsymbol{\delta}\}}
\sum_{\mathbf{k}_1\mathbf{k}_2\mathbf{k}_3\mathbf{k}_4}\!\!\!\!
\delta_{\mathbf{k}_1+\mathbf{k}_2,\mathbf{k}_3+\mathbf{k}_4}'\,
e^{-i(\mathbf{k}_1-\mathbf{k}_4)\cdot\boldsymbol{\delta}}\\
\times
u_\mu^*(\mathbf{k}_1)u_\nu^*(\mathbf{k}_2)
u_\nu(\mathbf{k}_3)u_\mu(\mathbf{k}_4)\,
\psi_{\mathbf{k}_1}^\dagger
\psi_{\mathbf{k}_2}^\dagger
\psi_{\mathbf{k}_3}^{\phantom{\dagger}}
\psi_{\mathbf{k}_4}^{\phantom{\dagger}}.\!\!\!\!
\end{multline}
Here $\mathbf{k}_n=(k_{x,n},k_{y,n})$, $n=1,2,3,4$, are summed over the first 
Brillouin zone, and the primed Kronecker-$\delta$ allows umklapp processes 
$\mathbf{k}_1+\mathbf{k}_2=\mathbf{k}_3+\mathbf{k}_4$ mod $(N_x,N_y)$.

Based on the physics of the FQH, we expect that the Hamiltonian in 
Eq.~\eqref{eq:Hlat-generic} could have a topological ground state at filling 
$\nu=1/q$ that resembles the Laughlin state.
The Laughlin state expansion in non-interacting many-body states is known and 
hence all we need for a model FCI state is an appropriate map between the FCI 
single-particle orbitals and those of the LLL.
The relevance of these trial wave functions to the actual FCI ground state 
is demonstrated in Sec.~\ref{sec:results}.
We now examine the details of this construction.

For a flat-band single-particle lattice Hamiltonian,
the many-body Hilbert space is spanned by the Slater determinant
states $|\{X,k_y\}\rangle$. The 1D index $j^{X,k_y}$ 
defined in Eq.~\eqref{eq:j-Xky} provides a formal mapping between the 
many-body basis states $|\{X,k_y\}\rangle$ on the lattice and 
$|\{j\}\rangle$ in the continuum. The lattice analogue of 
$|\Psi;s,r\rangle$ states can be constructed:
\begin{equation}\label{eq:lat-state}
|\Psi;s,r\rangle_\mathrm{lat}
=\sum_{\{X,k_y\}}|\{X,k_y\}\rangle\,
\langle\{j^{X,k_y}\}|\Psi;s,r\rangle,
\end{equation}
where $\langle\{j^{X,k_y}\}|\Psi;s,r\rangle$ are the FQH amplitudes of the 
states defined in Eq.~\eqref{eq:psi-s-r},
and $\{X,k_y\}$ is summed over all $N_e$-particle configurations 
in the Wannier basis.

Naively, this seems to be the end of the story.
However, the state $|\Psi;s,r\rangle_\mathrm{lat}$ defined above is \emph{not} 
covariant under a single-particle gauge transform on $|X,k_y\rangle$, as the 
continuum states $|j\rangle$ do \emph{not} transform accordingly. We have to 
fix the phase $e^{i\Phi_y(X,k_y)}$ of the Wannier states $|X,k_y\rangle$ in 
conformity with the phase of the $|j\rangle$ states. This is not 
surprising: following Ref.~\onlinecite{Qi11:Wavefunction}, up to now we have 
only established a mapping between the state \emph{labels} $(X,k_y)$ and $j$, 
rather than a mapping between the actual \emph{states}.
In the following, we seek the guidelines for choosing $e^{i\Phi_y(X,k_y)}$ 
based on the similarity with the LLL.
If we do not properly fix the $e^{i\Phi_y(X,k_y)}$ phase, the overlaps with 
the exact ground states can be tiny.

\subsection{Connection between Wannier States}

For the LLL orbitals $|j\rangle$ defined in Eq.~\eqref{eq:LLL-wf}, 
we find in Appx.~\ref{sec:FQH-Hamiltonian} that the adjacent orbitals satisfy
\begin{equation}\label{eq:j-connection}
\langle j|e^{-i\mathbf{G}_2\cdot\widetilde{\mathbf{r}}}|j+1\rangle
=e^{-|\mathbf{G}_2|^2l_B^2/4}\in\mathbb{R}_+.
\end{equation}
Here, $\mathbf{G}_2=2\pi(\hat{e}_y-\cot\theta\,\hat{e}_x)/L_2$ is the reciprocal 
lattice vector defined in Eq.~\eqref{eq:reciprocal-G} in Sec.~\ref{sec:FQH}, 
and the position operator $\widetilde{\mathbf{r}}$ takes values in the continuum.
We can interpret $e^{-i\mathbf{G}_2\cdot\widetilde{\mathbf{r}}}$ as a translation by 
$-\mathbf{G}_2$ in the momentum space.

This condition spells out the gauge choice of the LLL orbitals.
In the continuum limit, it gives the parallel transport between the $|j\rangle$ 
and $|j+1\rangle$ orbitals.
We need to find a consistent gauge for the Wannier states.
Recall that $\widehat{Y}$ implements translation in the momentum space along $k_y$, 
which is the direction reciprocal to $\mathbf{b}_2$.
Therefore, on a lattice, the quantity analogous to 
$\langle j|e^{-i\mathbf{G}_2\cdot\widetilde{\mathbf{r}}}|j+1\rangle$ is the 
connection between adjacent Wannier states
\begin{equation}
\langle X,k_y|\widehat{Y}|X',k_y'\rangle.
\end{equation}
Here, in the same spirit as the orthogonality fix for $|X,k_y\rangle$, we use 
the \emph{unitary} projected position operator $\widehat{Y}$, defined in 
Eq.~\eqref{eq:unitary-Y}, and $(X',k_y')$ is the label of the Wannier state 
\emph{next} to $|X,k_y\rangle$, defined by
\begin{equation}\label{eq:jprime}
j^{X',k_y'}=j^{X,k_y}+C.
\end{equation}
Plugging in the definition of $j^{X,k_y}$ in Eq.~\eqref{eq:j-Xky}, we find 
$k_y'=k_y+1$ mod $N_y$.
If going from $k_y$ to $k_y+1$ crosses the boundary of 
the principal Brillouin zone [Eq.~\eqref{eq:pBZ}], the 
Wannier centers $(X,k_y)$ and $(X',k_y')$ are located in different 
unit cells ($X'=X+C$), otherwise they are in the same unit cell ($X'=X$).
Notice that for $|C|>1$, Eq.~\eqref{eq:jprime} breaks the $N_xN_y$ Wannier 
states into $|C|$ groups, and it implies a $|C|$-layer FQH 
analogy.~\cite{Barkeshli11:Nematic}

\begin{figure}[tb]
\centering
\includegraphics[width=2.7in]{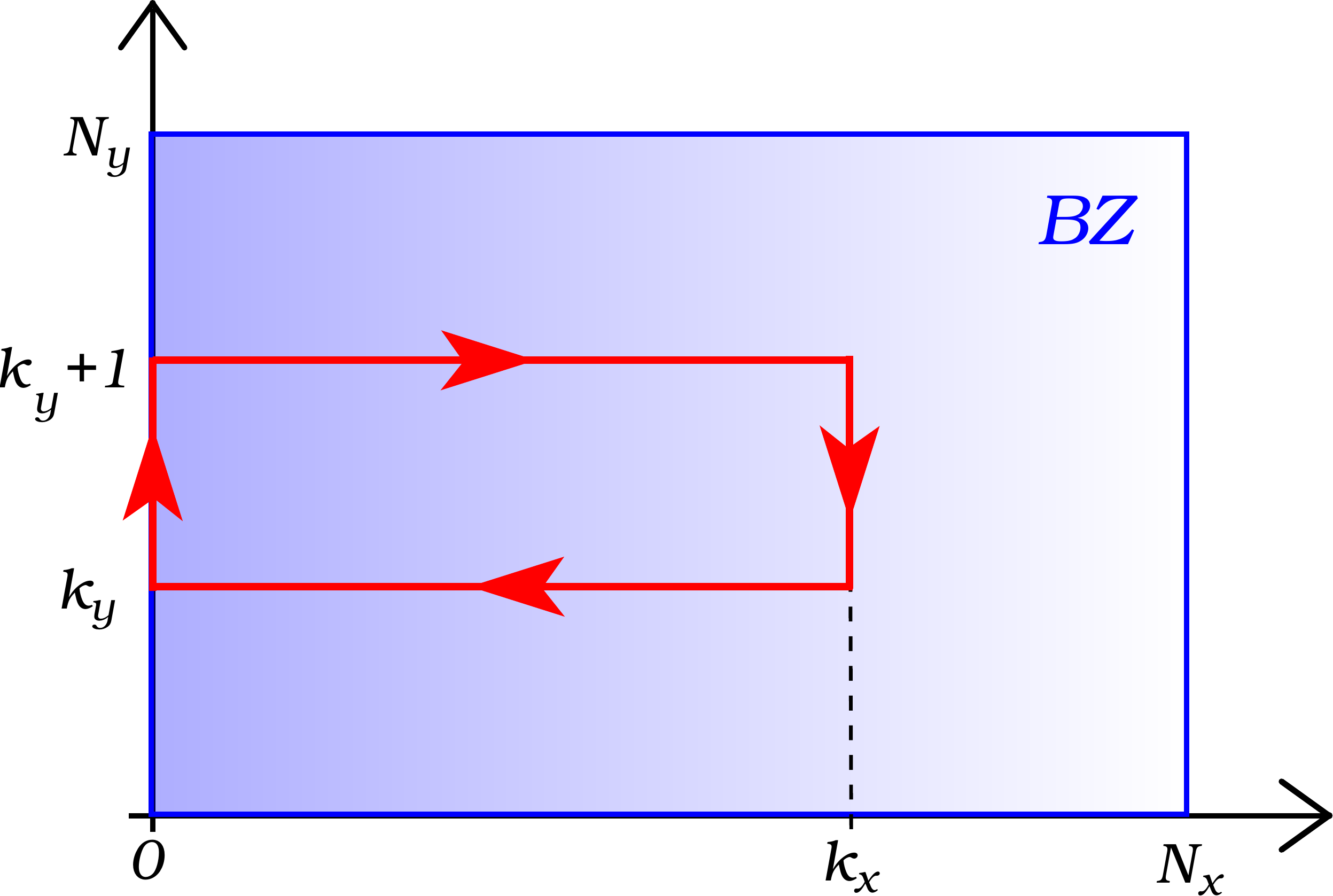}%
\caption{\label{fig:loop}
The loop $(0,k_y) {(0,k_y+1)} {(k_x,k_y+1)} (k_x,k_y) (0,k_y)$ in the 
$N_x\times N_y$ lattice Brillouin zone.
The Wilson loop around this path is defined in Eq.~\eqref{eq:W-rect} as
$W_{\protect\rule{5pt}{2pt}}(k_x,k_y)$. Notice that at $k_x=N_x$, the 
connections at the two vertical edges cancel each other due to periodic 
boundary, leading to $W_{\protect\rule{5pt}{2pt}}(N_x,k_y)=W_x(k_y)/W_x(k_y+1)$.
}
\end{figure}

We can calculate the Wannier connection 
$\langle X,k_y|\widehat{Y}|X',k_y'\rangle$ by expanding the Wannier states in 
Bloch basis using Eq.~\eqref{eq:wannier-eigenstate}.
Since the operator $\widehat{Y}$ is diagonal in $k_x$, the two Bloch basis 
expansions is reduced into a single sum over $k_x$.
Consider a generic term $k_x$ in the sum.
The $A_x$ factors from the Wannier states and the $A_y$ factor from 
$\widehat{Y}$ can be collected into a Wilson line going along 
$(0,k_y+1)(k_x,k_y+1)(k_x,k_y)(0,k_y)$, which equals $A_y(0,k_y)$ times the 
Wilson loop around the path shown in Fig.~\ref{fig:loop}.
The $\lambda_x$ factors from the Wannier states are 
$[\lambda_x(k_y+1)/\lambda_x(k_y)]^{k_x}$.
If the two Wannier states are not in the same unit cell, they must be in 
adjacent unit cells $X'=X+C$ [due to Eq.~\eqref{eq:jprime}], and we have 
another factor $e^{i2\pi k_x C/N_x}$.

Putting all these together, for $(X',k_y')$ defined in Eq.~\eqref{eq:jprime},
we have the expression
\begin{equation}\label{eq:wannier-connection}
\langle X,k_y|\widehat{Y}|X',k_y'\rangle
=\frac{e^{i\Phi_y(X',k_y')}}{e^{i\Phi_y(X,k_y)}}
A_y(0,k_y)\,\mathcal{U}_y(k_y),
\end{equation}
where the gauge-invariant quantity $\mathcal{U}_y(k_y)$ is defined as
\begin{equation}\label{eq:U-def}
\mathcal{U}_y(k_y)=\frac{1}{N_x}\sum_{k_x}^{N_x}
\frac{W_{\protect\rule{5pt}{2pt}}(k_x,k_y)}{\overline{W}_{\protect\rule{5pt}{2pt}}(k_x,k_y)}.
\end{equation}
Here, $W_{\protect\rule{5pt}{2pt}}(k_x,k_y)$ is the unitary Wilson loop in 
Fig.~\ref{fig:loop},
\begin{equation}\label{eq:W-rect}
W_{\protect\rule{5pt}{2pt}}(k_x,k_y)=
\frac{\prod_\kappa^{k_x}A_x(\kappa,k_y)}{\prod_\kappa^{k_x}A_x(\kappa,k_y+1)}
\frac{A_y(k_x,k_y)}{A_y(0,k_y)},
\end{equation}
while $\overline{W}_{\protect\rule{5pt}{2pt}}(k_x,k_y)=[\mu_x(k_y)]^{k_x}$ is 
given by $\mu_x(k_y)$ defined over the principal Brillouin zone [pBZ, the set 
of $\kappa$ satisfying $C\kappa+\delta_y\in\range{N_y}$],
\begin{equation}\label{eq:mu}
\mu_x(k_y)=
\begin{cases}
\displaystyle \frac{\lambda_x(k_y)}{\lambda_x(k_y+1)}
& \text{if }k_y+1\in\mathrm{pBZ},\\
\displaystyle e^{i2\pi C/N_x}\frac{\lambda_x(k_y)}{\lambda_x(k_y+1)}
& \text{otherwise.}
\end{cases}
\end{equation}
We emphasize that the above definition requires first shifting $k_y$ back to 
the pBZ.
The two cases in Eq.~\eqref{eq:mu} correspond to whether or not going from 
$k_y$ to $k_y+1$ crosses the boundary of the pBZ.

\begin{figure}[tb]
\centering
\includegraphics[width=2.7in]{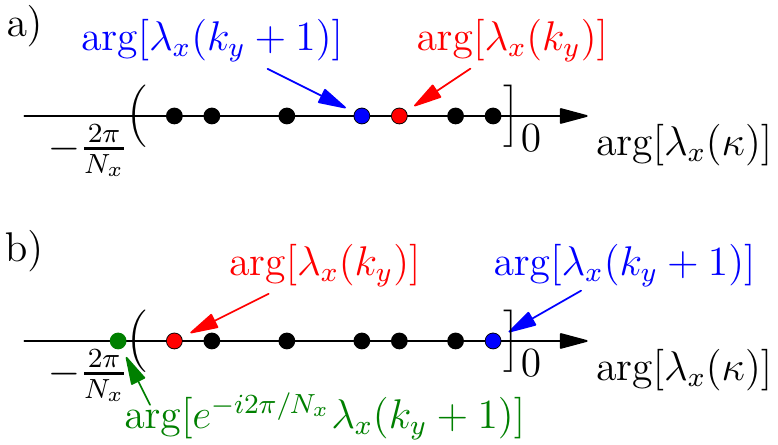}%
\caption{\label{fig:mu}
Flow of $\mathrm{arg}[\lambda_x(\kappa)]$ in the interval $(-2\pi/N_x,0]$.
The solid dots represents the $N_y$ values of $\mathrm{arg}[\lambda_x(\kappa)]$ for 
the $N_y$ Wannier centers in each unit cell. Here we show the case of $C=+1$: 
$\mathrm{arg}[\lambda_x(\kappa)]$ is a monotonically decreasing function of 
$\kappa$ in the pBZ.
Panel a) When $k_y$ and $k_y+1$ can be put in the 
principal Brillouin zone simultaneously, we have 
$\mathrm{arg}[\lambda_x(k_y)/\lambda_x(k_y+1)]\in(0,2\pi/N_x)$.
Panel b) When going from $k_y$ to $k_y+1$ crosses the boundary of the principal 
Brillouin zone, we have 
$\mathrm{arg}[e^{i2\pi/N_x}\lambda_x(k_y)/\lambda_x(k_y+1)]\in(0,2\pi/N_x)$.
Summarizing both cases, we have $\mathrm{arg}[\mu_x(k_y)]\in(0,2\pi/N_x)$, 
for $\mu_x(k_y)$ defined in Eq.~\eqref{eq:mu}.
}
\end{figure}

We now try to understand $\mu_x(k_y)$ and 
$\overline{W}_{\protect\rule{5pt}{2pt}}(k_x,k_y)$ physically.
Without loss of generality, we consider the case of $C=+1$.
Since $[\lambda_x(k_y)]^{N_x}=W_x(k_y)$ [Eq.~\eqref{eq:lambda-W}], we have for 
both cases of Eq.~\eqref{eq:mu},
\begin{equation}\label{eq:mu-W}
[\mu_x(k_y)]^{N_x}
=\left[\frac{\lambda_x(k_y)}{\lambda_x(k_y\!+\!1)}\right]^{N_x}
=W_{\protect\rule{5pt}{2pt}}(N_x,k_y).
\end{equation}
Now we need to take the $N_x$-th root of this equation.
Special care needs to be taken with the branch choice.
Recall from Sec.~\ref{sec:orthogonality} that $\mathrm{arg}[\lambda_x(\kappa)]$ 
lies in $(-2\pi/N_x,0]$ by definition.
Since we assume the absence of ``zigzag'' patterns in the winding of $W_x(\kappa)$, 
the motion of $\mathrm{arg}[\lambda_x(\kappa)]$ in the interval $(-2\pi/N_x,0]$ 
must be monotonic when $\kappa\in\mathrm{pBZ}$, and 
$\mathrm{arg}[\lambda_x(\kappa)]$ jumps from the left boundary of the interval 
to the right boundary when $\kappa$ crosses the boundary of the pBZ.
As illustrated in Fig.~\ref{fig:mu}, for all values of $k_y$, we have
$\mathrm{arg}[\mu_x(k_y)]\in(0,2\pi/N_x)$.
On the other hand, since the phase angle of the Wilson loop is given by the 
curvature enclosed in the loop, the argument angle 
$\mathrm{arg}[W_{\protect\rule{5pt}{2pt}}(N_x,k_y)]\in[0,2\pi)$ is the total curvature 
enclosed in the $N_x$ plaquettes in the row between $k_y$ and $k_y+1$.
Then, the $N_x$-th root of Eq.~\eqref{eq:mu-W} gives
\begin{equation}\label{eq:mu-arg}
\mathrm{arg}[\mu_x(k_y)]=\frac{1}{N_x}\mathrm{arg}[W_{\protect\rule{5pt}{2pt}}(N_x,k_y)],
\end{equation}
Therefore the argument angle 
$\mathrm{arg}[\overline{W}_{\protect\rule{5pt}{2pt}}(k_x,k_y)]=k_x\,\mathrm{arg}[\mu_x(k_y)]$
is $k_x/N_x$ times the total curvature enclosed in the $N_x$ plaquettes in the row between 
$k_y$ and $k_y+1$.
When the curvature is constant across the $N_x$ plaquettes at $k_y$, 
we have $\overline{W}_{\protect\rule{5pt}{2pt}}(k_x,k_y)=W_{\protect\rule{5pt}{2pt}}(k_x,k_y)$. 
The phase of each term in $\mathcal{U}_y(k_y)$
[Eq.~\eqref{eq:U-def}] can hence be interpreted as a measure of curvature 
fluctuations in the $k_x$ direction.

\subsection{Phase Fixing: Explicit Prescription}\label{sec:phase-fixing}

Compared with 
$\langle j|e^{-i\mathbf{G}_2\cdot\widetilde{\mathbf{r}}}|j+1\rangle \in\mathbb{R}_+$
in the LLL [Eq.~\eqref{eq:j-connection}],
a major difference in the Chern band is that the connection 
$\langle X,k_y|\widehat{Y}|X',k_y'\rangle$ between the Wannier states \emph{cannot}
be gauge fixed to a real number.
To see this, we take the product of 
Eq.~\eqref{eq:wannier-connection} over all $(X,k_y)$. Thanks to the 
periodicity of $e^{i\Phi_y(X,k_y)}$ [Eqs.~\eqref{eq:Phi-periodicity-x} 
and~\eqref{eq:Phi-periodicity-y}], the product takes a simple form:
\begin{equation}\label{eq:prod-Wannier-connection}
\prod_X^{N_x}\prod_{k_y}^{N_y}\langle X,k_y|\widehat{Y}|X',k_y'\rangle
=\left[W_y(0)\prod_{k_y}^{N_y}\mathcal{U}_y(k_y)\right]^{N_x},
\end{equation}
Without further constraints from symmetry, the expression on the right hand 
side is a complex, rather than a positive real number. Even if we ignore the 
curvature fluctuations, in general we still \emph{cannot} make all 
$\langle X,k_y|\widehat{Y}|X',k_y'\rangle\in\mathbb{R}_+$. This is fundamentally 
different from the LLL [Eq.~\eqref{eq:j-connection}].

Hence, we relax this condition and look for $e^{i\Phi_y(X,k_y)}$ that makes 
the \emph{phase} of $\langle X,k_y|\widehat{Y}|X',k_y'\rangle$
\emph{independent} from $X$ and $k_y$.~\footnote{The absolute value of 
$\langle X,k_y|\widehat{Y}|X',k_y'\rangle$ does not depend on the choice of 
the phase $e^{i\Phi_y(X,k_y)}$. It is given by $|\,\mathcal{U}_y(k_y)|$.}
This phase is given by the $N_\phi$-th root of the right hand side of 
Eq.~\eqref{eq:prod-Wannier-connection}.
Define the phase $U_y(k_y)=\mathcal{U}_y(k_y)/|\,\mathcal{U}_y(k_y)|$.
Similar to the phase $\lambda_y(k_x)$ for the connection $A_y(k_x,k_y)$
[Eq.~\eqref{eq:lambda-W}], we can define a phase $\omega_y$ with argument 
angle in $(-\pi/N_y,\pi/N_y]$ by
\begin{equation}\label{eq:omega-y}
(\omega_y)^{N_y}=\prod_\kappa^{N_y}U_y(\kappa).
\end{equation}
Then, to have the phase of $\langle X,k_y|\widehat{Y}|X',k_y'\rangle$ 
independent from $X$ and $k_y$, we can use
\begin{equation}\label{eq:wannier-connection-final}
\langle X,k_y|\widehat{Y}|X',k_y'\rangle
=\lambda_y(0)\,\omega_y\,|\,\mathcal{U}_y(k_y)|.
\end{equation}
Here, we have picked a specific branch of the $N_\phi$-th root of (the phase 
of) Eq.~\eqref{eq:prod-Wannier-connection}.
We could have picked another gauge by putting on the right hand side of 
Eq.~\eqref{eq:wannier-connection-final} an extra factor $e^{i2\pi w/N_\phi}$, 
with $w\in\range[(]{N_\phi}$. However, this makes no difference to the resulting set 
of many-body states $\{|\Psi;s,r\rangle_\mathrm{lat}\}$ 
[Eq.~\eqref{eq:lat-state}]. The reason is the following.
Without loss of generality, we consider the effect of this extra factor 
$e^{i2\pi w/N_\phi}$ in the case with $C>0$.
When we gauge fix the Wannier states one by one, using 
Eq.~\eqref{eq:wannier-connection-final} but with an extra $e^{i2\pi w/N_\phi}$ 
between each pair of adjacent Wannier states,
the Wannier state $|X,k_y\rangle$ receives an extra factor 
$e^{i2\pi w j^{X,k_y}/N_\phi}$.
Therefore, the many-body state in Eq.~\eqref{eq:lat-state} becomes
\begin{multline}\label{eq:lat-state-w}
|\Psi;s,r\rangle_\mathrm{lat}^w
=\sum_{\{X,k_y\}}|\{X,k_y\}\rangle\\
e^{i2\pi w\sum j^{X,k_y}/N_\phi}
\langle\{j^{X,k_y}\}|\Psi;s,r\rangle,
\end{multline}
Plugging in Eq.~\eqref{eq:psi-s-shift}, we find
\begin{equation}
|\Psi;s,r\rangle_\mathrm{lat}^w=e^{i2\pi w(\kappa_y+rN_e)/N_\phi}
|\Psi;s+w,r\rangle_\mathrm{lat}.
\end{equation}
The $w$-dependence is reduced to an overall phase factor, and a shift in the 
$s$ index, which can be absorbed by a reshuffling of $s$.
Therefore, we can safely set $w$ to zero, and use 
Eq.~\eqref{eq:wannier-connection-final} to fix the phase of the Wannier states.

Plugging this prescription into Eq.~\eqref{eq:wannier-connection}, we find
\begin{equation}\label{eq:Phi-recursive}
e^{i\Phi_y(X',k_y')-i\Phi_y(X,k_y)}=
\frac{\lambda_y(0)}{A_y(0,k_y)}\frac{\omega_y}{U_y(k_y)}.
\end{equation}
Notices that the quantities on the right hand side are all unitary.
Together with the initial condition $e^{i\Phi_y(0,0)}=1$, 
Eq.~\eqref{eq:Phi-recursive} recursively specifies the choice of 
$e^{i\Phi_y(X,k_y)}$ for all the $N_\phi$ states.
Thanks to Eqs.~\eqref{eq:lambda-W} and~\eqref{eq:omega-y}, this choice
of $e^{i\Phi_y(X,k_y)}$ does not depend on $X$, and we can drop the $X$ 
argument and write $e^{i\Phi_y(k_y)}$ instead.
We note that any choice of $e^{i\Phi_y(k_y)}$ without an $X$-dependence 
\emph{that is periodic in $k_y$} can be achieved by simply modifying the gauge 
of the single-particle Bloch states along $k_x=0$. This changes the phase of 
the Wannier states by $e^{i\eta(0,k_y)}$ as shown in 
Eq.~\eqref{eq:wannier-gauge-transform}.

\subsection{Translational Invariance}\label{sec:translational-inv}

We now examine the translational symmetry of the many-body states 
$|\Psi;s,r\rangle$.
For the moment we do not specialize to the phase choice in 
Eq.~\eqref{eq:Phi-recursive} and consider the constraint from translational 
invariance on a generic $e^{i\Phi_y(X,k_y)}$.

Define the $N_e$-particle center-of-mass translation operators on the lattice
\begin{equation}
\begin{aligned}
T_\mathrm{lat,cm}^x&=\sum_{\{x,y,\alpha\}}
|\{x+1,y,\alpha\}\rangle\langle\{x,y,\alpha\}|,\\
T_\mathrm{lat,cm}^y&=\sum_{\{x,y,\alpha\}}
|\{x,y+1,\alpha\}\rangle\langle\{x,y,\alpha\}|.
\end{aligned}
\end{equation}

\begin{widetext}
The action on the wave function of the many-body lattice state 
$|\Psi;s,r\rangle_\mathrm{lat}$ defined in Eq.~\eqref{eq:lat-state} is found 
to be
\begin{align}
\label{eq:lat-sym-x-1}
\langle\{X,k_y\}|T_\mathrm{lat,cm}^x|\Psi;s,r\rangle_\mathrm{lat}
&=e^{i\sum[\Phi_y(X,k_y)-\Phi_y(X-1,k_y)]}
\langle\{X-1,k_y\}|\Psi;s,r\rangle_\mathrm{lat},\\
\label{eq:lat-sym-y}
\langle\{X,k_y\}|T_\mathrm{lat,cm}^y|\Psi;s,r\rangle_\mathrm{lat}
&=e^{-i2\pi\sum k_y/N_y}
\langle\{X,k_y\}|\Psi;s,r\rangle_\mathrm{lat},
\end{align}
where $\sum$ means summation over all the $N_e$ phases of different quantum 
numbers $\{X,k_y\}$ of the $N_e$ electron wave function.
The translational invariance in the $y$ direction is already apparent:
Eq.~\eqref{eq:FQH-wf-sym-j} dictates that the non-vanishing components of 
$\langle\{X,k_y\}|\Psi;s,r\rangle_\mathrm{lat}$ have the same value of 
$\sum k_y=C[\kappa_y+(r-\delta_y)N_e]\text{ mod }N_y$. This is the total 
momentum $K_y$.
The prefactor in Eq.~\eqref{eq:lat-sym-y} is the same for all the 
non-vanishing components, i.e. the wave function is translationally invariant 
in the $y$ direction.
The situation in the $x$ direction is more involved. Plugging 
Eqs.~\eqref{eq:FQH-wf-sym-x} and~\eqref{eq:j-Xky} into 
Eq.~\eqref{eq:lat-sym-x-1}, we have 
\begin{equation}\label{eq:lat-sym-x-2}
\langle\{X,k_y\}|T_\mathrm{lat,cm}^x|\Psi;s,r\rangle_\mathrm{lat}
=e^{-i2\pi (\kappa_x+sN_e)/N_x}
e^{i\sum[\Phi_y(X,k_y)-\Phi_y(X-1,k_y)]}
\langle\{X,k_y\}|\Psi;s,r\rangle_\mathrm{lat}.
\end{equation}
\end{widetext}
In general, the exponential prefactor is different for each component (Slater 
determinants of different sets $\{X,k_y\}$), 
spoiling the translational invariance.
To restore this symmetry, $e^{i\Phi_y(X,k_y)-i\Phi_y(X-1,k_y)}$
has to be \emph{independent} from $X$ and $k_y$. 
Translational invariance is part of the main reason why we also have asked for 
the phase of $\langle X,k_y|\widehat{Y}|X',k_y'\rangle$ to be independent from 
$X$ and $k_y$ in the previous section.
The periodic boundary condition on 
$e^{i\Phi_y(X,k_y)}$ then guarantees that the exponential prefactor is an $N_x$-th 
root of unity, and thus the state recovers translational invariance in the $x$ 
direction.

\subsection{Total Momenta and the Folding Picture}\label{sec:folding}

We now consider the phase choice given in Sec.~\ref{sec:phase-fixing}.
Plugging Eq.~\eqref{eq:Phi-recursive} into Eq.~\eqref{eq:lat-sym-x-2}, we 
have 
\begin{multline}
\langle\{X,k_y\}|T_\mathrm{lat,cm}^x|\Psi;s,r\rangle_\mathrm{lat}\\
=e^{-i2\pi (\kappa_x+sN_e)/N_x}
\langle\{X,k_y\}|\Psi;s,r\rangle_\mathrm{lat},
\end{multline}
where the relative momentum $\kappa_x$ of the FQH state $|\Psi\rangle$ is 
defined by the eigenvalue of the relative translation operator 
$T_\mathrm{rel}^x$ in Eq.~\eqref{eq:kappa-x-y}.
Hence the FQH-analogue many-body wave functions 
$\langle\{X,k_y\}|\Psi;s,r\rangle_\mathrm{lat}$ constructed through 
the Wannier functions using the above prescription are indeed translationally 
invariant. The total momentum of the $N_e$ particles in state 
$|\Psi;s,r\rangle_\mathrm{lat}$ is given by the wave numbers 
\begin{equation}\label{eq:lattice-K}
\begin{aligned}
K_x&=\kappa_x+sN_e\text{ mod }N_x,\\
K_y&=C[\kappa_y+(r-\delta_y)N_e]\text{ mod }N_y.
\end{aligned}
\end{equation}

Ref.~\onlinecite{Bernevig12:Counting} obtained the counting for FCI 
model states from the corresponding counting for FQH model states by first 
folding the $N\times N$ relative Brillouin zone down to $N_{0x}\times N_{0y}$, and 
then unfolding to the $N_x\times N_y$ lattice Brillouin zone. 
We now show that our procedure precisely reproduces this folding picture. 

First consider the case $C=+1$.
Recall that $\mathrm{GCD}(N_e,N_x)=N_{0x}$,
$N_x=q_xN_{0x}$, $\mathrm{GCD}(N_e,N_y)=N_{0y}$, $N_y=q_yN_{0y}$. For 
$s\in\range{q_x}$, the $q_x$ values of $K_x$ are all 
distinct and can be written as $\kappa_x+t_xN_{0x}$ mod $N_x$ with 
$t_x\in\range{q_x}$. This clearly implements the folding rule in the $x$ 
direction. In the $y$ direction, for $r\in\range{q/q_x}$, the $q/q_x$ values 
of $K_y$ form nothing but a $q/(q_xq_y)$-fold replica of the $q_y$ values of 
$\kappa_y+t_yN_{0x}$ mod $N_y$ with $t_y\in\range{q_y}$. 
This corresponds to the folding rule in the $y$ direction, producing $q/(q_xq_y)$ 
states in each momentum sector.
The shift parameter $\delta_y$ can be absorbed into $r$ and thus only 
reshuffles the order of $K_y$ values.
For the case of $C=-1$, the minus sign from $C$ in $K_y$ 
[Eq.~\eqref{eq:lattice-K}] can be absorbed as a reshuffling of $r$. This is 
explained in details later, in Eq.~\eqref{eq:r-rbar}.
The results are thus in full agreement with 
Ref.~\onlinecite{Bernevig12:Counting}.

In Ref.~\onlinecite{Wang12:MR}, the generalized Pauli 
principle~\cite{Bernevig08:Jack,Bernevig08:Jack2} was invoked 
through the Wannier mapping~\cite{Qi11:Wavefunction} to determine the 
\emph{total} number of Moore-Read FCI quasihole states. Using our updated 
formalism, the number of quasihole states in each momentum sector could be 
found as well, and it is in agreement with the earlier results obtained in 
Ref.~\onlinecite{Wang12:MR}.

\subsection{Many-Body Amplitudes in the Bloch Basis}

We are now in a position to give the final formula for the 
amplitudes of the many-body lattice states constructed from the FQH states 
$|\Psi;s,r\rangle$,
\begin{widetext}
\begin{align}\label{eq:bloch-amplitude}
\langle\{k_x,k_y\}|\Psi;s,r\rangle_\mathrm{lat}
=\prod\left\{
\frac{(\omega_y)^{k_y}}{\prod_\kappa^{k_y}U_y(\kappa)}
\frac{[\lambda_y(0)]^{k_y}}
{\prod_\kappa^{k_y}A_y(0,\kappa)}
\frac{[\lambda_x(k_y)]^{k_x}}{\prod_\kappa^{k_x}A_x(\kappa,k_y)}
\right\}
\frac{1}{\sqrt{N_x}^{N_e}}\sum_{\{X\}}
e^{-i2\pi\sum k_x X/N_x}
\langle\{j^{X,k_y}\}|\Psi;s,r\rangle.
\end{align}
Here the product outside the curly braces is over the $(k_x,k_y)$ 
configurations of the $N_e$ particles, and 
$\langle\{j^{X,k_y}\}|\Psi;s,r\rangle$ are the many-body amplitudes of the 
recombined FQH states [Eq.~\eqref{eq:psi-s-r}].
\end{widetext}
The above formula is the central result of this paper.
This prescription is not limited to a specific model wave function.
It applies to \emph{any FQH state} that can be expressed in a second-quantized 
basis; a real-space wave function is not actually needed.
We note that the absolute value of each component in the Bloch basis does not 
depend on the gauge choice of the Wannier states, thanks to the $X$-independence 
of $e^{i\Phi_y(X,k_y)}$.

\section{Inversion Symmetry}\label{sec:inversion}

Although not a vital element of a topological phase, the inversion symmetry is 
respected by FQH model wave-functions. Many Chern insulator 
models are inversion symmetric as well.
For such models, the acceptable many-body wave functions for a \emph{featureless}
liquid state must respect inversion symmetry.
To this end, in the following we specialize to inversion-symmetric Chern 
insulators and show that our construction always preserves inversion symmetry 
on the many-body level.

\subsection{Inversion Symmetry of the FQH States}\label{sec:FQH-inversion}

First we briefly discuss the FQH inversion symmetry in the continuum.
Inversion is implemented by the unitary operator $\mathcal{P}$:
\begin{equation}
\mathcal{P}|\widetilde{x},\widetilde{y}\rangle
=|{-\widetilde{x}},-\widetilde{y}\rangle. 
\end{equation}
Evidently $\mathcal{P}^2=1$.
Using the definition of $\langle \widetilde{x},\widetilde{y}|j\rangle$ in 
Eq.~\eqref{eq:LLL-wf}, we 
can show that the annihilation operator $\Psi_j$ of state $|j\rangle$ satisfies 
$\mathcal{P}\Psi_j\mathcal{P}=\Psi_{-j}$, and thus the occupation-number basis 
states satisfy
\begin{equation}\label{eq:inv-j}
\mathcal{P}|\{j\}\rangle=|\{{-j}\}\rangle.
\end{equation}
It is straight-forward to show that the FQH Hamiltonian $H$ commutes with 
$\mathcal{P}$, and that inversion flips the direction of translation operators,
\begin{equation}\label{eq:inv-T}
\mathcal{P}T\mathcal{P}=T^\dagger,
\end{equation}
where $T$ can be any of single-particle or many-body translation operator, 
such as $T_\mathrm{cm}^x$, $S_x$, or $R_y$.

Inversion does not commute with the relative translation operators. Therefore, 
in general the operator $\mathcal{P}$ mixes different center-of-mass 
multiplets. However, the ground state energy level in an Abelian topological 
phase is populated exclusively by the $q$-fold degenerate ground state, and
the operator $\mathcal{P}$ must operate within the $q$-fold 
center-of-mass multiplet, since $[H,\mathcal{P}]=0$.

Consider the action of $\mathcal{P}$ on the state $|\Psi\rangle$ that we pick 
from the $q$-fold ground states. This state has total momentum 
$(\kappa_x,\kappa_y)$. By the above argument, $\kappa_x$ must be inversion 
symmetric, while $\kappa_y$ changes to $-\kappa_y$ under inversion.
Inversion on $|\Psi\rangle$ can be compensated by a center-of-mass translation.
We can find a unique $t\in\range{q}$ such that
\begin{equation}
tN_e=-2\kappa_y\text{ mod }N_\phi.
\end{equation}
Then, $|\Psi\rangle$ must be an eigenstate of $\mathcal{P}(T_\mathrm{cm}^x)^t$. 
This composite operator satisfies $[\mathcal{P}(T_\mathrm{cm}^x)^t]^2=1$ 
thanks to Eq.~\eqref{eq:inv-T}, and thus the eigenvalue $\zeta_\Psi$ has to be 
$\pm 1$:
\begin{equation}\label{eq:inversion-psi}
\mathcal{P}|\Psi\rangle=\zeta_\Psi (T_\mathrm{cm}^x)^t|\Psi\rangle.
\end{equation}

We move on to discuss the recombined states $|\Psi;s,r\rangle$.
Plugging Eq.~\eqref{eq:inversion-psi} into Eq.~\eqref{eq:psi-s-r}, we have
\begin{align}
\nonumber
&\mathcal{P}|\Psi;s,r\rangle\\
\nonumber
=&\frac{1}{\sqrt{q_x}}\sum_m^{q_x}e^{i2\pi m(\kappa_x+sN_e)/(Nq_x)}
(S_x)^{-m}(T_\mathrm{cm}^x)^{t-r}\zeta_\Psi|\Psi\rangle\\
\label{eq:FQH-inv}
=&\zeta_\Psi|\Psi;\bar{s},\bar{r}\rangle,
\end{align}
Here $\bar{r}$ is defined by
\begin{equation}
\label{eq:rbar}
\bar{r}=t-r,
\end{equation}
and $\bar{s}\in\range{q_x}$ by
\begin{equation}
\label{eq:s-sbar}
-(\kappa_x+sN_e)=\kappa_x+\bar{s}N_e\text{ mod }Nq_x.
\end{equation}
It is easy to show that there exists a unique value of $\bar{s}\in\range{q_x}$ 
satisfying the above constraint.
From $\bar{r}=t-r$, we can show that
\begin{equation}\label{eq:r-rbar}
-(\kappa_y+rN_e)=\kappa_y+\bar{r}N_e\text{ mod }N_{0x}N_y.
\end{equation}
In components, Eq.~\eqref{eq:FQH-inv} reads
\begin{equation}\label{eq:FQH-wf-inv}
\langle \{-j\}|\Psi;s,r\rangle=
\zeta_\Psi\langle\{j\}|\Psi;\bar{s},\bar{r}\rangle.
\end{equation}

\subsection{Inversion Operator on the Lattice}

We now extend the inversion operator $\mathcal{P}$ to the lattice. As shown in 
Appx.~\ref{sec:inv-operator}, the exponentiated position operators transforms 
under inversion by
\begin{align}\label{eq:x-y-inversion}
\mathcal{P}\widehat{x}\mathcal{P}&=\widehat{x}^\dagger,&
\mathcal{P}\widehat{y}\mathcal{P}&=\widehat{y}^\dagger.
\end{align}
Since the single-particle Hamiltonian is inversion symmetric, the unitary 
operator $\mathcal{P}$ must take the Bloch state $|k_x,k_y\rangle$ to 
$|{-k_x},-k_y\rangle$, up to a $k$-dependent
phase factor $e^{i\xi_{k_x,k_y}}$\footnote{In the case of many occupied bands, 
the inversion operator is a matrix, but in the absence of additional 
symmetries in 2D, it can always be represented in a diagonal 
$[U(1)]^{N_\mathrm{occupied}}$ form.~\cite{Hughes11:InvSymm}}
\begin{equation}\label{eq:inversion-sewing}
|{-k_x},-k_y\rangle=e^{i\xi_{k_x,k_y}}\mathcal{P}|k_x,k_y\rangle.
\end{equation}
Considering that $\mathcal{P}^2=1$, we have
\begin{equation}\label{eq:inv-xi}
e^{-i\xi_{k_x,k_y}}=e^{i\xi_{-k_x,-k_y}}.
\end{equation}
These phase factors constitute the inversion sewing matrix; the explicit 
expressions can be found in Appx.~\ref{sec:inv-operator}.
Since we work in periodic gauge, the phase factor must be periodic as well,
\begin{equation}\label{eq:periodic-xi}
e^{i\xi_{k_x+mN_x,k_y+nN_y}}=e^{i\xi_{k_x,k_y}},
\end{equation}
for $m,n\in\mathbb{Z}$. If both $N_x$ and $N_y$ are even, there are four 
points $(\bar{k}_x,\bar{k}_y)$ in the Brillouin zone that are inversion 
symmetric (up to a reciprocal vector), namely $(0,0)$, $(N_x/2,0)$, $(0,N_y/2)$, 
and $(N_x/2,N_y/2)$. Thanks to Eqs.~\eqref{eq:inv-xi} 
and~\eqref{eq:periodic-xi}, we have $e^{i\xi_{k_x,k_y}}=\pm 1$ if $(k_x,k_y)$ 
is inversion symmetric, 
and $e^{i\xi_{k_x,k_y}}$ at these points are just the inversion eigenvalues of 
the lowest band.
The periodic boundary condition is crucial in relating $e^{i\xi_{k_x,k_y}}$ to 
inversion eigenvalue if $(k_x,k_y)\neq(0,0)$.

\subsection{Inversion Symmetry of the Connections}\label{sec:wilson-inversion}
We now look for the implication of the inversion symmetry in the Berry 
connection and the Wilson loop. The Berry connection satisfies
\begin{align}
\mathcal{A}_x(k_x,k_y)
&=\langle k_x,k_y|\mathcal{P}^2\widehat{x}\mathcal{P}^2|k_x+1,k_y\rangle\\
\nonumber
&=e^{i\xi_{k_x,k_y}-i\xi_{k_x+1,k_y}}
\langle{-k_x},-k_y|\widehat{x}^\dagger|{-k_x-1},-k_y\rangle\\
\nonumber
&=e^{i\xi_{k_x,k_y}-i\xi_{k_x+1,k_y}}
[\mathcal{A}_x(-k_x-1,-k_y)]^*,
\end{align}
or for the unitary connections,
\begin{align}
\label{eq:Ax-inversion}
A_x(k_x,k_y)A_x(-k_x-1,-k_y)&=e^{i\xi_{k_x,k_y}-i\xi_{k_x+1,k_y}},\\
\label{eq:Ay-inversion}
A_y(k_x,k_y)A_y(-k_x,-k_y-1)&=e^{i\xi_{k_x,k_y}-i\xi_{k_x,k_y+1}}.
\end{align}
These are the lattice versions of the transformation of the Berry connections 
in the continuum [Eq.~(73) in Ref.~\onlinecite{Hughes11:InvSymm}.]

We now derive the lattice version of some well-known inversion properties of 
the Wilson loop.~\cite{Hughes11:InvSymm}
Due to the periodicity of the phase factors $e^{i\xi_{k_x,k_y}}$, 
the unitary discrete Wilson loop satisfies
\begin{equation}\label{eq:Wilson-inversion}
W_x(k_y)W_x(-k_y)=W_y(k_x)W_y(-k_x)=1.
\end{equation}
Therefore, at inversion symmetric momentum $(\bar{k}_x,\bar{k_y})$, we have
\begin{align}
W_x(\bar{k}_y)&=\pm 1,&
W_y(\bar{k}_x)&=\pm 1.
\end{align}
At inversion symmetric $\bar{k}_y$, the Wannier center must be either on the 
Bravais lattice, or at a mid-bond point of the Bravais lattice.

If $N_x$ is even, we can go farther and express $W_x(\bar{k}_y)$ as a product 
of the two inversion eigenvalues at $\bar{k}_y$:~\cite{Alexandradinata12:InvSymm}
\begin{align}
W_x(\bar{k}_y)&=\prod_\kappa^{N_x/2}
A_x(\kappa,\bar{k}_y)A_x(-\kappa-1,\bar{k}_y)\\
&=e^{i\xi_{0,\bar{k}_y}-i\xi_{N_x/2,\bar{k}_y}}.
\end{align}
Similarly, if $N_y$ is even, we can show that
\begin{equation}\label{eq:Wy-xi}
W_y(\bar{k}_x)=e^{i\xi_{\bar{k}_x,0}-i\xi_{\bar{k}_x,N_y/2}}.
\end{equation}

For the moment, we consider the continuum limit with large and even $N_x,N_y$.
Ref.~\onlinecite{Hughes11:InvSymm} showed that the product of the four 
inversion eigenvalues at $(k_x,k_y)=(0,0)$, $(N_x/2,0)$, $(0,N_y/2)$, and 
$(N_x/2,N_y/2)$ is equal to $(-1)^C=-1$ for $|C|=1$. Therefore, in addition to 
Eq.~\eqref{eq:Wilson-inversion}, we now have
\begin{align}\label{eq:W-inv-Chern}
W_x(0)W_x(N_y/2)=W_y(0)W_y(N_x/2)=-1.
\end{align}
The case where $N_x$ or $N_y$ is odd needs special treatment, as we will see 
soon.

\begin{figure*}[t]
\centering
\includegraphics[width=6.5in]{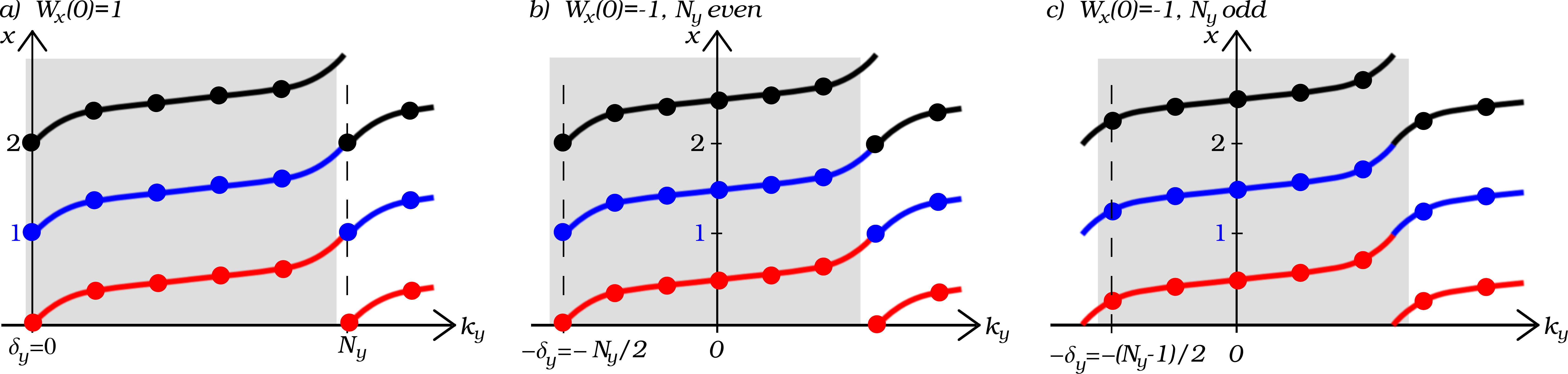}%
\caption{\label{fig:delta-y-inv}
The shift parameter $\delta_y$ fixed by inversion for $C=+1$. The three panels 
correspond to the three cases in Eq.~\eqref{eq:inv-delta}.
Compared with Fig.~\ref{fig:delta-y}, here we highlight the discrete nature of 
the finite-size system. The wave number $k_y$ takes $N_y$ discrete values in 
each Brillouin zone.
The solid dots represents the $N_y$ Wannier centers in each unit cell, colored 
coded by the unit cell index $X$.
The gray shade shows the principal Brillouin zone $Ck_y+\delta_y\in\range{N_y}$. 
}
\end{figure*}

These properties of the Wilson loop strongly constrain the shift parameter 
$\delta_y$ for the Wannier-LLL mapping introduced in Sec.~\ref{sec:wannier-j}.
Recall that we pick the argument angle 
$\mathrm{arg}[W_x(k_y)]\in(-2\pi,0]$. The shift $\delta_y\in\range{N_y}$ is 
defined as the number of $W_x(k_y)$ with 
$\mathrm{arg}[W_x(k_y)]>\mathrm{arg}[W_x(0)]$.
We have shown that the Wilson loops $W_x(k_y)$ come in complex-conjugate 
pairs at momenta coupled by inversion, and in particular, $W_x(0)=\pm 1$.
As illustrated in Fig.~\ref{fig:delta-y-inv}, with inversion, the shift 
$\delta_y$ is fixed to
\begin{equation}\label{eq:inv-delta}
\delta_y=
\begin{cases}
0 & \text{if }W_x(0)=1;\\
N_y/2 & \text{if }W_x(0)=-1\text{ and }N_y\text{ even};\\
(N_y-1)/2 & \text{if }W_x(0)=-1\text{ and }N_y\text{ odd}.
\end{cases}
\end{equation}

A key quantity in our gauge choice Eq.~\eqref{eq:Phi-recursive} for the 
Wannier states is $\lambda_y(0)\,\omega_y$. We now examine the constraint from 
inversion symmetry on it.
Since $[\lambda_y(0)]^{N_y}=W_y(0)$ [Eq.~\eqref{eq:lambda-W}]
and $W_y(0)$ is fixed by inversion symmetry to $\pm 1$, we have
\begin{equation}\label{eq:lambda-y-Ny}
[\lambda_y(0)]^{2N_y}=1.
\end{equation}
As shown in Appx.~\ref{sec:U-inversion}, the curvature fluctuations are also 
subject to the inversion symmetry,
\begin{equation}\label{eq:U-inversion}
U_y(k_y)U_y(-k_y-1)=1.
\end{equation}
Recall from Eq.~\eqref{eq:omega-y} that
$(\omega_y)^{N_y}=\prod_\kappa^{N_y}U_y(\kappa)$.
We can recombine the $U_y$ factors in $[\prod_\kappa^{N_y}U_y(\kappa)]^2$ into 
pairs, and get
\begin{equation}
(\omega_y)^{2N_y}=1.
\end{equation}
Combining this with Eq.~\eqref{eq:lambda-y-Ny}, we have
\begin{equation}\label{eq:inv-lambda-y-omega}
[\lambda_y(0)\,\omega_y]^{2N_y}=1.
\end{equation}

\subsection{Inversion Symmetry of the Wannier States}

In Appx.~\ref{sec:Xky-inversion-raw}, we find the action of the inversion 
operator $\mathcal{P}$ on the Wannier states $|X,k_y\rangle$:
\begin{align}\label{eq:Xky-inversion-raw}
&e^{i\xi_{0,k_y}}e^{-i\Phi_y(X,k_y)}\mathcal{P}|X,k_y\rangle\\
\nonumber
&=
\begin{cases}
e^{-i\Phi_y(-X,-k_y)}\,|{-X},-k_y\rangle
& \text{if }W_x(k_y)=1,\\
e^{-i\Phi_y(-X-1,-k_y)}\,|{-X-1},-k_y\rangle
& \text{otherwise.}
\end{cases}
\end{align}
Considering that we place the $X$-th unit cell over the interval $[X,X+1)$,
this rule of transformation is not surprising: for example, inversion 
takes an on-site Wannier state centered at $x=2+0$ to a state centered at 
$x=-2+0$, while it takes a Wannier state centered at $x=2+0.3$ to $x=-3+0.7$, 
to be in $[X,X+1)$.

In Sec.~\ref{sec:phase-fixing}, we make the gauge choice with 
$e^{i\Phi_y(X,k_y)}$ independent from $X$. We now drop the $X$ argument and 
write $e^{i\Phi_y(k_y)}$.
This enables us to further simplify the $e^{i\Phi_y(k_y)}$ factors in 
Eq.~\eqref{eq:Xky-inversion-raw}.
The starting point is the recursive definition of $e^{i\Phi_y(k_y)}$ in 
Eq.~\eqref{eq:Phi-recursive}.
Define the shorthand notation
\begin{equation}
\widetilde{A}_y(k_y)=A_y(0,k_y)U_y(k_y).
\end{equation}
Then Eq.~\eqref{eq:Phi-recursive} could be rewritten as 
\begin{equation}\label{eq:Phi-recursive-simple}
e^{i\Phi_y(k_y+1)-i\Phi_y(k_y)}=
\frac{\lambda_y(0)\,\omega_y}{\widetilde{A}_y(k_y)},
\end{equation}
thanks to the periodicity of $e^{i\Phi_y(k_y)}$ in $k_y$.
We would now like to relate, by recursion of 
Eq.~\eqref{eq:Phi-recursive-simple}, $e^{i\Phi_y(k_y)}$ with 
$e^{i\Phi_y(-k_y)}$.
The result would be a product of $\lambda_y(0)\,\omega_y/\widetilde{A}_y(\kappa)$ 
along the path connecting $-k_y$ to $k_y$.
Thanks to the periodicity of $e^{i\Phi_y(k_y)}$ in $k_y$, we can focus on the 
first Brillouin zone $k_y\in\range{N_y}$. Since $-k_y\leq k_y$, successive 
application of Eq.~\eqref{eq:Phi-recursive-simple} leads to
\begin{multline}\label{eq:Phi-recursion-ky}
e^{i\Phi_y(0,k_y)-\Phi_y(0,-k_y)}
=\prod_{\kappa=-k_y}^{k_y-1}
\frac{\lambda_y(0)\,\omega_y}{\widetilde{A}_y(\kappa)}\\
=[\lambda_y(0)\,\omega_y]^{2k_y}
\Bigg/\,\prod_{\kappa=0}^{k_y-1}
\widetilde{A}_y(\kappa)\widetilde{A}_y(-\kappa-1).
\end{multline}
The product of $\widetilde{A}_y$ factors in the rewritten form on the second 
line shows the inversion symmetry structure explicitly: from the inversion 
transformation of $A_y(k_x)$ in Eq.~\eqref{eq:Ay-inversion} and that of 
$U_y(k_y)$ in Eq.~\eqref{eq:U-inversion}, we have
\begin{equation}
\widetilde{A}_y(k_y)\widetilde{A}_y(-k_y-1)=e^{i\xi_{0,k_y}-i\xi_{0,k_y+1}}.
\end{equation}
Plugging this into Eq.~\eqref{eq:Phi-recursion-ky}, we get
the relative phase between the Wannier states which are inversion partners
\begin{equation}\label{eq:Phi-inversion}
e^{i\Phi_y(k_y)-i\Phi_y(-k_y)}
=e^{i\xi_{0,k_y}-i\xi_{0,0}}\,[\lambda_y(0)\,\omega_y]^{2k_y}.
\end{equation}
Therefore, the inversion operation on the Wannier states 
[Eq.~\eqref{eq:Xky-inversion-raw}] can be simplified to
\begin{multline}\label{eq:Xky-inversion}
e^{i\xi_{0,0}}[\lambda_y(0)\,\omega_y]^{-2k_y}
\mathcal{P}|X,k_y\rangle\\
=\begin{cases}
|{-X},-k_y\rangle & \text{if }W_x(k_y)=1,\\
|{-X-1},-k_y\rangle & \text{otherwise.}
\end{cases}
\end{multline}
This concludes our analysis of the action of inversion on the one-body states.

\subsection{Inversion Symmetry of the Many-Body States}

When the one-body model is inversion symmetric,
we require the FCI many-body wave functions on lattice to inherit the 
inversion symmetry of the FQH states on a torus. First we check whether the 
inversion transformation is compatible with the Wannier-LLL mapping on the 
\emph{index} level. Under inversion, the LLL state index transforms 
by $j\rightarrow -j$ mod $N_\phi=N_xN_y$, while as shown in 
Eq.~\eqref{eq:Xky-inversion}, the Wannier state index transforms by
\begin{equation}\label{eq:Xky-label-inversion}
(X,k_y)\rightarrow
\begin{cases}
(-X,-k_y) & \text{if }W_x(k_y)=1,\\
(-X-1,-k_y) & \text{otherwise.}
\end{cases}
\end{equation}
In Appx.~\ref{sec:jXky-inversion}, we show that the 1D index $j^{X,k_y}$ 
defined by Eq.~\eqref{eq:j-Xky} transforms by
\begin{equation}\label{eq:jXky-inversion}
j^{X,k_y}\rightarrow
\begin{cases}
-j^{X,k_y}-1 & \text{if }W_x(0)=-1\text{ and }N_y\text{ odd};\\
-j^{X,k_y} & \text{otherwise}.
\end{cases}
\end{equation}
Note that for a given system, all the $j^{X,k_y}$ indices obey the \emph{same} 
transformation rule. The transformation in the first case is \emph{not} a good 
FQH symmetry on the single-particle level. However, on the many-body level, a 
simultaneous change $j$ to $-j-1$ for \emph{all particles} is a good symmetry, 
implemented by $\mathcal{P}T_\mathrm{cm}^x$.

Finally, we are ready to investigate the inversion symmetry of the many-body 
lattice state $|\Psi;s,r\rangle_\mathrm{lat}$ defined in 
Eq.~\eqref{eq:lat-state}. 
Plugging Eqs.~\eqref{eq:Xky-inversion} and~\eqref{eq:jXky-inversion} into the 
Wannier-LLL mapping in Eq.~\eqref{eq:lat-state}, we can get the inverted 
amplitudes on the FQH side. Then, we can apply Eq.~\eqref{eq:FQH-wf-inv} and 
relate it to another state in the $q$-fold multiplet and map it back to the 
lattice. This identifies the inversion partner of the many-body lattice state 
and obtains the many-body matrix elements of the lattice operator $\mathcal{P}$ 
in the $q$-dimensional subspace. The details are presented in 
Appx.~\ref{sec:manybody-inversion}, while the gist is the following.
The $q$ lattice states $|\Psi;s,r\rangle_\mathrm{lat}$ come in inversion 
pairs:
\begin{multline}\label{eq:inv-psi-s-r-lat}
\mathcal{P}|\Psi;s,r\rangle_\mathrm{lat}\\
\propto\begin{cases}
|\Psi;\bar{s},\bar{r}-1\rangle_\mathrm{lat}
& \text{if }W_x(0)=-1\text{ and }N_y\text{ odd};\\
|\Psi;\bar{s},\bar{r}\rangle_\mathrm{lat}
& \text{otherwise}.
\end{cases}
\end{multline}
Here the indices $(\bar{s},\bar{r})$ are defined conjugate to $(s,r)$, in 
Eqs.~\eqref{eq:rbar} and~\eqref{eq:s-sbar}.
Recall that the total momentum of the $N_e$ particles in state 
$|\Psi;s,r\rangle$ is given by 
$(K_x,K_y)=[\kappa_x+sN_e,C\kappa_y+C(r-\delta_y)N_e]$ mod $(N_x,N_y)$
[Eq.~\eqref{eq:lattice-K}].
The total momentum of its inversion partner given in the 
above equation is $(-K_x,-K_y)$ mod $(N_x,N_y)$, as anticipated and shown in 
Appx.~\ref{sec:manybody-inversion}.

To summarize, our construction of lattice wave functions indeed preserves on 
the many-body level the inversion symmetry of the host Chern insulator.
In contrast,  the gauge and the original construction in 
Ref.~\onlinecite{Qi11:Wavefunction} 
does not preserve inversion symmetry unless $W_y(0)=1$, as shown in 
Appx.~\ref{sec:Qi}.

\section{Numerical Tests}\label{sec:results}

We perform the Wannier construction in the five fermionic lattice models 
studied previously at filling $\nu=1/3$: the checkerboard lattice 
model,~\cite{Sun11:Flatband}
the Haldane model on the honeycomb lattice,~\cite{Haldane88:Honeycomb}
a two-orbital model that resembles half (spin-up) of the 
mercury-telluride two-dimensional topological insulator,~\cite{Bernevig06:BHZ} 
the Kagome lattice model with spin-orbit coupling between nearest 
neighbors,~\cite{Tang11:Kagome} 
and the spin-polarized ruby lattice model.~\cite{Hu11:Ruby}
For each model, we use the parameter set specified 
previously~\cite{Regnault11:FCI,Wu12:Zoology} and focus on the lattice sizes 
and aspect ratios where the system has a threefold quasi-degenerate ground 
state in numerics well separated from the excited states.
These are all inversion symmetric models, for which we list the values 
of $W_x(0)$, $W_y(0)$, and $C$ for each model in Table~\ref{tab:models}.

\begin{table}[b]
\caption{\label{tab:models}
Values of the Chern number $C$, the Wilson loops $W_x(0)$ and $W_y(0)$, and 
the twist angle $\theta$ of the corresponding FQH torus,
for the five lattice models at the studied parameters.
}
\begin{ruledtabular}
\begin{tabular}{ccccc}
Model & $C$ & $W_x(0)$ & $W_y(0)$ & $\theta$ \\\hline
Checkerboard & $-1$ & $+1$ & $-1$ & $\pi/2$ \\
Haldane & $-1$ & $+1$ & $-1$ & $\pi/3$ \\
Kagome & $+1$ & $+1$ & $+1$ & $\pi/3$ \\
Ruby & $-1$ & $-1$ & $+1$ & $\pi/3$ \\
Two-orbital & $+1$ & $-1$ & $-1$ & $\pi/2$
\end{tabular}
\end{ruledtabular}
\end{table}

We obtain the ``Laughlin'' FCI states on the lattice using the formalism 
described in this paper, i.e. replacing $|\Psi\rangle$ in 
Eqs.~\eqref{eq:psi-s-r} and~\eqref{eq:bloch-amplitude} by the Laughlin FQH 
states on the torus in the continuum.
The threefold Laughlin FQH states are obtained by numerically diagonalizing 
the FQH model Hamiltonian with only the first pseudopotential, 
$V_m=\delta_{m,1}$. Details of the continuum FQH Hamiltonian can be found in 
Appx.~\ref{sec:FQH-Hamiltonian}.
As detailed in Sec.~\ref{sec:wannier-j}, the continuum torus for the FQH 
calculations has the same aspect ratio as the lattice system, and the twist 
angle $\theta$ of the five models are given in the last column of 
Table~\ref{tab:models}.
The three Laughlin states are exact replicas except for a shift in the total 
momentum $\kappa_y$. As shown in Sec.~\ref{sec:recombination}, they can be 
recombined into $|\Psi;s,r\rangle$ and then mapped to the lattice using the 
Wannier basis. We will refer to the resulting lattice states as the Laughlin 
FCI states, as opposed to the Laughlin FQH states in the continuum.

For all of the five models, we find that the Laughlin FCI states are indeed 
translationally invariant and inversion symmetric.
In contrast, if we follow the prescription in 
Ref.~\onlinecite{Qi11:Wavefunction}, the resulting many-body states break 
the inversion symmetry for the checkerboard lattice model, the two-orbital
model, and the Haldane model. As explained in Appx.~\ref{sec:Qi}, the issue 
responsible for this
is that the Wilson loop $W_y(0)$ equals $-1$ in these models at the studied 
parameter set. 

\subsection{Overlap}\label{sec:overlap}

The relevance of the Laughlin FCI states can be checked by computing the 
overlap with the ground states obtained from exact diagonalization of the FCI 
models at filling $\nu=1/3$.~\cite{Regnault11:FCI,Wu12:Zoology}.

The number and momenta of Laughlin FCI states (shown in 
Sec.~\ref{sec:folding}) match those of the exact ground states obtained by 
exact diagonalization.
When both $N_x$ and $N_y$ are divisible by $3$, all of the threefold states
have total momentum zero.
In such cases with more than one state in a certain momentum sector, the 
states can mix and we do not have a one-one correspondence between the two 
groups of states to compute the overlaps.
Instead, we first build the projector into the subspace spanned by the exact 
diagonalization states, then the projector into the subspace spanned by the 
Laughlin FCI states, and finally, we define the overlap to be the trace of the 
product of the two projectors, divided by the dimension of the subspace, i.e.
\begin{equation}
\frac{1}{3}\sum_{i}\sum_{s,r}
\Big|{}_\mathrm{lat}\!\langle\mathrm{ED};i|\Psi;s,r\rangle_\mathrm{lat}
\Big|^2,
\end{equation}
where $|\mathrm{ED};i\rangle_\mathrm{lat}$, $i=1,2,3$ are the threefold ground 
states obtained from exact diagonalization, and the indices $(s,r)$ are 
summed over the threefold Laughlin FCI states in the same momentum sector.
This naturally generalizes the usual definition of overlap as the absolute 
square of the inner product.
When the Laughlin FCI states appear at different momenta, we take the usual 
overlap with the exact diagonalization states.

\begin{figure}[tb]
\centering
\includegraphics[]{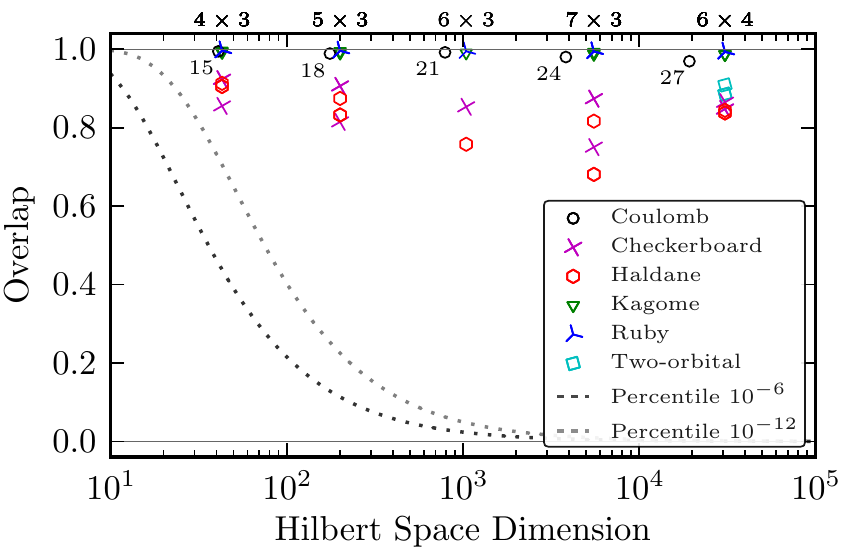}%
\caption{\label{fig:overlap}
Overlap between variational states and the exact ground states. 
The size of the lattice is labeled by $N_x\times N_y$ at top.
The plot includes the checkerboard lattice model (purple crosses), the 
Haldane model (red hexagons), the Kagome lattice model (green triangles), the 
ruby lattice model (blue Y-shapes), and the two-orbital model (cyan 
squares).
The Wannier construction is performed using the $|X,k_y\rangle$ basis.
At $6\times 3$, the three states in a multiplet are in the same momentum 
sector, and thus there is one overlap value for each model.
At other lattice sizes, two out of the three states have identical overlap due 
to inversion symmetry.
Also shown is the overlap between the FQH ground state of the Coulomb 
interaction at filling $\nu=1/3$ and the Laughlin model state, denoted by 
black circles and labeled by the number of flux $N_\phi$.
As a baseline, we also plot the top $10^{-6}$ and $10^{-12}$ percentiles 
(gray lines) of the overlap values \emph{between random unit vectors} as a 
function of the Hilbert space dimension.
The overlap axis range is set to $[0,1]$ to contrast with Fig.~\ref{fig:overlap-xlq}.
}
\end{figure}

Shown in Fig.~\ref{fig:overlap} against the Hilbert space dimension are the 
overlap values between the Laughlin FCI states obtained by the Wannier 
construction and the exact ground states at filling $\nu=1/3$, of the five 
models on a lattice of size $4\times 3$, $5\times 3$, $6\times 3$, 
$7\times 3$, and $6\times 4$.
The only exception is the two-orbital model: it does not have a 
\emph{gapped} threefold quasi-degenerate ground state
when $N_y=3$.~\cite{Wu12:Zoology}
At $6\times 3$ the threefold states have the same total momentum,
and thus there is only one overlap value for each model. At other lattice 
sizes, two of the threefold states form an inversion pair and have the same 
overlap value.
Also shown in Fig.~\ref{fig:overlap} are the overlap values between the 
Laughlin state and the FQH ground state of the Coulomb interaction on a torus 
with unity aspect ratio, pierced by $N_\phi=15,18,21,24,27$ fluxes.

We find that at comparable Hilbert space sizes, the overlap for the Kagome 
lattice model and the ruby lattice model are comparable with or \emph{even higher} 
than the overlap for the Coulomb interaction in the continuum, while the 
overlap for the Haldane model, the two-orbital model, and the checkerboard 
lattice model have slightly lower overlap than the Coulomb interaction.
We note one possible reason for the higher overlaps in the FCI systems. The 
lattice interactions used in the FCI models are Hubbard interactions between 
the nearest neighbors. Their short-ranged nature is preserved by the mapping 
to the LLL via the \emph{localized} Wannier states, and thus the FCI 
interaction in the Wannier/LLL basis maybe be closer to the Laughlin 
pseudopotential Hamiltonian, compared with the long-range Coulomb interaction 
in the continuum.

We emphasize that the overlap with the Laughlin FCI state should be 
interpreted in the context of the size of the Hilbert space.
In Appx.~\ref{sec:random-overlap} we find that the probability 
for two random complex unit vectors in a space of dimension $d$ to have an 
overlap larger than a given value is \emph{exponentially small} at large $d$. 
This is illustrated in Fig.~\ref{fig:overlap} by the plot of the top 
$10^{-6}$ and $10^{-12}$ percentile values of the overlap between random 
vectors, as a function of the Hilbert space dimension.

We observe a slight decreasing trend in the overlap as the system size 
increases, much slower than the exponential decay of the random overlap.
There is also an upward kink when going from $7\times 3$ to 
$6\times 4$ for the checkerboard lattice and the Haldane models. This agrees 
with the finding in Ref.~\onlinecite{Regnault11:FCI} that the topological 
phase is more stable when the aspect ratio is closer to unity.

\begin{figure}[tb]
\centering
\includegraphics[]{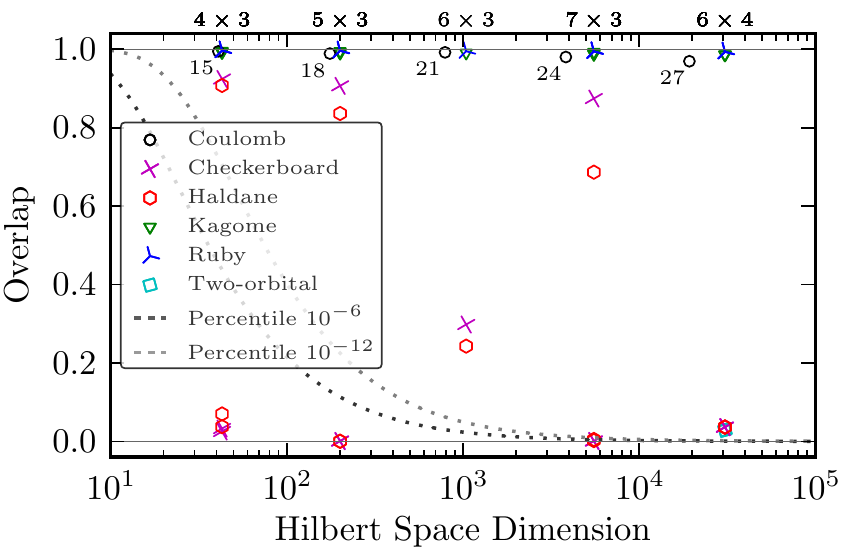}%
\caption{\label{fig:overlap-xlq}
Overlaps between variational states and the exact ground states using the 
prescription in Ref.~\onlinecite{Qi11:Wavefunction}.
Inversion breaking for the checkerboard lattice model and the Haldane model 
is clearly visible at $4\times 3$, and present (although not discernible in 
this figure) at other lattice sizes.
Inversion breaking for these models is also present at $7\times 3$ and 
$6\times 4$, even though obscured by the finite resolution of the plot. 
The inversion symmetry of the two-orbital model at $6\times 4$ is 
also broken.}
\end{figure}

We have also tested the performance of the proposal in 
Ref.~\onlinecite{Qi11:Wavefunction}.
While the original proposal was given in the continuum limit 
$N_x,N_y\rightarrow\infty$ (formulated using derivatives such as 
$\langle k_x,k_y|\partial_{k_x}|k_x,k_y\rangle$) and on a cylinder,
as detailed in Appx.~\ref{sec:Qi}, we extend this proposal to handle 
finite-size lattice models with multiple sublattices and periodic boundaries,
but do not change the gauge used there.
We repeat the overlap calculations using this alternative prescription. The 
results are compiled in Fig.~\ref{fig:overlap-xlq}.
The overlap values for the Kagome lattice model and the ruby lattice model are 
comparable with those calculated using our proposal.
However, the overlap values for the checkerboard lattice model, the two-orbital
model (which is the model in the original 
discussion~\cite{Qi11:Wavefunction}), and the Haldane model are much lower 
than the values obtained from our prescription.
At $6\times 4$, the overlap values for these three models fall
below $0.04$. Also, the constructed many-body states break the inversion 
symmetry in these models.
The culprit is explained in Appx.~\ref{sec:wannier-continuum} and 
Appx.~\ref{sec:Qi}.
In short, when $W_y(0)\neq 1$, the gauge choice $a_y=0$ used in 
Ref.~\onlinecite{Qi11:Wavefunction} does not mimic the behavior of the LLL 
orbitals: enforcing $a_y=0$ in the interior of the Brillouin zone mandates 
that the connection on the Brillouin zone boundary has to take care of the 
gauge-invariant Wilson loop, and deviates from $a_y=0$.
The prescription in Ref.~\onlinecite{Qi11:Wavefunction} does not properly 
handle the cases where the Wilson loop $W_y(0)$ is not unity, which include 
the above three lattice models at the studied parameters, as shown in 
Table~\ref{tab:models}.

\subsection{Mapping Parameters Revisited}\label{sec:delta-y-theta}

In the previous section, the shift parameter $\delta_y$ for the Wannier-LLL 
mapping and the twist angle $\theta$ of the continuum torus used in the FQH 
calculations were set according to the discussions in 
Sec.~\ref{sec:wannier-j}.
We now revisit the choice of $\delta_y$ and $\theta$ by checking explicitly 
how it affects the overlaps with the FCI ground states.

\begin{figure}[tb]
\centering
\includegraphics[]{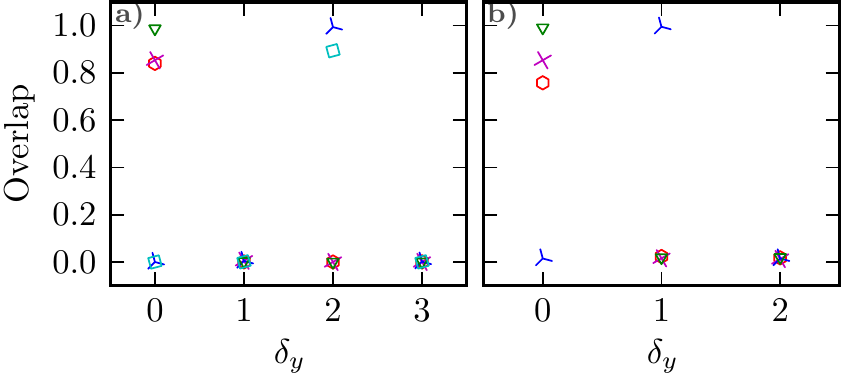}%
\caption{\label{fig:overlap-delta}
Overlaps between the Laughlin FCI states and the exact diagonalization states, 
as a function of the shift parameter $\delta_y$ in the Wannier-LLL 
mapping. Please refer to the legend and caption of Fig.~\ref{fig:overlap} for 
the annotation of scatter symbols.
The left (right) panel shows the calculations on a $6\times 4$
($6\times 3$) lattice. For clarity, for each model at each $\delta_y$ we show 
the average overlap of the threefold states. The two-orbital model is not 
included in the $6\times 3$ calculation, as it does not have a gapped 
topological ground state at this lattice size.
The values of $\delta_y$ that produce significant overlaps are in full 
agreement with Eq.~\eqref{eq:inv-delta}.
}
\end{figure}

First we perform the Wannier constructions for the five models using the 
$\theta$ angle specified in Table~\ref{tab:models}, but we vary the shift 
parameter $\delta_y$ over $\range{N_y}$.
In Fig.~\ref{fig:overlap-delta}, we show the dependence of the overlaps on 
$\delta_y$ for $6\times 4$ and $6\times 3$ lattices.
At $N_y=4$ (even), for the three models with $W_x(0)=+1$, namely (see 
Table~\ref{tab:models}), the Kagome lattice model, the checkerboard lattice 
model, and the Haldane model, the Laughlin FCI states have significant overlaps
with the exact diagonalization ground states only at $\delta_y=0$; for 
the two models with $W_x(0)=-1$, namely, the ruby lattice model and the 
two-orbital model, the significant overlaps appear at $\delta_y=2=N_y/2$.
At $N_y=3$ (odd), the significant overlaps for the three models with 
$W_x(0)=+1$ are still at $\delta_y=0$, but for the ruby lattice models with 
$W_x(0)=-1$, the peak in overlap is
shifted to $\delta_y=1=(N_y-1)/2$.\footnote{The two-orbital model is not 
included in the $6\times 3$ calculations, because this model does not have a 
gapped topological ground state at this lattice size.}
The above results are in full agreement with the choice of $\delta_y$ in 
Eq.~\eqref{eq:inv-delta}.

\begin{figure}[tb]
\centering
\includegraphics[]{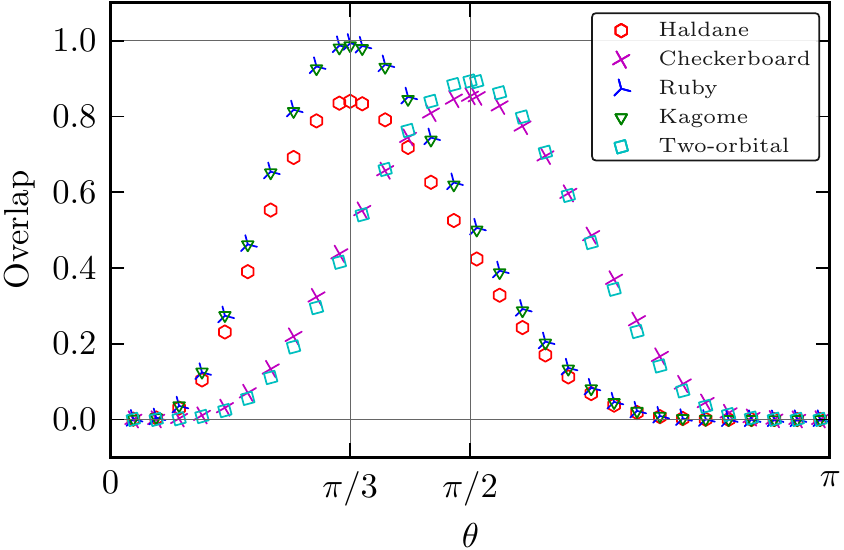}%
\caption{\label{fig:overlap-angle}
Overlaps between the Laughlin FCI states and the exact diagonalization states 
on a $6\times 4$ lattice, as a function of the twist angle $\theta$ of the 
continuum torus for the FQH calculations. 
For clarity, for each model at each $\delta_y$ we show the average overlap of 
the threefold states. For each model, a peak in overlap is clearly visible at 
the value of $\theta$ given in Table~\ref{tab:models}, in accordance with the 
discussion in Sec.~\ref{sec:wannier-j}.
}
\end{figure}

We now put $\delta_y$ to the values specified by Eq.~\eqref{eq:inv-delta}, but 
vary the twisted angle $\theta$ of the continuum torus for FQH calculations.
The overlaps as a function of $\theta$ are shown in Fig.~\ref{fig:overlap-angle}.
For each of the five models, we find a clear peak in overlap centered around 
the value of $\theta$ suggested in Sec.~\ref{sec:wannier-j}, as shown in 
Table~\ref{tab:models}.
This confirms that our choice of the twist angle $\theta$ is appropriate.

\subsection{Gauge Optimization}

\begin{figure}[tb]
\centering
\includegraphics[]{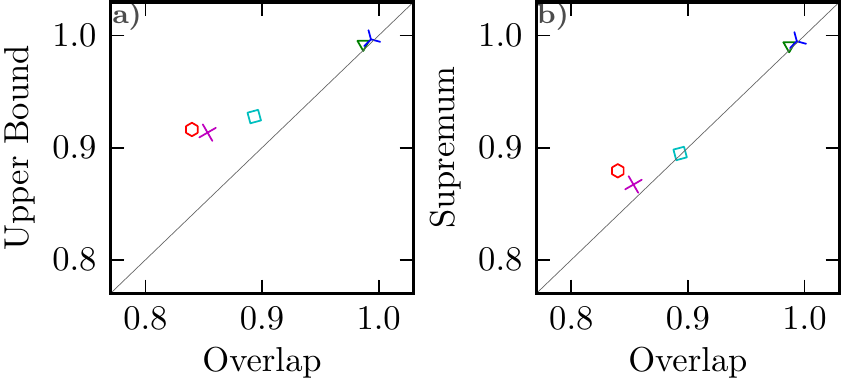}%
\caption{\label{fig:optimization}
Overlap optimization for the Laughlin lattice states by tuning the 
single-particle gauge. The calculations are performed on a $6\times 4$ lattice.
Please refer to the legend and caption of Fig.~\ref{fig:overlap} for 
the annotation of scatter symbols.
Shown in a) is the overlap upper bound~\eqref{eq:drop-phase} obtained by 
removing the relative phase at each component between the exact ground state 
and the Laughlin lattice state, against the actual overlap.
Shown in b) is the supremum (the least upper bound) of the overlap, obtained 
by a brute force optimization of the single-particle gauge of the Bloch states, 
against the actual overlap.
}
\end{figure}

As detailed in Sec.~\ref{sec:phase-fixing}, the phase choice 
$e^{i\Phi_y(X,k_y)}$ of the Wannier basis is based on the similarity with the 
LLL orbital behavior.
There are other possible alternative gauge choices.
We now try to understand how close our gauge fixing procedure gets to several 
brute-force optimized gauges.

We first examine the limit where the phases of the Slater determinant weights 
in the variational trial states are tuned to match perfectly with the phases in 
the exact diagonalization ground states.
Eq.~\eqref{eq:bloch-amplitude} shows that the \emph{absolute value} of 
each component $\langle\{k_x,k_y\}|\Psi;s,r\rangle_\mathrm{lat}$ of the 
many-body state in the Bloch basis does not depend on the parallel-transport 
gauge in the $k_x$ direction, nor on the gauge choice of the Wannier states if 
we limit ourselves to a set of $e^{i\Phi_y(X,k_y)}$ phases that have no real $X$ 
dependence.
Inspired by this, we remove the relative phase from the Bloch-basis overlap 
formula between the Laughlin lattice state $|\Psi;s,r\rangle_\mathrm{lat}$ and 
the corresponding exact diagonalization ground state 
$|\mathrm{ED};s,r\rangle$, and calculate
\begin{equation}\label{eq:drop-phase}
\!\!\!\!\left[\sum_{\{k_x,k_y\}}\!
\Big|{}_\mathrm{lat}\!\langle\mathrm{ED};s,r|\{k_x,\!k_y\}\rangle
\langle\{k_x,\!k_y\}|\Psi;s,r\rangle_\mathrm{lat}\Big|\right]^2\!\!.
\end{equation}
This real number serves as \emph{an upper bound} to the actual overlap.
It neglects the phases of the Slater determinant weights in the many-body wave 
functions.\footnote{
Notice that this quantity cannot be generalized to the case with more than 
one state in a momentum sector, for example the Laughlin FCI state on a 
$6\times 3$ lattice, as the absolute value operation breaks the orthogonality 
between the degenerate states at the same momentum.}
In Fig.~\ref{fig:optimization}a), the numerical values of this upper bound are 
shown against the overlap values from our prescription for the five lattice 
models.
We find that the upper bound is still far from unity for the three 
under-performing models, namely, the checkerboard lattice model, the Haldane 
model, and the two-orbital model.
This can be interpreted in two possible ways:
The $X$ dependence of $e^{i\Phi_y(X,k_y)}$ may be essential for the
three models, or, the ground states of the three models may be simply more 
complicated than the Laughlin lattice states.
Our results are not sufficient to distinguish between the two scenarios.

The upper bound is higher than the actual overlaps from our prescription,
which would naively suggest that our prescription is not optimal.
This is not true. To reach the upper bound calculated in 
Eq.~\eqref{eq:drop-phase}, we need to tune the phases of all the Slater 
determinant weights in the many-body state. 
The number of such phases is on the order of $(N_xN_y)$-choose-$N_e$.
We in no way have the freedom to tune these phases independently.
The most we are allowed to do is tune the phases of the $N_xN_y$ Wannier 
states independently.
The \emph{least} upper bound, i.e. the supremum, is hence found 
by tuning the single-particle gauge to reach maximum overlap.
In general, this variational overlap value cannot reach the upper bound in 
Eq.~\eqref{eq:drop-phase}.

The actual supremum can be found by a brute-force optimization.
We introduce $N_x\times N_y$ variational gauge degrees of freedom 
$e^{i\Upsilon_{k_x,k_y}}$ for the Bloch states $|k_x,k_y\rangle$. This 
transforms the Bloch amplitudes by
\begin{equation}\nonumber
\langle\{k_x,k_y\}|\Psi;s,r\rangle_\mathrm{lat}\Rightarrow
e^{-i\sum\Upsilon_{k_x,k_y}}\langle\{k_x,k_y\}|\Psi;s,r\rangle_\mathrm{lat}.
\end{equation}
We optimize the average overlap with the threefold exact ground states by 
tuning these $N_xN_y$ parameters.
As explained in Sec.~\ref{sec:gauge-freedom}, $e^{i\Upsilon_{0,k_y}}$ covers 
the degrees of freedom carried by the Wannier phase choice $e^{i\Phi_y(k_y)}$.
We are not able to perform the optimization on $e^{i\Phi_y(X,k_y)}$ with 
non-trivial $X$-dependence due to the prohibiting computational complexity.
We use the Nelder-Mead (downhill simplex) algorithm~\cite{Nelder65} to find 
the local overlap maximum around our prescription. The convergence threshold 
is $10^{-4}$ on the overlap (output) and the phase angles of the gauge 
transform (input).

In Fig.~\ref{fig:optimization}b), the suprema are shown against the overlap 
values from our prescription. 
Comparing with Fig.~\ref{fig:optimization}a), we find that the supremum is 
much lower than the upper bound~\eqref{eq:drop-phase} for the three 
under-performing models.
This reduction is not surprising, since the variational degrees of freedom are 
far fewer than the number of many-body amplitudes.
The proximity of the data points to the diagonal line in 
Fig.~\ref{fig:optimization}b) suggests that our prescription is very close to 
the optimal one if we refrain from introducing $X$-dependent phases in 
$|X,k_y\rangle$.
From the values of $e^{i\Upsilon_{k_x,k_y}}$ at the supremum, we find that to 
improve our prescription, we need to move slightly away from the parallel 
transport gauge [Sec.~\ref{sec:gauge-freedom}] in the $k_x$ direction.
This may be attributed to the curvature fluctuations. We leave explorations in 
this direction for future work.

\subsection{Entanglement Spectrum}

Overlap only provides an overall, sometimes 
inaccurate~\cite{Simon07:Gaffnian,Regnault09:Jain} estimate on the 
similarity between two states. To compare the correlations in the Laughlin FCI 
states and the ground states, we now turn to their particle entanglement spectra.

First introduced by Li and Haldane,~\cite{Li08:ES} the entanglement spectrum 
examines the information imprinted in a many-body state.
By effectively performing a spectral decomposition of the many-body state,
the entanglement spectrum reveals the excitations supported by the state.
Of particular interest in our study is the 
particle entanglement spectrum (PES), which encodes the characteristics of the 
quasihole excitations from the ground state.~\cite{Sterdyniak11:PES}
We divide the $N_e$ particles into two groups $A$ and $B$ of $N_A$ and $N_B$ 
particles respectively, trace out the 
degrees of freedom carried by the particles in $B$, and examine
of the negative logarithm of the eigenvalues of the reduced density matrix 
$\rho_A$.

The Laughlin FQH state exhibits a highly non-trivial pattern in the 
PES:~\cite{Li08:ES,Sterdyniak11:PES,Bernevig12:Counting}
In each momentum sector, the number of entanglement levels at finite 
entanglement energy matches the counting of the Laughlin quasiholes of $N_A$ 
particles in $qN_e$ fluxes.
The total number of levels, i.e. the number of non-zero modes of the reduced 
density matrix, is much smaller than the dimension of the reduced 
density matrix. 

\begin{figure}[tb]
\centering
\includegraphics[]{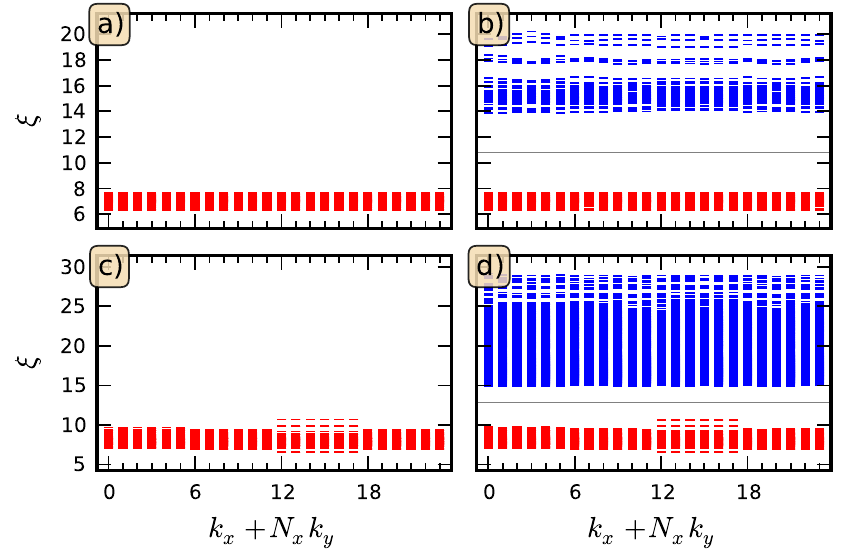}%
\caption{\label{fig:kagome-ent}
Particle entanglement spectrum of the Kagome lattice model of $N=8$ particles 
on the $N_x\times N_y=6\times 4$ lattice. Shown in a) and b) are the spectra 
of the Laughlin FCI state and the exact ground state, respectively, with $N_A=3$. 
Shown in c) and d) are the corresponding spectra with 
$N_A=4$. In each momentum sector, we calculate the number of 
admissible configurations $n$, and color the lowest $n$ entanglement energy 
levels in red and the rest in blue.
The structure shown in the left panel exhibits the defining characteristic of 
the Laughlin state.
}
\end{figure}

The PES of various FCI ground states at filling 
$\nu=1/3$ has a low entanglement energy structure similar to the PES of the 
Laughlin state.~\cite{Regnault11:FCI,Bernevig12:Counting,Wu12:Zoology}
It also displays a non-universal high entanglement energy structure separated 
by an entanglement gap from the Laughlin-like structure.
The number of levels below the gap at each total momentum matches the 
number of quasiholes folded to the lattice Brillouin 
zone.~\cite{Bernevig12:Counting}
An ideal Laughlin state in the FCI would exhibit an infinite entanglement gap.

For the Laughlin FCI model state obtained from the Wannier construction, we 
find a large number of zero modes in the reduced density matrix -- corresponding 
to an infinite entanglement gap.
The number of the finite entanglement energy levels is given exactly by the counting 
of Laughlin quasiholes folded to the lattice Brillouin zone.
Writing the Laughlin FQH state in the Wannier basis is just a unitary
transformation. Thus it does not modify the rank of the reduced density 
matrix, and it preserves the entanglement spectrum of the Laughlin FQH state.
An example is shown in Fig.~\ref{fig:kagome-ent}.
The entanglement energy levels below the gap of the exact diagonalization 
ground state exhibit almost identical structure in the spectra of the Laughlin 
FCI state and the ground state.
The Laughlin FCI state captures a very important feature of the 
correlations in the FCI ground states. This further corroborates the 
proposal~\cite{Regnault11:FCI,Bernevig12:Counting}
that the quasihole counting in the PES can be interpreted as a signature of a 
FQH-type of topological phase.

\section{Isotropy/Nematicity of the Wannier Construction}\label{sec:nematicity}

An undesirable feature of the Wannier construction is that it seems to 
explicitly break the isotropy between the two axes of the lattice system when 
the original Hamiltonian is not fourfold rotationally 
symmetric.~\footnote{Even when the $C_4$ symmetry is present in the hopping 
amplitudes locally, the lattice aspect ratio has to be fixed to specific 
values (square, for example) for the symmetry to survive at finite size.}
The formalism singles out one direction in the 2D lattice along which the 
single-particle states (both on lattice and in the LLL) localize.
The two alternative choices of localizing along the $x/y$ direction
are not guaranteed to produce the same set of $q$-fold many-body states for 
the FCI. 

In the following, we show that the weights of the corresponding Bloch-basis 
Slater determinants in the two sets of many-body states constructed from the 
two alternatives have the same absolute value, but the phases may differ.
However, in the limit of flat curvature, the two alternatives produce 
exactly the same set of $q$-fold many-body states.
(In this paper we do not explore other possible localization 
directions of the Wannier basis.)

So far we have been using single-particle states $|X,k_y\rangle$ that are 
plane waves in the $\hat{e}_y$ direction.
We now examine the alternative choice of using single-particle states 
$|Y,k_x\rangle$ that are \emph{localized} in the $\hat{e}_y$ direction, and 
the corresponding alternative basis in the LLL.

\subsection{FQH States in a Rotated Gauge}

\begin{figure}[tb]
\centering
\includegraphics[width=3.2in]{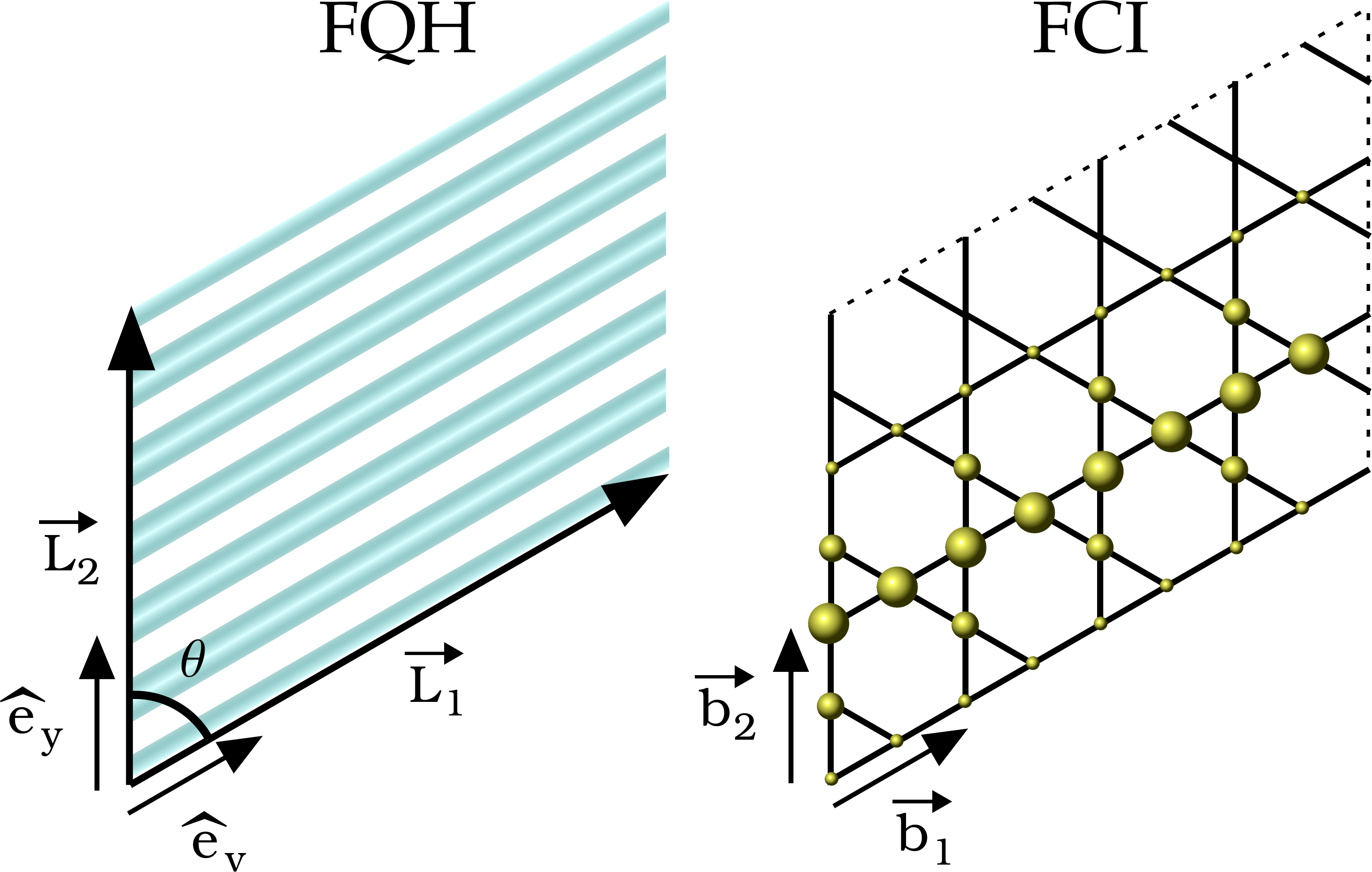}%
\caption{\label{fig:orbitals-x}
Single-particle orbitals in the lowest Landau level (LLL) and the Chern band 
in the alternative gauge choice.
In the left panel we show the Landau orbitals in the LLL on a torus. The two 
fundamental cycles are marked by $\mathbf{L}_1$ and $\mathbf{L}_2$, and the 
twist angle is labeled by $\theta$.
The Landau orbitals are plane waves in the $\hat{e}_v$ direction that are 
localized in the direction perpendicular to propagation. 
In the right panel we show a 1D Wannier state localized in the $\mathbf{b}_2$ 
direction and in the lowest band of the Kagome lattice model. The size of the 
spheres depicts the weights of the Wannier state on each lattice site.
}
\end{figure}

We look for a one-body basis in the LLL that are localized in the $\hat{e}_y$
direction, as illustrated in Fig.~\ref{fig:orbitals-x}.
The ``natural'' gauge in this case is aligned along the $\hat{e}_v$ direction.
We setup a rotated coordinate system $(\widetilde{u},\widetilde{v})$ with the 
$\widetilde{v}$ axis pointing in the direction of $\hat{e}_v$:
\begin{equation}
\begin{aligned}
\widetilde{u}(\widetilde{x},\widetilde{y})
&=\widetilde{x}\cos\theta-\widetilde{y}\sin\theta,\\
\!\!\widetilde{v}(\widetilde{x},\widetilde{y})
&=\widetilde{x}\sin\theta+\widetilde{y}\cos\theta,
\end{aligned}
\end{equation}
where $\theta$ is the angle between the two fundamental cycles of the torus.
Then the new gauge is the Landau gauge in the rotated coordinates
$\mathbf{A}'(\widetilde{x},\widetilde{y})
=B\,\widetilde{u}(\widetilde{x},\widetilde{y})\,\hat{e}_v$.
This is related to the original Landau gauge $\mathbf{A}=B\widetilde{x}\,\hat{e}_y$ 
by the gauge transform $\mathbf{A}'-\mathbf{A}=\nabla \chi$, with
$\nabla=(\partial_{\widetilde{x}},\partial_{\widetilde{y}})$ and
\begin{equation}\label{eq:gauge-transform}
\chi(\widetilde{x},\widetilde{y})
=\frac{B}{2}[\widetilde{u}(\widetilde{x},\widetilde{y})\,
\widetilde{v}(\widetilde{x},\widetilde{y})-\widetilde{x}\widetilde{y}].
\end{equation}

Since the guiding-center momentum is gauge-covariant, the guiding-center 
translation operator in the primed gauge is given by
\begin{equation}\label{eq:T-gauge-transform}
T_i'(\mathbf{a})=e^{ie\chi/\hbar}T_i(\mathbf{a})e^{-ie\chi/\hbar}.
\end{equation}
From this we can obtain the primed variants of the many-body translation 
operators discussed earlier.

We denote by $|j\rangle_v^\prime$ and ${}_v^{\backprime}\!\langle j|$ the 
single-particle states in this gauge 
that satisfy the guiding-center periodic boundary conditions 
$T_i'(\mathbf{L}_1)=T_i'(\mathbf{L}_2)=1$.
The prime highlights the alternative gauge choice $\mathbf{A}'$, while the 
superscript $v$ emphasizes that this state is propagating in the $\hat{e}_v$ 
direction (and localized in the $\hat{e}_y$ direction).
The wave functions of these states resemble those in the unprimed gauge
[Eq.~\eqref{eq:LLL-wf}],
\begin{widetext}
\begin{equation}\label{eq:LLL-wf-v-prime}
\phi_j^{v\prime}(\widetilde{x},\widetilde{y})=
\frac{1}{(\sqrt{\pi}L_1 l_B)^{1/2}}
\sum_n^{\mathbb{Z}}
\exp\left[
-2\pi(j+nN_\phi)\frac{\widetilde{u}(\widetilde{x},\widetilde{y})+
i\widetilde{v}(\widetilde{x},\widetilde{y})}{L_1}
+i\frac{\pi}{N_\phi}\frac{L_2e^{i\theta}}{L_1}(j+nN_\phi)^2
\right]
e^{-[\widetilde{u}(\widetilde{x},\widetilde{y})]^2/(2l_B^2)}.
\end{equation}
\end{widetext}
The state $|j\rangle_v'$ is localized along the line
$\widetilde{y}=\widetilde{x}\cot\theta+jL_2/N_\phi$. The center line moves in 
the \emph{positive} $\hat{e}_y$ direction when we increase $j$.

The Slater determinant basis $|\{j\}\rangle_v'$ for the $N_e$-electron 
Hilbert space generates a representation of the many-body translation 
operators. In particular, we find
\begin{equation}\label{eq:j-vprime-translation}
\begin{aligned}
T_\mathrm{rel}^{x\prime}|\{j\}\rangle_v'&=e^{-i2\pi\sum j/N}|\{j\}\rangle_v',\\
T_\mathrm{rel}^{y\prime}|\{j\}\rangle_v'&=|\{j-q\}\rangle_v',\\
T_\mathrm{cm}^{x\prime}|\{j\}\rangle_v'&=
e^{i2\pi\sum j/N_\phi}|\{j\}\rangle_v',\\
T_\mathrm{cm}^{y\prime}|\{j\}\rangle_v'&=|\{j+1\}\rangle_v'.
\end{aligned}
\end{equation}
In contrast to the unprimed transformations in Eq.~\eqref{eq:j-translation}, 
the translations in the $\mathbf{L}_1$ (rather than $\mathbf{L}_2$) direction 
are diagonal, and the eigenvalues of the diagonal operators differ by 
a complex conjugate.

We can diagonalize the interaction system with the help of the many-body 
translational symmetries. For obvious reasons the eigenstates are just the 
gauge-transform of the states discussed in Sec.~\ref{sec:recombination}.
Recall the state $|\Psi\rangle$ diagonal in 
$(T_\mathrm{rel}^x,T_\mathrm{cm}^y)$ with wave number $(\kappa_x,\kappa_y)$.
Denote by $|\Psi\rangle'$ the gauge-transformed state. Then we have
\begin{equation}
\begin{aligned}
T_\mathrm{rel}^{x\prime}|\Psi\rangle'&=e^{i2\pi \kappa_x/N}|\Psi\rangle',\\
T_\mathrm{cm}^{y\prime}|\Psi\rangle'&=e^{-i2\pi \kappa_y/N_\phi}|\Psi\rangle'.
\end{aligned}
\end{equation}
And the simultaneous eigenstates of $S_x'$, $R_y'$, and the Hamiltonian
can be constructed from $|\Psi\rangle'$ in a parallel manner to the 
construction of $|\Psi;s,r\rangle$ state in Sec.~\ref{sec:recombination}. The 
result is of course given by the gauge transform $|\Psi;s,r\rangle'$.

We can express the $|\Psi;s,r\rangle'$ states by the amplitudes 
${}_v^\backprime\!\langle\{j\}|\Psi;s,r\rangle'$. We emphasize that
\begin{equation}\label{eq:prime-is-not-unprime}
{}_v^\backprime\!\langle\{j\}|\Psi;s,r\rangle'
\neq \langle\{j\}|\Psi;s,r\rangle,
\end{equation}
because the single-particle state $|j\rangle_v'$ is \emph{not} the gauge 
transform of the state $|j\rangle$.
We now seek a link between the two unequal amplitudes in 
Eq.~\eqref{eq:prime-is-not-unprime}.
First, we can transform $|j\rangle_v'$ back to the unprimed gauge, 
using $\chi(\widetilde{x},\widetilde{y})$ given by Eq.~\eqref{eq:gauge-transform}.
The resulting states $|j\rangle_v$ are given by the wave function
\begin{equation}\label{eq:LLL-wf-v}
\phi_j^v(\widetilde{x},\widetilde{y})
=e^{i[\widetilde{x}\widetilde{y}-
\widetilde{u}(\widetilde{x},\widetilde{y})\,
\widetilde{v}(\widetilde{x},\widetilde{y})]/(2l_B^2)}
\phi_j^{v\prime}(\widetilde{x},\widetilde{y}).
\end{equation}
These state are still propagating in the $\hat{e}_v$ direction, as signified by 
the super/sub-script $v$.
We thus arrive at an alternative single-particle basis $|j\rangle_v$ of the 
LLL in the Landau gauge.

Thanks to the invariance of the inner product under the gauge transform, the 
many-body amplitudes satisfy
\begin{equation}\label{eq:prime-is-unprime}
{}_v^\backprime\!\langle\{j\}|\Psi;s,r\rangle'={}_v\!\langle\{j\}|\Psi;s,r\rangle.
\end{equation}
The last ingredient is the expansion of the state $|j\rangle_v$ in the 
$|j\rangle$ basis.
We have already seen some hint from the two representations of the many-body 
translation operators. Comparing Eq.~\eqref{eq:j-translation} and 
Eq.~\eqref{eq:j-vprime-translation}, we find that the Slater determinant state 
formed by a \emph{Fourier series} of the Landau orbitals $|m\rangle$ transforms in 
the same way as $|\{j\}\rangle_v$ under the many-body translation operators.
We can actually prove that 
\begin{equation}\label{eq:j-jv}
|j\rangle_v
=\frac{e^{i(\frac{\pi}{4}-\frac{\theta}{2})}}{\sqrt{N_\phi}}\sum_m^{N_\phi}
e^{-i2\pi jm/N_\phi}|m\rangle.
\end{equation}
This behavior under a Fourier transform is expected of the shifted Gaussian 
form of the single-particle wave function.
The details are presented in Appx.~\ref{sec:j-jv}.

\subsection{Alternative Wannier Construction}

The Wannier basis $|Y,k_x\rangle$ localized in the $\hat{e}_y$ direction (see 
Fig.~\ref{fig:orbitals-x}) can 
be obtained trivially from $|X,k_y\rangle$ by 
exchanging all $x$ and $y$ and replacing $C$ by $-C$. Using this substitution 
rule, we can define $\delta_x$, $\omega_x$, $U_x(k_x)$ alongside their sub-$y$ 
counterparts.

We link the $|Y,k_x\rangle$ state to the LLL 
orbital $|j^{Y,k_x}\rangle_v'$. The index mapping is given by
\begin{equation}
j_v^{Y,k_x}=YN_x-Ck_x+\delta_x.
\end{equation}
Here we put a subscript $v$ to distinguish it from the $j^{X,k_y}$ mapping 
defined earlier.
Then we can transcribe the FQH states $|\Psi;s,r\rangle'$ to the lattice. 
The resulting many-body lattice states $|\Psi;s,r\rangle_\mathrm{lat}'$ are 
translationally invariant. Despite the name, the new lattice states 
$|\Psi;s,r\rangle_\mathrm{lat}'$ are \emph{not the gauge transform} of 
the states $|\Psi;s,r\rangle_\mathrm{lat}$, since they are constructed from 
different Wannier mappings not related by any gauge transform on a lattice at 
zero magnetic field.
We emphasize that only in the LLL does the prime denote the rotated Landau gauge.

The total momenta of $|\Psi;s,r\rangle_\mathrm{lat}'$ are given by
\begin{equation}\label{eq:lattice-K-prime}
\begin{aligned}
K_x&=C[\kappa_x+(s+\delta_x)N_e]\text{ mod }N_x,\\
K_y&=\kappa_y+rN_e\text{ mod }N_y.
\end{aligned}
\end{equation}
The counting of the $q$ states in each total momentum sector is exactly 
the same as the counting of the unprimed lattice states.
The amplitudes in the Bloch basis are given by
\begin{widetext}
\begin{equation}\label{eq:bloch-amplitude-prime}
\langle\{k_x,k_y\}|\Psi;s,r\rangle_\mathrm{lat}'
=\prod\left\{
\frac{(\omega_x)^{k_x}}{\prod_\kappa^{k_x}U_x(\kappa)}
\frac{[\lambda_x(0)]^{k_x}}
{\prod_\kappa^{k_x}A_x(\kappa,0)}
\frac{[\lambda_y(k_x)]^{k_y}}{\prod_\kappa^{k_y}A_y(k_x,\kappa)}
\right\}
\frac{1}{\sqrt{N_y}^{N_e}}\sum_{\{Y\}}
e^{-i2\pi\sum k_y Y/N_y}\,
{}_v^{\backprime}\!\langle\{j_v^{Y,k_x}\}|\Psi;s,r\rangle'.
\end{equation}

By comparison, recall that
\begin{equation}\tag{\ref{eq:bloch-amplitude}}
\langle\{k_x,k_y\}|\Psi;s,r\rangle_\mathrm{lat}
=\prod\left\{
\frac{(\omega_y)^{k_y}}{\prod_\kappa^{k_y}U_y(\kappa)}
\frac{[\lambda_y(0)]^{k_y}}
{\prod_\kappa^{k_y}A_y(0,\kappa)}
\frac{[\lambda_x(k_y)]^{k_x}}{\prod_\kappa^{k_x}A_x(\kappa,k_y)}
\right\}
\frac{1}{\sqrt{N_x}^{N_e}}\sum_{\{X\}}
e^{-i2\pi\sum k_x X/N_x}\,
\langle\{j^{X,k_y}\}|\Psi;s,r\rangle.
\end{equation}
\end{widetext}

\subsection{Lattice Amplitudes}\label{sec:isotropy-proof}

Now we are ready to check the isotropy/nematicity of the lattice states 
obtained by the Wannier constructions.

To avoid the unnecessary clutter, here we focus on the simplest case with 
$C=+1$ and $\delta_x=\delta_y=0$. The generic case is treated in 
Appx.~\ref{sec:isotropy-generic} and the conclusion is the same.

In this case, the Wannier mappings are reduced to
\begin{align}
j^{X,k_y}&=XN_y+k_y,&
j_v^{Y,k_x}&=YN_x-k_x.
\end{align}
Recall from Sec.~\ref{sec:wannier-j} the subtle point that the above mapping 
holds only if $(k_x,k_y)$ is in the principal Brillouin zone, in this case 
$(-N_x~..~0]\times\range{N_y}$.
To minimize the trouble, we set all the $N_e$ momenta in $\{k_x,k_y\}$ on the 
left hand side of Eqs.~\eqref{eq:bloch-amplitude} 
and~\ref{eq:bloch-amplitude-prime} to the principal Brillouin zone. 

Note that the fractional factors enclosed by the curly braces in those 
equations are invariant under a shift of Brillouin zones: for example, 
$(\omega_x)^{k_x}/\prod_\kappa^{k_x}U_x(\kappa)$ does not change if we shift 
$k_x\rightarrow k_x+N_x$. Therefore, we can always pull $(k_x,k_y)$ back to 
the first Brillouin zone $\range{N_x}\times\range{N_y}$ when evaluating the 
prefactors in the curly braces.

Thanks to $C=+1$, $\delta_x=\delta_y=0$, the dependence of $(K_x,K_y)$ on 
$(s,r)$ is identical for both $|\Psi;s,r\rangle_\mathrm{lat}$ and 
$|\Psi;s,r\rangle_\mathrm{lat}'$. Hence we try to relate the 
${}_v^{\backprime}\!\langle\{j\}|\Psi;s',r'\rangle'$
and
$\langle\{j\}|\Psi;s,r\rangle$ with
\begin{equation}\label{eq:srprime}
(s',r')=(s,r).
\end{equation}

\begin{widetext}
We first plug the Fourier transform in Eq.~\eqref{eq:j-jv} into 
Eq.~\eqref{eq:prime-is-unprime}.
We can rewrite each sum over $m$ in the Fourier transform as a double sum over 
all unit cells $X$ and all $l_y$ points in the principal Brillouin zone 
$[0,N_y)$.
We find that
\begin{equation}
{}_v^\backprime\!\langle\{j\}|\Psi;s',r'\rangle'
=\frac{e^{i\gamma}}{\sqrt{N_\phi}^{N_e}}
\sum_{\{X\}}e^{-i2\pi \sum k_x X/N_x}
\sum_{\{l_y\}}
e^{i2\pi \sum l_y Y/N_y}
e^{-i2\pi \sum k_xl_y/N_\phi}\,
\langle\{j^{X,l_y}\}|\Psi;s,r\rangle,
\end{equation}
where $e^{i\gamma}=e^{-iN_e(\frac{\pi}{4}-\frac{\theta}{2})}$ is a constant 
phase factor.
Here we have used $e^{i2\pi XYN_xN_y/N_\phi}=1$, thanks to $N_xN_y=N_\phi$ for 
$|C|=1$.
Therefore, we have
\begin{equation}
\frac{1}{\sqrt{N_y}^{N_e}}\sum_{\{Y\}}
e^{-i2\pi\sum k_y Y/N_y}\,
{}_v^{\backprime}\!\langle\{j_v^{Y,k_x}\}|\Psi;s',r'\rangle'
=e^{i\gamma}e^{-i2\pi\sum k_x k_y/N_\phi}
\frac{1}{\sqrt{N_x}^{N_e}}
\sum_{\{X\}}e^{-i2\pi \sum k_x X/N_x}
\langle\{j^{X,k_y}\}|\Psi;s,r\rangle,
\end{equation}
Here the sum over each $Y$ produces a $\delta^{\text{mod }N_y}_{k_y,l_y}$.
Since both $k_y$ and $l_y$ are limited to $\range{N_y}$, this is just 
$\delta_{k_y,l_y}$ and thus we can remove the sum over $\{l_y\}$.
Putting the above equation into Eq.~\eqref{eq:bloch-amplitude-prime}, we find 
that
\begin{multline}\label{eq:bloch-amplitude-prime-transform}
\langle\{k_x,k_y\}|\Psi;s',r'\rangle_\mathrm{lat}'
=e^{i\gamma}\prod\left\{
\frac{(\omega_x)^{k_x}}{\prod_\kappa^{k_x}U_x(\kappa)}
\frac{[\lambda_x(0)]^{k_x}}
{\prod_\kappa^{k_x}A_x(\kappa,0)}
\frac{[\lambda_y(k_x)]^{k_y}}{\prod_\kappa^{k_y}A_y(k_x,\kappa)}
e^{-i2\pi Ck_xk_y/N_\phi}
\right\}\\
\frac{1}{\sqrt{N_x}^{N_e}}\sum_{\{X\}}
e^{-i2\pi\sum k_x X/N_x}\,
\langle\{j^{X,k_y}\}|\Psi;s,r\rangle.
\end{multline}
\end{widetext}

The above equation holds (up to an inconsequential overall constant phase 
factor) for generic $(C,\delta_x,\delta_y)$ values, as proved in 
Appx.~\ref{sec:isotropy-generic}.

Disregarding the overall constant phase $e^{i\gamma}$, the only difference 
between $\langle\{k_x,k_y\}|\Psi;s',r'\rangle_\mathrm{lat}'$ and
$\langle\{k_x,k_y\}|\Psi;s,r\rangle_\mathrm{lat}$ [Eq.~\eqref{eq:bloch-amplitude}]
is the phase enclosed by the curly braces.
Therefore, the two alternative choices of the Wannier 
localization direction produce two sets of many-body lattice states with
exactly the same absolute value for each Bloch-basis component,
\begin{equation}
\Big|\langle\{k_x,k_y\}|\Psi;s',r'\rangle_\mathrm{lat}'\Big|
=\Big|\langle\{k_x,k_y\}|\Psi;s,r\rangle_\mathrm{lat}\Big|.
\end{equation}

We now examine the phase part. Define the shorthand notations for the phase 
prefactors in Eqs.~\eqref{eq:bloch-amplitude} and~\eqref{eq:bloch-amplitude-prime}:
\begin{equation}
\begin{aligned}
e^{i\phi_{k_x,k_y}}&\equiv
\frac{(\omega_y)^{k_y}}{\prod_\kappa^{k_y}U_y(\kappa)}
\frac{[\lambda_y(0)]^{k_y}}
{\prod_\kappa^{k_y}A_y(0,\kappa)}
\frac{[\lambda_x(k_y)]^{k_x}}{\prod_\kappa^{k_x}A_x(\kappa,k_y)},\\
e^{i\phi_{k_x,k_y}'}&\equiv
\frac{(\omega_x)^{k_x}}{\prod_\kappa^{k_x}U_x(\kappa)}
\frac{[\lambda_x(0)]^{k_x}}
{\prod_\kappa^{k_x}A_x(\kappa,0)}
\frac{[\lambda_y(k_x)]^{k_y}}{\prod_\kappa^{k_y}A_y(k_x,\kappa)}.
\end{aligned}
\end{equation}
Intuitively, $e^{i\phi_{k_x,k_y}}$ comes from 
the parallel transport in the Brillouin zone along
\begin{equation}\nonumber
(0,0)\leftarrow(0,k_y)\leftarrow(k_x,k_y),
\end{equation}
while the phase $e^{i\phi_{k_x,k_y}'}$
comes from the parallel transport along 
\begin{equation}\nonumber
(0,0)\leftarrow(k_x,0)\leftarrow(k_x,k_y).
\end{equation}

As these phases are not identical for all Slater determinants, the Wannier 
wave functictions exhibit a certain degree of nematicity.
However, in the limit of flat Berry curvature, we expect that the two 
different routes of parallel transport picks up a relative phase of 
$e^{i2\pi k_xk_y/N_\phi}$ for $C=1$, which is canceled exactly by the 
$e^{-i2\pi k_xk_y/N_\phi}$ phase from the basis transform from $|j\rangle_v$ 
to $|j\rangle$.

The algebra indeed works out.
As explained earlier, we can shift $(k_x,k_y)$ to 
$\range{N_x}\times\range{N_y}$ when evaluating the phase prefactor.
For flat curvature, we have $\omega_{x,y}=U_{x,y}=1$, and the Wilson loop 
around the rectangle is
\begin{multline}
\prod_\kappa^{k_x}A_x(\kappa,0)
\prod_\kappa^{k_y}A_y(k_x,\kappa)
\left[
\prod_\kappa^{k_y}A_y(0,\kappa)
\prod_\kappa^{k_x}A_x(\kappa,k_y)
\right]^{\dagger}\\
=e^{i2\pi Ck_xk_y/N_\phi}.
\end{multline}
Recall that $\lambda_x(k_y)$ and $\lambda_y(k_x)$ are related to the Wilson 
loops $W_x(k_y)$ and $W_y(k_x)$ by Eq.~\eqref{eq:lambda-W}, and thus
\begin{equation}
\left[\frac{\lambda_y(k_x)}{\lambda_y(0)}\right]^{k_y}
=\left[\frac{\lambda_x(0)}{\lambda_x(k_y)}\right]^{k_x}
=e^{i2\pi Ck_xk_y/N_\phi}.
\end{equation}
Putting all these together, we find as claimed
\begin{equation}
e^{i\phi_{k_x,k_y}}=e^{i\phi_{k_x,k_y}'}
e^{-i2\pi Ck_xk_y/N_\phi},
\end{equation}
and thus the Wannier construction \emph{is} indeed isotropic in the limit of 
flat Berry curvature:
\begin{equation}
|\Psi;s,r\rangle_\mathrm{lat}
\propto|\Psi;s',r'\rangle_\mathrm{lat}'.
\end{equation}

\subsection{Numerical Check}\label{sec:overlap-nematicity}

\begin{figure}[tb]
\centering
\includegraphics[]{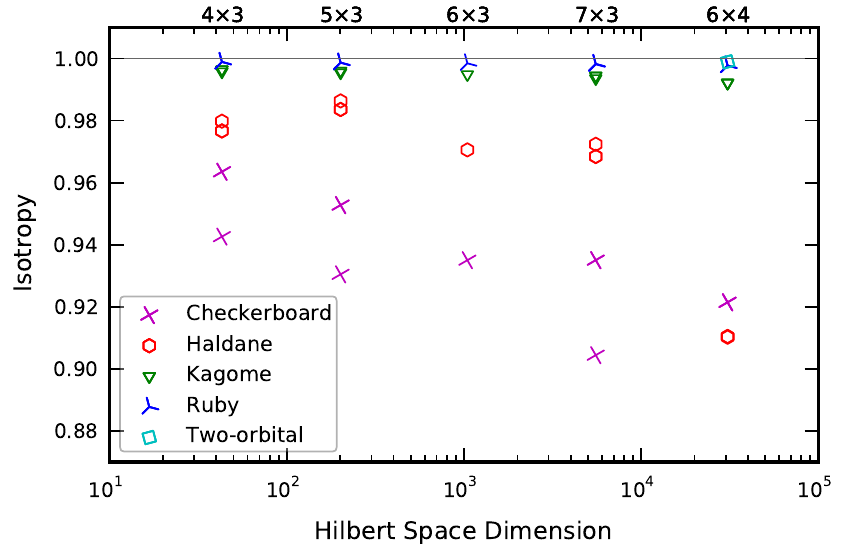}%
\caption{\label{fig:overlap-nematicity}
Isotropy of the Wannier construction,
measured by the overlap between the two sets of Laughlin lattice states 
constructing using the $|X,k_y\rangle$ and $|Y,k_x\rangle$ Wannier bases.
Please refer to the caption and legend of Fig.~\ref{fig:overlap} for the 
annotation of scatter groups and lines.
Inversion symmetry symmetry can be observed similar to Fig.~\ref{fig:overlap}.
}
\end{figure}

Realistic lattice models inevitably introduce fluctuations in the Berry 
curvature. The fluctuations persist in the thermodynamic limit 
$N_e\rightarrow\infty$.
We note that the phase difference between 
$\langle\{k_x,k_y\}|\Psi;s,r\rangle_\mathrm{lat}'$ and
$\langle\{k_x,k_y\}|\Psi;s,r\rangle_\mathrm{lat}$
does not vanish in the continuum limit if the curvature is not flat.
It cannot be eliminated by any point group symmetry, either.

A detailed analysis of the impact on the isotropy of the Wannier construction 
is hard to carry out. But we can still examine the isotropy or nematicity 
numerically. We find strong isotropy in the five FCI models despite 
fluctuating Berry curvature.
Shown in Fig.~\ref{fig:overlap-nematicity} is the isotropy measured by 
the overlap between the Laughlin lattice states constructed using the 
$|X,k_y\rangle$ and the $|Y,k_x\rangle$ Wannier bases. 
The isotropy is higher than $0.997$ for the ruby lattice model and the 
two-orbital model at all system sizes checked.
A decreasing trend is observable for the checkerboard lattice model and the 
Haldane model, but the slope is much slower than that of the overlap values 
with the exact ground states.

\section{Discussion and Conclusion}\label{sec:conclusion}

In this paper, we have studied the Wannier construction of FQH states for
fractional Chern insulators. 
Despite being first proposed more than a year ago~\cite{Qi11:Wavefunction}, no 
overlap studies with the exact diagonalization have been reported.
If one did the comparison naively, low overlap would be obtained (see 
Fig.~\ref{fig:overlap-xlq}).
We have proposed a new, related prescription which fixes several outstanding 
issues in Ref.~\onlinecite{Qi11:Wavefunction}.
In particular, we highlight the gauge freedom in the 1D localized Wannier 
states and the restriction imposed by the Wilson loops.
The key point is that the relative phase between adjacent Wannier states must 
follow the same pattern as the LLL orbitals in the Landau gauge, otherwise the 
states obtained from the FQH-FCI mapping are not relevant to the FCI problem.

We describe and justify in details a finite-size prescription for a generic 
lattice model with multiple sublattices.
We provide an explicit, step-by-step recipe to construct the counterpart of 
any FQH wave function on a Chern insulator.
Our prescription is tailored for the torus geometry to make 
contact with existing numerical studies. It preserves the full translational 
invariance in both directions, as well as the inversion symmetry.
Moreover, in the limit of flat Berry curvature, the constructed many-body 
states are the same using the Wannier basis localized in either $x$ or $y$ 
direction.

We find the major obstacle to the Wannier construction to be the fluctuations 
in the Berry curvature.
We try to accommodate the fluctuations in the localization direction in the 
gauge choice of the Wannier basis.
The fluctuations in the other direction keeps the Wannier centers from evenly 
spaced like the LLL orbitals, and we find no way to ameliorates the situation 
without spoiling translational invariance.
We have no real control over the nematicity developed from the curvature 
fluctuations.
This complication from curvature is an important feature that sets the FCI 
apart from the FQH in the continuum.

In an earlier paper~\cite{Bernevig12:Counting}, two of us discovered a folding 
rule to relate the counting of the FQH states on a torus to the counting of 
the FCI states at each total momentum. It was established by arguing that the 
FCI phase on a lattice is in the same universality class as the FQH phase on a 
torus with $\delta$-function lattice pinning potentials. 
In this paper, the Wannier construction reorganizes the degenerate FQH model 
states and produces a set of model lattice states. These states have exactly 
the same counting in each total momentum sector of the energy and the 
entanglement spectra as predicted by the counting rule in 
Ref.~\onlinecite{Bernevig12:Counting}. 
This provides a concrete implementation of the folding rule that connects
the FCI and the FQH effects.

We perform the Wannier construction for the Laughlin state numerically in 
the five lattice models known to support a fractionalized phase.
We find that the model state obtained from the Wannier construction and the 
actual FCI ground state have consistently large overlap in all cases, and 
their entanglement spectra exhibit very similar structures.
The many-body physics in Chern insulators at filling $\nu=1/3$ is indeed a 
close parallel to the Laughlin-type of FQH physics in the continuum with a 
strong magnetic field.

Using these five examples, we demonstrate that our prescription for the 
single-particle gauge is at or very close to the local optimum in terms of the 
overlap with the exact ground states. We also show that the our prescription 
has strong isotropy by checking the overlap between the many-body states 
constructed using the Wannier bases localized in either $x$ or $y$ direction.

Our results provide another comparison across the array of known FCI models.
Among the five models that we have checked, the ground states of the 
Kagome lattice model and the ruby lattice model have the largest overlap with 
the Laughlin model state.
This is consistent with the finding in Ref.~\onlinecite{Wu12:Zoology} that 
these two models host the most stable Laughlin-type FCI phase among the five.

We note a few interesting directions for future work.
Two of us recently studied the thin-torus limit of the FCI.~\cite{Bernevig12:TT}
In that limit, the projected interaction is dominated by classical 
electrostatic terms and allows an exact or perturbative solution.
This may provide a better understanding of the varied performance of the 
Wannier construction in different models.

In this paper we focus the numerical tests on the Laughlin state.
The general formalism of the Wannier construction, however, is able to
handles more complicated FQH states. 
Lattice analogs of composite fermion states~\cite{Jain89:CF} and non-Abelian 
FQH states such as Moore-Read~\cite{Moore91:MR} and Read-Rezayi 
states~\cite{Read99:RR} have been identified in a few Chern 
insulators.~\cite{Wang12:MR,Bernevig12:Counting,Wu12:Zoology,Liu12:CF}
Moreover, it was proposed that
topological flat bands with Chern 
number $|C|\geq 2$~\cite{Wang11:Dice,Wang12:C2,Yang12:Flatband,
Trescher12:Flatband,Liu12:HigherC}
may be may be understood as a multi-component FQH system.~\cite{Barkeshli11:Nematic}
It would be interesting to see how well these states could be captured by some 
established FQH trial wave functions mapped to the lattice by the Wannier 
construction.

Time-reversal invariant fractional $\mathbb{Z}_2$ topological 
insulators have been constructed from two copies of 
FCI and the field theory describing their low-energy excitations have been 
established.~\cite{Neupert11:Z2,Santos11:BF}
More recently, the algebraic structure of the projected density operators for 
topological insulators in three~\cite{Neupert12:3DAlgebra} or 
higher dimensions~\cite{Estienne12:DAlgebra} have been studied.
With necessary generalizations, the gauge-fixed Wannier construction may be 
useful for the study of the wave functions of these more complicated systems.
In particular, the case of 2D FTI (quantum spin Hall) can be tackled similar 
to the FCI problem, as the FTI system becomes decoupled in the Wannier basis.

Another interesting approach to build trial wave functions for the FCI is the 
parton construction~\cite{Lu11:Parton,McGreevy11:Parton}, 
initially pioneered by Wen~\cite{Wen91:Parton,Wen99:Parton} to construct the 
FQH wave functions in the continuum.
This alternative is plagued by a critical issue at the moment.
To achieve fractional filling $\nu=1/q$ of electrons, the number of 
partons in a fully filled band should be reduced accordingly.
For the original FQH system, this is not a problem since the size of the 
parton LLL orbital increases automatically due to the decrease of charge.
On a lattice, this change has to be put in by hand through a unit-cell enlargement.
Refs.~\onlinecite{Lu11:Parton,McGreevy11:Parton} discussed how to preserve the 
lattice symmetries in the electronic wave functions, but their solutions 
involve partonic bands that are \emph{drastically different} from the original 
electronic band: even the band \emph{topologies} may be totally different. 
It would be interesting to find a cure to this unsettling problem,
to develop a concrete prescription for correct torus degeneracies,
and to compare with the Wannier construction and the exact ground states.

\section{Acknowledgements}

We wish to thank A.~Alexandradinata, X.~Dai, C.~Fang, F.D.M.~Haldane, 
X.L.~Qi, and B.~Yang for helpful comments and inspiring discussions.
BAB was supported by Princeton Startup Funds, NSF CAREER DMR-095242, 
ONR-N00014-11-1-0635, DARPA-N66001-11-1-4110, Packard Foundation, Keck grant, 
and NSF-MRSEC DMR-0819860 at Princeton University.
NR was supported by NSF CAREER DMR-095242, ONR-N00014-11-1-0635, Packard 
Foundation, and Keck grant.
YLW was supported by NSF CAREER DMR-095242.
BAB thanks Technion, Israel, and Ecole Normale Superieure, Paris, 
for generous hosting during the stages of this work.
YLW thanks the Institute of Physics, Chinese Academy of Sciences for 
hospitality.

\appendix

\section{FQH Projected Hamiltonian}\label{sec:FQH-Hamiltonian}

In this Appendix we give the expression of the FQH interacting Hamiltonian 
projected to the lowest Landau level (LLL) on a twisted torus.
This is a trivial generalization of the formula given in 
Ref.~\onlinecite{Yoshioka83:Torus}.

In the limit of large cyclotron energy, we keep in the Hilbert space only the 
states in the LLL. Since the kinetic energy is fully quenched,
the Hamiltonian is simply the projected density-density interaction term $V$.

Define the field operator $\psi_j$ to annihilate the Landau orbital state 
$|j\rangle$.
The LLL-projected density operator can be written in momentum space as
\begin{align}\label{eq:projected-density-1}
\rho_\mathbf{q}
&=\sum_{j_1,j_2}
\langle j_1|e^{-i\mathbf{q}\cdot\widetilde{\mathbf{r}}}|j_2\rangle
\psi_{j_1}^\dagger\psi_{j_2}\\
&=\int \mathrm{d}\widetilde{\mathbf{r}}\,
e^{-i\mathbf{q}\cdot\widetilde{\mathbf{r}}}\sum_{j_1,j_2}
\phi_{j_1}^*(\widetilde{\mathbf{r}})
\phi_{j_2}(\widetilde{\mathbf{r}})
\psi_{j_1}^\dagger\psi_{j_2},
\end{align}
where the reciprocal lattice vector $\mathbf{q}$ is given by
$\mathbf{q}=q_1\mathbf{G}_1+q_2\mathbf{G}_2$ with $(q_1,q_2)\in\mathbb{Z}^2$,
the continuous coordinate $\widetilde{\mathbf{r}}$ is integrated over the 
principal region of the torus, and the orbital index $j_1$, $j_2$ are summed 
over $\range{N_\phi}$.
Following the detailed steps laid out in Ref.~\onlinecite{Yoshioka10:Book}, 
we can simplify the above formula and get
\begin{equation}\label{eq:projected-density}
\rho_\mathbf{q}=e^{-|\mathbf{q}|^2l_B^2/4}\sum_j^{N_\phi}
e^{-i2\pi q_1(j+q_2/2)/N_\phi}\psi_j^\dagger\psi_{j+q_2}.
\end{equation}
In the second-quantized form, the projected two-body interaction can be 
written as
\begin{equation}
V=\frac{1}{2L_1L_2\sin\theta}
\sum_\mathbf{q}V_\mathbf{q}\rho_{-\mathbf{q}}\rho_\mathbf{q},
\end{equation}
where the reciprocal lattice vector 
$\mathbf{q}=q_1\mathbf{G}_1+q_2\mathbf{G}_2$ is summed 
over the infinite grid $(q_1,q_2)\in\mathbb{Z}^2$.
Plugging in $\rho_\mathbf{q}$, we find
\begin{multline}
V=\frac{1}{2L_1L_2\sin\theta}
\sum_{\mathbf{q}}e^{-|\mathbf{q}|^2l_B^2/2}\,V_\mathbf{q}
\sum_{j_1}^{N_\phi}\sum_{j_2}^{N_\phi}\\
e^{i2\pi q_1(j_1-j_2-q_2)/N_\phi}\,
\psi^\dagger_{j_1}\psi^\dagger_{j_2}
\psi^{\phantom{\dagger}}_{j_2+q_2}\psi^{\phantom{\dagger}}_{j_1-q_2},
\end{multline}
where 
$V_\mathbf{q}=\int\mathrm{d}\mathbf{r}e^{-i\mathbf{q}\cdot\mathbf{r}}V(\mathbf{r})$
is the Fourier transform of the interaction
potential $V(\mathbf{r})$.~\cite{Haldane85:TorusBZ}
The Fourier coefficient $V_\mathbf{q}$ can be expressed in terms of the Haldane 
pseudopotentials $V_m$,~\cite{Haldane83:Sphere}
\begin{equation}
V_\mathbf{q}=4\pi l_B^2\sum_{m=0}^{\infty}V_m L_m(|\mathbf{q}|^2l_B^2),
\end{equation}
where $L_m(x)$ are the Laguerre polynomials.

As a side note, we point out an interesting link between adjacent $|j\rangle$ 
orbitals. Combining Eqs.~\eqref{eq:projected-density-1} 
and~\eqref{eq:projected-density} at $\mathbf{q}=\mathbf{G}_2$, we find
\begin{equation}
\langle j|e^{-i\mathbf{G}_2\cdot\widetilde{\mathbf{r}}}|j+1\rangle
=e^{-|\mathbf{G}_2|^2l_B^2/4}\in\mathbb{R}_+.
\end{equation}
This can be viewed as the defining property of the particular gauge choice 
of $\phi_j(\widetilde{\mathbf{r}})$ made in Eq.~\eqref{eq:LLL-wf}.

\section{Orthogonality Problem}\label{sec:orthogonality-appx}

In this Appendix, we show that the eigenstates of the projected position 
operator $\widehat{\mathcal{X}}$ at finite size are \emph{not} orthogonal to each 
other due to the non-unitarity of the Berry connections. 

For simplicity, we focus on the case of a single occupied band; the 
problem is generic for any number of occupied bands.
In this case, the Berry connection $\mathcal{A}_x(k_x,k_y)$ is Abelian, and 
the project position operator in the $x$ direction is given by 
Eq.~\eqref{eq:non-unitary-XY}:
\begin{equation}
\widehat{\mathcal{X}}=\sum_{k_x,k_y}|k_x,k_y\rangle
\mathcal{A}_x(k_x,k_y)\langle k_x+1,k_y|,
\end{equation}
where the Berry connection is defined as [Eq.~\eqref{eq:discrete-A-nonunitary}]
\begin{equation}\nonumber
\mathcal{A}_x(k_x,k_y)
=\!\sum_\alpha e^{-i2\pi\epsilon^x_\alpha/N_x}
u_{m,\alpha}^*(k_x,k_y)u_{n,\alpha}(k_x\!+\!1,k_y).
\end{equation}
Generically, at finite size we have $|\mathcal{A}_x(k_x,k_y)|<1$; only in the 
thermodynamic limit does this become an equality.

Since $\widehat{\mathcal{X}}$ is diagonal in $k_y$, we can solve the eigenvalue 
problem at each $k_y$ independently. The $N_x$ eigenvalues of 
$\widehat{\mathcal{X}}$ at $k_y$ are
\begin{equation}
e^{-i2\pi X/N_x}D(k_y)\lambda_x(k_y),
\end{equation}
labeled by $X\in\range{N_x}$. Here, the unitary phase factor $\lambda_x(k_y)$ 
is the $N_x$-th root of the \emph{phase} part of the Wilson loop along a fixed 
$k_y$, defined in Eq.~\eqref{eq:lambda-W}, while the positive real number 
$D(k_y)$ is the $N_x$-th root of the absolute value of the Wilson loop,
\begin{equation}
D(k_y)=\left|\prod_\kappa^{N_x}\mathcal{A}_x(\kappa,k_y)\right|^{1/N_x}.
\end{equation}
The corresponding eigenstates are
\begin{equation}\nonumber
|X,k_y\rangle=\mathcal{N}_{k_y}\!
\sum_{k_x}^{N_x}e^{-i2\pi k_x X/N_x}
\frac{[D(k_y)\lambda_x(k_y)]^{k_x}}{\prod_\kappa^{k_x}\mathcal{A}_x(\kappa,k_y)}
|k_x,k_y\rangle,
\end{equation}
and the normalization factor is given by
\begin{equation}
\mathcal{N}_{k_y}=\left\{\sum_{k_x}^{N_x}
\frac{[D(k_y)]^{2k_x}}{\prod_\kappa^{k_x}|\mathcal{A}_x(\kappa,k_y)|^2}
\right\}^{-1/2}.
\end{equation}
Therefore,
\begin{multline}
\langle X_1,k_y|X_2,k_y\rangle
=[\mathcal{N}_{k_y}]^*\mathcal{N}_{k_y}\\
\times\sum_{k_x}^{N_x}
e^{i2\pi k_x(X_1-X_2)}\frac{[D(k_y)]^{k_x}}{\prod_\kappa^{k_x}|A_x(\kappa,k_y)|}.
\end{multline}
Evidently, in general the Wannier states localized in different unit cells at 
each $k_y$ are \emph{not} orthogonal at finite size, preventing a nice mapping 
into the orthogonal LLL orbitals.
The Wannier orbitals become orthogonal quickly in the thermodynamic limit due 
to the oscillatory nature of the sum, which averages to zero if $X_1\neq X_2$.

\section{Inversion Symmetry}

\subsection{Inversion Operator}\label{sec:inv-operator}

We derive the inversion transformation of the exponentiated position operators 
$\widehat{x}$ and $\widehat{y}$ in Eq.~\eqref{eq:x-y-inversion} and provide an 
explicit expression for the sewing matrix elements $e^{i\xi_{k_x,k_y}}$ 
defined in Eq.~\eqref{eq:inversion-sewing}.

By properly choosing the reference point in each unit cell, we can put the 
inversion centers on the Bravais lattice $(x,y)\in\mathbb{Z}^2$.
Inversion takes the $\alpha$ orbital 
located at $(x+\epsilon^x_\alpha,y+\epsilon^y_\alpha)$ to 
$(-x-\epsilon^x_\alpha,-y-\epsilon^y_\alpha)$. For the insulator to be inversion 
symmetric, we must have an orbital at 
$(-x-\epsilon^x_\alpha,-y-\epsilon^y_\alpha)$ before taking inversion. So there 
must exist a Bravais lattice shift 
$(\Delta^x_\alpha,\Delta^y_\alpha)\in\mathbb{Z}^2$ 
and sublattice displacements 
$(\epsilon^x_\beta,\epsilon^y_\beta)\in\mathbb{R}^2$ such that
\begin{equation}
\label{eq:inversion-displacement}
\begin{aligned}
-x-\epsilon^x_\alpha&=-x-\Delta^x_\alpha+\epsilon^x_\beta\\
-y-\epsilon^y_\alpha&=-y-\Delta^y_\alpha+\epsilon^y_\beta
\end{aligned}
\end{equation}
Note that $(\Delta^x_\alpha,\Delta^y_\alpha)$ depends only on $\alpha$, not on 
$\beta$. Different orbitals $\beta$ may share the same value of 
$(\epsilon^x_\beta,\epsilon^y_\beta)$.
In orbital space, inversion is implemented by
\begin{equation}
\mathcal{P}|x,y,\alpha\rangle=\sum_\beta \mathbb{P}_{\alpha\beta}
|{-x-\Delta^x_\alpha},-y-\Delta^y_\alpha,\beta\rangle,
\end{equation}
where the matrix $\mathbb{P}$ is unitary and satisfies $\mathbb{P}^2=1$. 

\begin{widetext}
The position operator $\widehat{y}$ transforms under inversion
\begin{align}
\mathcal{P}\widehat{y}\mathcal{P}
&=\sum_{x,y}
\sum_\alpha e^{-i2\pi(y+\epsilon^y_\alpha)/N_y}
\sum_{\beta\gamma}\mathbb{P}_{\alpha\beta}
|{-x-\Delta^x_\alpha},-y-\Delta^y_\alpha,\beta\rangle
\langle{-x-\Delta^x_\alpha},-y-\Delta^y_\alpha,\gamma|
\mathbb{P}_{\gamma\alpha}\\
&=\sum_{x,y}
\sum_\alpha e^{-i2\pi(-y-\Delta^y_\alpha+\epsilon^y_\alpha)/N_y}
\sum_{\beta\gamma}\mathbb{P}_{\alpha\beta}
|x,y,\beta\rangle\langle x,y,\gamma|
\mathbb{P}_{\gamma\alpha}\\
&=\sum_{x,y}
\sum_{\beta\gamma} e^{i2\pi(y+\epsilon^y_\beta)/N_y}
\left(\sum_\alpha
\mathbb{P}_{\gamma\alpha}\mathbb{P}_{\alpha\beta}\right)
|x,y,\beta\rangle\langle x,y,\gamma|\\
&=\widehat{y}^\dagger.
\end{align}
In the second step we changed the summation dummy index 
$(x,y)\rightarrow(-x-\Delta^x_\alpha,-y-\Delta^y_\alpha)$, and in the third 
step we made use of Eq.~\eqref{eq:inversion-displacement}. Similarly, we can 
show that $\mathcal{P}\widehat{x}\mathcal{P}=\widehat{x}^\dagger$.
In momentum/orbital space, we have
\begin{align}
\mathcal{P}|k_x,k_y,\alpha\rangle
&=\frac{1}{\sqrt{N_xN_y}}\sum_{x,y}
e^{i2\pi(k_xx/N_x+k_yy/N_y)}\sum_\beta\mathbb{P}_{\alpha\beta}
|{-x-\Delta^x_\alpha},-y-\Delta^y_\alpha,\beta\rangle\\
&=e^{-i2\pi(k_x\Delta^x_\alpha/N_x + k_y\Delta^y_\alpha/N_y)}
\sum_\beta\mathbb{P}_{\alpha\beta}|{-k_x},-k_y,\beta\rangle.
\end{align}
\end{widetext}

The sewing matrix elements 
\begin{equation}
e^{i\xi_{k_x,k_y}}=\langle k_x,k_y|\mathcal{P}|{-k_x},-k_y\rangle
\end{equation}
can be expressed in terms of 
$\mathbb{P}_{\alpha\beta}$,
\begin{multline}
e^{i\xi_{k_x,k_y}}=\sum_\alpha\sum_\beta
e^{i2\pi(k_x\Delta^x_\alpha/N_x+k_y\Delta^y_\alpha/N_y)}\\
\times u^*_\beta(k_x,k_y)\mathbb{P}_{\alpha\beta}u_\alpha(-k_x,-k_y).
\end{multline}

\subsection{Inversion Symmetry in Curvature Fluctuations}
\label{sec:U-inversion}

We study the inversion property of the curvature fluctuation 
$U_y(k_y)=\mathcal{U}_y(k_y)/|\mathcal{U}_y(k_y)|$, where $\mathcal{U}_y(k_y)$ is 
defined in Eq.~\eqref{eq:U-def} as:
\begin{equation}\tag{\ref{eq:U-def}}
\mathcal{U}_y(k_y)=\frac{1}{N_x}\sum_{k_x}^{N_x}
\frac{W_{\protect\rule{5pt}{2pt}}(k_x,k_y)}{\overline{W}_{\protect\rule{5pt}{2pt}}(k_x,k_y)}.
\end{equation}

First, we study the Wilson loop $W_{\protect\rule{5pt}{2pt}}(k_x,k_y)$.
The inversion transformation of the Berry connection in
Eqs.~\eqref{eq:Ax-inversion} and~\eqref{eq:Ay-inversion} relates
$W_{\protect\rule{5pt}{2pt}}(k_x,k_y)$ to the Wilson loop around 
$(N_x-k_x,-k_y)(N_x,-k_y)(N_x,-k_y-1)(N_x-k_x,-k_y-1)(N_x-k_x,-k_y)$, except 
for the corner case $W_{\protect\rule{5pt}{2pt}}(0,k_y)=1$. Due to the \emph{loop} 
structure, all the sewing matrix elements cancel completely, and we have
\begin{equation}\label{eq:Wsq-inversion}
W_{\protect\rule{5pt}{2pt}}(k_x,k_y)
\frac{W_{\protect\rule{5pt}{2pt}}(N_x-k_x,-k_y-1)}{W_{\protect\rule{5pt}{2pt}}(N_x,-k_y-1)}=1.
\end{equation}
It is easy to check that the corner case $k_x=0$ is incorporated as well.

Then we study the denominator 
$\overline{W}_{\protect\rule{5pt}{2pt}}(k_x,k_y)=[\mu_x(k_y)]^{k_x}$, with $\mu_x(k_y)$ 
defined by Eq.~\eqref{eq:mu}.
We can prove that
\begin{equation}\label{eq:Wbar-inversion}
\overline{W}_{\protect\rule{5pt}{2pt}}(k_x,k_y)=\overline{W}_{\protect\rule{5pt}{2pt}}(k_x,-k_y-1).
\end{equation}
Setting $k_x=N_x$ in Eq.~\eqref{eq:Wsq-inversion}, we have [recall that 
$W_{\protect\rule{5pt}{2pt}}(0,-k_y-1)=1$]
\begin{equation}
W_{\protect\rule{5pt}{2pt}}(N_x,k_y)=W_{\protect\rule{5pt}{2pt}}(N_x,-k_y-1).
\end{equation}
Using Eq.~\eqref{eq:mu-arg} to fix the branch choice, we can take the $N_x$-th 
root of this equation and get
\begin{equation}
\mu_x(k_y)=\mu_x(-k_y-1).
\end{equation}
Taking the $k_x$-th power of the above equation proves Eq.~\eqref{eq:Wbar-inversion}.

Combine Eq.~\eqref{eq:Wsq-inversion} with Eq.~\eqref{eq:Wbar-inversion}, we 
can show that the curvature fluctuation $\mathcal{U}_y(k_y)$ satisfy
\begin{align}
\nonumber
[\mathcal{U}_y(k_y)]^*\!
&=\frac{1}{N_x}\sum_{k_x}^{N_x}\frac{\left[W_{\protect\rule{5pt}{2pt}}(k_x,k_y)\right]^*}
{\left[\overline{W}_{\protect\rule{5pt}{2pt}}(k_x,k_y)\right]^*}\\
\nonumber
&=\frac{1}{N_x}\!\sum_{k_x}^{N_x}\frac{W_{\protect\rule{5pt}{2pt}}(N_x-k_x,-k_y-1)}
{W_{\protect\rule{5pt}{2pt}}(N_x,-k_y-1)/\,\overline{W}_{\protect\rule{5pt}{2pt}}(k_x,-k_y-1)}\\
\nonumber
&=\frac{1}{N_x}\!\sum_{k_x}^{N_x}\frac{W_{\protect\rule{5pt}{2pt}}(N_x-k_x,-k_y-1)}
{\overline{W}_{\protect\rule{5pt}{2pt}}(N_x,-k_y-1)/\,\overline{W}_{\protect\rule{5pt}{2pt}}(k_x,-k_y-1)}\\
\nonumber
&=\frac{1}{N_x}\sum_{k_x}^{N_x}\frac{W_{\protect\rule{5pt}{2pt}}(N_x-k_x,-k_y-1)}
{W_{\protect\rule{5pt}{2pt}}(N_x-k_x,-k_y-1)}\\
\label{eq:U-raw-inv}
&=\mathcal{U}_y(-k_y-1).
\end{align}
Here we have used 
$W_{\protect\rule{5pt}{2pt}}(N_x,k_y)=\overline{W}_{\protect\rule{5pt}{2pt}}(N_x,k_y)$,
which follows from the definition of 
$\overline{W}_{\protect\rule{5pt}{2pt}}(k_x,k_y)=[\mu_x(k_y)]^{k_x}$ and Eq.~\eqref{eq:mu-W}.
This proves Eq.~\eqref{eq:U-inversion}, namely,
\begin{equation}\nonumber
U_y(k_y)U_y(-k_y-1)=1.
\end{equation}

\subsection{Inversion Transformation of the Wannier States}
\label{sec:Xky-inversion-raw}

We prove the inversion transformation of $|X,k_y\rangle$ in 
Eq.~\eqref{eq:Xky-inversion-raw}.

Using $[\lambda_x(k_y)]^{N_x}=W_x(k_y)$ [Eq.~\eqref{eq:lambda-W}],
we can rewrite the inversion transformation of $W_x(k_y)$ in 
Eq.~\eqref{eq:Wilson-inversion} as
\begin{equation}
\left[\lambda_x(k_y)\lambda_x(-k_y)\right]^{N_x}=1.
\end{equation}
Taking the $N_x$-th root, we have
\begin{equation}\label{eq:lambda0-inversion}
\lambda_x(k_y)=e^{-i2\pi n/N_x}/\lambda_x(-k_y),
\end{equation}
where $n\in\mathbb{Z}$ is to be determined. Recall that \emph{by definition}, 
both $\lambda_x(k_y)$ and $\lambda_x(-k_y)$ must have their argument angles 
in $(-2\pi/N_x,0]$ mod $2\pi$. So there are the following two cases:
\begin{description}
\item[``on-site''] If $W_x(k_y)=1$, we have 
$\lambda_x(k_y)=\lambda_x(-k_y)=1$, and thus 
\begin{equation}\label{eq:lambda-inv-onsite}
\lambda_x(k_y)=1/\lambda_x(-k_y).
\end{equation}
The Wannier centers are located on the Bravais lattice.
\item[``off-site''] If $W_x(k_y)\neq 1$, we have
\begin{equation}\label{eq:lambda-inv-offsite}
\lambda_x(k_y)=e^{-i2\pi/N_x}/\lambda_x(-k_y),
\end{equation}
since both $\lambda_x(k_y)$ and $\lambda_x(-k_y)$ have argument angle in 
$(-2\pi/N_x,0)$ mod $2\pi$.
In particular, for $W_x(k_y)=-1$, we have 
$\lambda_x(k_y)=\lambda_x(-k_y)=e^{-i\pi/N_x}$.
\end{description}

\begin{widetext}
From the inversion transformation of the Berry connection $A_x(k_y)$ in 
Eq.~\eqref{eq:Ax-inversion} we can get
\begin{equation}
\left.
\prod_{\kappa}^{N_x}A_x(\kappa,k_y)
\right/
\prod_{\kappa}^{N_x-k_x}A_x(\kappa,k_y)
=\prod_{\kappa=N_x-k_x}^{N_x-1}A_x(\kappa,k_y)
=e^{i\xi_{N_x-k_x,k_y}}\;e^{-i\xi_{N_x,k_y}}
\left/\,\prod_{\kappa}^{k_x}A_x(\kappa,-k_y)\right..
\end{equation}
Therefore, we have the inversion transformation of the Wannier states
\begin{align}
e^{i\xi_{0,k_y}}e^{-i\Phi_y(X,k_y)}\mathcal{P}|X,k_y\rangle
&=\frac{1}{\sqrt{N_x}}\sum_{k_x}^{N_x}
e^{-i2\pi k_xX/N_x}
\frac{\left[\lambda_x(k_y)\right]^{k_x}}
{\prod_{\kappa}^{k_x}A_x(\kappa,k_y)}\;
e^{i\xi_{N_x,k_y}-i\xi_{k_x,k_y}}|{-k_x},-k_y\rangle\\
&=\frac{1}{\sqrt{N_x}}\sum_{k_x}^{N_x}
e^{i2\pi k_x X/N_x}
\frac{\left[\lambda_x(k_y)\right]^{N_x-k_x}}
{\prod_{\kappa}^{N_x-k_x}A_x(\kappa,k_y)}\;
e^{i\xi_{N_x,k_y}-i\xi_{N_x-k_x,k_y}}|k_x,-k_y\rangle\\
&=\frac{1}{\sqrt{N_x}}\sum_{k_x}^{N_x}
e^{i2\pi k_x X/N_x}
\left[\lambda_x(k_y)\right]^{-k_x}
\frac{\prod_{\kappa}^{N_x}A_x(\kappa,k_y)}
{\prod_{\kappa}^{N_x-k_x}A_x(\kappa,k_y)}\;
e^{i\xi_{N_x,k_y}-i\xi_{N_x-k_x,k_y}}|k_x,-k_y\rangle\\
&=\frac{1}{\sqrt{N_x}}\sum_{k_x}^{N_x}
e^{-i2\pi k_x(-X)/N_x}
\frac{\left[\lambda_x(k_y)\right]^{-k_x}}
{\prod_{\kappa}^{k_x}A_x(\kappa,-k_y)}
|k_x,-k_y\rangle\\
&=
\begin{cases}
e^{-i\Phi_y(-X,-k_y)}\,|{-X},-k_y\rangle
& \text{if }W_x(k_y)=1,\\
e^{-i\Phi_y(-X-1,-k_y)}\,|{-X-1},-k_y\rangle
& \text{otherwise.}
\end{cases}
\end{align}
\end{widetext}
Here in the second step we performed a change of summation variable 
$k_x\rightarrow N_x-k_x$ and made use of the invariance of each term under 
$k_x\rightarrow k_x+N_x$, while in the last step we used 
Eqs.~\eqref{eq:lambda-inv-onsite} and~\eqref{eq:lambda-inv-offsite}.

\subsection{Inversion Transformation of the 1D Index}\label{sec:jXky-inversion}

We prove the inversion transformation rule of $j^{X,k_y}$ in 
Eq.~\eqref{eq:jXky-inversion}, namely,
\begin{equation}\tag{\ref{eq:jXky-inversion}}
j^{X,k_y}\rightarrow
\begin{cases}
-j^{X,k_y}-1 & \text{if }W_x(0)=-1\text{ and }N_y\text{ odd};\\
-j^{X,k_y} & \text{otherwise}.
\end{cases}
\end{equation}

We will use the inversion transformation of $(X,k_y)$ in 
Eq.~\eqref{eq:Xky-label-inversion}, namely,
\begin{equation}\nonumber
(X,k_y)\rightarrow
\begin{cases}
(-X,-k_y) & \text{if }W_x(k_y)=1,\\
(-X-1,-k_y) & \text{otherwise.}
\end{cases}
\end{equation}
A discussion of the location of $W_x(k_y)=1$ is thus necessary.
At a generic $k_y$, due to the absence of any protection by symmetry, 
in general we can assume $W_x(k_y)\neq 1$.
With inversion symmetry, We can safely assume that $W_x(k_y)=1$ happens only 
at $k_y=0$ or $\frac{N_y}{2}$, otherwise the flow of the Wannier center would 
exhibit ``zigzag'' patterns.

We only elaborate the proof for $C=+1$; the situation when $C=-1$ is 
completely analogous.
We discuss the three cases of $\delta_y$ in Eq.~\eqref{eq:inv-delta} one by 
one, namely,
\begin{equation}\nonumber
\delta_y=
\begin{cases}
0 & \text{if }W_x(0)=1;\\
N_y/2 & \text{if }W_x(0)=-1\text{ and }N_y\text{ even};\\
(N_y-1)/2 & \text{if }W_x(0)=-1\text{ and }N_y\text{ odd}.
\end{cases}
\end{equation}
Recall that the 1D label $j^{X,k_y}=XN_y+Ck_y+\delta_y$ is defined with 
$Ck_y+\delta_y\in\range{N_y}$.
As emphasized in Sec.~\ref{sec:wannier-j}, before performing the mapping, we 
need to first shift $k_y$ to the principal Brillouin zone (pBZ).
The pBZ is defined by $Ck_y+\delta_y\in\range{N_y}$. The three cases for the 
shift parameter $\delta_y$ and the corresponding pBZ are illustrated in 
Fig.~\ref{fig:delta-y-inv}.  

In the first case, $W_x(0)=1$. In this case we have $\delta_y=0$, and the 
principal Brillouin zone is $\mathrm{pBZ}=\range{N_y}$.
If $k_y=0$, we have $W_x(0)=1$ and $-k_y=0\in\mathrm{pBZ}$, and thus under 
inversion
\begin{equation}\nonumber
j^{X,0}=XN_y\rightarrow -XN_y=-j^{X,0}.
\end{equation}
If $k_y\in\range[(]{N_y}$, we have $W_x(k_y)\neq 1$ and 
$N_y-k_y\in\mathrm{pBZ}$, and thus under inversion 
\begin{multline}\nonumber
j^{X,k_y}=XN_y+k_y\\
\rightarrow(-X-1)N_y+N_y-k_y=-j^{X,k_y}.
\end{multline}

In the second case, $W_x(0)=-1$ and $N_y$ is even. We have 
$\delta_y=\frac{N_y}{2}$, and $\mathrm{pBZ}=[-\frac{N_y}{2}~..~\frac{N_y}{2})$.
Due to $|C|=1$, we have $W_x(-\frac{N_y}{2})=-1/W_x(0)=1$ 
[Eq.~\eqref{eq:W-inv-Chern}].
If $k_y=-\frac{N_y}{2}$, we have $W_x(-\frac{N_y}{2})=1$ and 
$-N_y-k_y=-\frac{N_y}{2}\in\mathrm{pBZ}$, and thus under inversion
\begin{multline}\nonumber
j^{X,-\frac{N_y}{2}}=XN_y-\frac{N_y}{2}+\frac{N_y}{2}\\
\rightarrow -XN_y+\frac{N_y}{2}-\frac{N_y}{2}=-j^{X,-\frac{N_y}{2}}.
\end{multline}
If $k_y\in(-\frac{N_y}{2}~..~\frac{N_y}{2})$, we have $W_x(k_y)\neq 1$ and 
$-k_y\in\mathrm{pBZ}$, and thus under inversion
\begin{multline}\nonumber
j^{X,k_y}=XN_y+k_y+\frac{N_y}{2}\\
\rightarrow (-X-1)N_y+(-k_y)+\frac{N_y}{2}=-j^{X,k_y}.
\end{multline}

In the third case, $W_x(0)=-1$ and $N_y$ is odd. We have $W_x(k_y)\neq 1$ and 
$\delta_y=\frac{N_y-1}{2}$, and 
$\mathrm{pBZ}=[-\frac{N_y-1}{2}~..~\frac{N_y-1}{2}]$ (illustrated in 
Fig.~\ref{fig:delta-y-inv}c). For any 
$k_y\in\mathrm{pBZ}$, we have $-k_y\in\mathrm{pBZ}$, and thus
\begin{multline}\nonumber
j^{X,k_y}=XN_y+k_y+\frac{N_y-1}{2}\\
\rightarrow (-X-1)N_y+(-k_y)+\frac{N_y-1}{2}=-j^{X,k_y}-1.
\end{multline}

Summarizing the three cases, we have proved 
Eq.~\eqref{eq:jXky-inversion}. 
We emphasize that for a given system [and thus given $W_x(0)$ and $N_y$], all 
the $j^{X,k_y}$ indices obey the \emph{same} transformation rule, but the 
transformation of $(X,k_y)$ indices still comes in two cases 
[Eq.~\eqref{eq:Xky-label-inversion}].

\subsection{Inversion Symmetry of the Many-Body States}
\label{sec:manybody-inversion}

We study the inversion operation on the many-body states 
$|\Psi;s,r\rangle_\mathrm{lat}$. From Eq.~\eqref{eq:lat-state} we have
\begin{multline}\label{eq:lat-state-inv}
\langle\{X,k_y\}\mathcal{P}|\Psi;s,r\rangle_\mathrm{lat}
=\sum_{\{X_1,k_{y,1}\}}\\
\langle\{X,k_y\}|\mathcal{P}|\{X_1,k_{y,1}\}\rangle\,
\langle\{j^{X_1,k_{y,1}}\}|\Psi;s,r\rangle.
\end{multline}
Plugging into the above equation the inversion transformation of the Wannier 
states in Eq.~\eqref{eq:Xky-inversion}, the sum over $(X_1,k_{y,1})$ reduces 
to \emph{a single term}, with $(X_1,k_{y,1})$ equals either $(-X,-k_y)$ or 
$(-X-1,-k_y)$, depending on the condition in Eq.~\eqref{eq:Xky-inversion}, 
namely, whether $W_x(0)=-1$ and $N_y$ is odd.
For both cases, the Wannier-basis inversion matrix element has the same value
\begin{equation}\label{eq:Xky-inversion-amplitude}
\langle\{X,k_y\}|\mathcal{P}|\{X_1,k_{y,1}\}\rangle=
e^{i\xi_{0,0}}[\lambda_y(0)\,\omega_y]^{-2k_y}.
\end{equation}
And we want to relate the FQH amplitude factor 
$\langle\{j^{X_1,k_{y,1}}\}|\Psi;s,r\rangle$ back to 
$\langle\{j^{X,k_y}\}|\Psi;s,r\rangle$, but this depends on the inversion
transformation $j^{X,k_y}\rightarrow j^{X_1,k_{y,1}}$.
As shown in the previous section of this Appendix, we need to discuss two 
cases in Eq.~\eqref{eq:jXky-inversion} for the inversion transformation of the 
1D index $j^{X,k_y}$.

First, consider the case where $W_x(0)=1$ or $N_y$ is even. 
In this case, $j^{X,k_y}\rightarrow j^{X_1,k_{y,1}}=-j^{X,k_y}$ under inversion.
Also, according to Eq.~\eqref{eq:inv-delta}, the shift $\delta_y$ is either $0$ 
or $N_y/2$, and thus $N_y$ divides $2\delta_y$.
From Eqs.~\eqref{eq:lat-state-inv} and~\eqref{eq:Xky-inversion-amplitude} we 
then have the amplitudes of the inverted state
\begin{align}
&\langle\{X,k_y\}|\mathcal{P}|\Psi;s,r\rangle_\mathrm{lat}\\
\nonumber
&=e^{iN_e\xi_{0,0}}
[\lambda_y(0)\,\omega_y]^{-2\sum k_y}
\langle\{-j^{X,k_y}\}|\Psi;s,r\rangle.
\end{align}
The inversion transformation of the FQH state in Eq.~\eqref{eq:FQH-wf-inv} 
then relates this to the amplitude of another FQH state 
$|\Psi;\bar{s},\bar{r}\rangle$, and we can map it back to the 
corresponding state on lattice
$|\Psi;\bar{s},\bar{r}\rangle_\mathrm{lat}$:
\begin{align}\label{eq:manybody-inv-amp}
&\langle\{X,k_y\}|\mathcal{P}|\Psi;s,r\rangle_\mathrm{lat}\\
\nonumber
&=\zeta_\Psi\,e^{iN_e\xi_{0,0}}
[\lambda_y(0)\,\omega_y]^{-2\sum k_y}
\langle\{X,k_y\}|\Psi;\bar{s},\bar{r}\rangle_\mathrm{lat}.
\end{align}
The indices $\bar{s}$, $\bar{r}$ are defined \emph{uniquely} by 
Eqs.~\eqref{eq:s-sbar} and~\eqref{eq:rbar}. The constant $\zeta_\Psi$ is the 
eigenvalue corresponding to $|\Psi\rangle$ of a composite operator 
[Eq.~\eqref{eq:inversion-psi}] and it takes value from $\pm 1$.
The inversion operation thus relates the components of
the many-body state $|\Psi;s,r\rangle_\mathrm{lat}$ with total momentum 
[Eq.~\eqref{eq:lattice-K}]
\begin{equation}\nonumber
(K_x,K_y)=[\kappa_x+sN_e,C\kappa_y+C(r-\delta_y)N_e]\text{ mod }(N_x,N_y),
\end{equation}
to those of the state $|\Psi;\bar{s},\bar{r}\rangle_\mathrm{lat}$ with total 
momentum
\begin{align}
(\,\overline{\!K}_x,\,\overline{\!K}_y)
&=[\kappa_x+\bar{s}N_e,C\kappa_y+C(\bar{r}-\delta_y)N_e]\\
&=(-K_x,-K_y)
\text{ mod }(N_x,N_y).
\end{align}
thanks to Eqs.~\eqref{eq:s-sbar} and~\eqref{eq:r-rbar} and the fact that $N_x$ 
divides $Nq_x$ and $N_y$ divides $2\delta_y$ as noted earlier.

The coefficients of the inversion transformation in 
Eq.~\eqref{eq:manybody-inv-amp} appear to vary for each component, as 
$\sum k_y$ is different for different components due to umklapp (mod $N_y$) 
processes. However, since with inversion $[\lambda_y(0)\,\omega]^{2N_y}=1$, 
according to Eq.~\eqref{eq:inv-lambda-y-omega},
the troublesome factor $[\lambda_y(0)\,\omega_y]^{2\sum k_y}$ depends only on 
$\sum k_y$ mod $N_y$, and thus \emph{all} the non-zero components 
$\langle\{X,k_y\}|\Psi;\bar{s},\bar{r}\rangle_\mathrm{lat}$ have the 
same value of $\sum k_y$ mod $N_y$, namely, $-K_y$.
This means that the coefficients in Eq.~\eqref{eq:manybody-inv-amp} are 
actually the same for each component.
Therefore, the inversion operator $\mathcal{P}$ indeed brings one state to 
another \emph{within} the $q$-fold multiplet,
\begin{align}\label{eq:inv-psi-s-r-lat-1}
&\mathcal{P}|\Psi;s,r\rangle_\mathrm{lat}\\
\nonumber
&=\zeta_\Psi\, e^{iN_e\xi_{0,0}}
[\lambda_y(0)\,\omega_y]^{2C[\kappa_y+(r-\delta_y)N_e]}\,
|\Psi;\bar{s},\bar{r}\rangle_\mathrm{lat}.
\end{align}

We now proceed to the other case where $W_x(0)=-1$ and $N_y$ is odd.
The majority part of the above argument carries over, except that now 
$j^{X,k_y}$ is mapped to $-j^{X,k_y}-1$ under inversion. The $-1$ can be 
pulled out as a translation $T_\mathrm{cm}^x$ and absorbed into the state 
$|\Psi;\bar{s},\bar{r}\rangle$ [see Eq.~\eqref{eq:psi-s-r}].
The amplitudes of the inverted state reads
\begin{align}\label{eq:manybody-inv-amp-corner}
&\langle\{X,k_y\}|\mathcal{P}|\Psi;s,r\rangle_\mathrm{lat}\\
\nonumber
&=\zeta_\Psi\,e^{iN_e\xi_{0,0}}
[\lambda_y(0)\,\omega_y]^{-2\sum k_y}
\langle\{X,k_y\}|\Psi;\bar{s},\bar{r}-1\rangle_\mathrm{lat}.
\end{align}
In the current case we have $\delta_y=-1-\delta_y$ mod $N_y$.
and thus the state $|\Psi;\bar{s},\bar{r}-1\rangle_\mathrm{lat}$ has the total 
momentum of the inversion partner of 
$|\Psi;s,r\rangle_\mathrm{lat}$,
\begin{align}
(\,\overline{\!K}_x,\,\overline{\!K}_y)
&=[\kappa_x+\bar{s}N_e,C\kappa_y+C(\bar{r}-1-\delta_y)N_e]\\
&=[-\kappa_x-sN_e,-C\kappa_y-C(r-\delta_y)N_e]\\
&=(-K_x,-K_y)
\text{ mod }(N_x,N_y).
\end{align}
Again, the inversion operator $\mathcal{P}$ brings one state to another 
\emph{within} the $q$-fold multiplet,
\begin{align}\label{eq:inv-psi-s-r-lat-2}
&\mathcal{P}|\Psi;s,r\rangle_\mathrm{lat}\\
\nonumber
&=\zeta_\Psi\, e^{iN_e\xi_{0,0}}
[\lambda_y(0)\,\omega_y]^{2C[\kappa_y+(r-\delta_y)N_e]}\,
|\Psi;\bar{s},\bar{r}-1\rangle_\mathrm{lat}.
\end{align}
The index $\bar{r}-1$ needs careful handling, since 
$|\Psi;\bar{s},\bar{r}\rangle$ acquires a phase when 
$\bar{r}\rightarrow\bar{r}+q/q_x$, as shown in 
Eq.~\eqref{eq:psi-r-quasiperiodic}. 

Combining Eqs.~\eqref{eq:inv-psi-s-r-lat-1} and~\eqref{eq:inv-psi-s-r-lat-2},
we have proved Eq.~\eqref{eq:inv-psi-s-r-lat}.

\section{Recombination of the LLL Orbitals}\label{sec:j-jv}

In this Appendix we prove the recombination formula for the LLL orbitals:
\begin{equation}\label{eq:j-jv-wf}
\phi_j^v(\widetilde{x},\widetilde{y})
=\frac{e^{i(\frac{\pi}{4}-\frac{\theta}{2})}}{\sqrt{N_\phi}}\sum_m^{N_\phi}
e^{-i2\pi jm/N_\phi}\phi_m(\widetilde{x},\widetilde{y}),
\end{equation}
where the orbital wave function $\phi_m(\widetilde{x},\widetilde{y})$ is given 
by Eq.~\eqref{eq:LLL-wf}, and $\phi_j^v(\widetilde{x},\widetilde{y})$ by 
Eqs.~\eqref{eq:LLL-wf-v-prime} and~\eqref{eq:LLL-wf-v}.

Consider the single-particle magnetic translation operator 
$T(\mathbf{L}_1/N_\phi)$ in the Landau gauge 
$\mathbf{A}=B\widetilde{x}\,\hat{e}_y$.
This operator commutes with $T(\mathbf{L}_1)$ and $T(\mathbf{L}_2)$, and thus 
is compatible with the periodic boundary conditions in the LLL.
From the explicit expressions, we can find the action on the Landau orbitals
\begin{align}\label{eq:T-LLL-wf}
T(\mathbf{L}_1/N_\phi)\phi_j(\widetilde{x},\widetilde{y})
&=\phi_{j+1}(\widetilde{x},\widetilde{y}),\\
\label{eq:T-LLL-wf-v}
T(\mathbf{L}_1/N_\phi)\phi_j^{v}(\widetilde{x},\widetilde{y})
&=e^{i2\pi j/N_\phi}\phi_j^{v}(\widetilde{x},\widetilde{y}).
\end{align}
The second equation shows that the operator $T(\mathbf{L}_1/N_\phi)$ has 
$N_\phi$ distinct 
eigenvalues within the LLL. Since the LLL as a Hilbert space has dimension 
$N_\phi$, each of the $N_\phi$ eigenspaces of $T(\mathbf{L}_1/N_\phi)$ within 
the LLL has to be non-degenerate. 
Moreover, from Eq.~\eqref{eq:T-LLL-wf} we can construct another state that has 
the same eigenvalue as $\phi_j^v(\widetilde{x},\widetilde{y})$,
\begin{multline}
T(\mathbf{L}_1/N_\phi)
\sum_m^{N_\phi}e^{-i2\pi jm/N_\phi}\phi_m(\widetilde{x},\widetilde{y})\\
=e^{i2\pi j/N_\phi}
\sum_m^{N_\phi}e^{-i2\pi jm/N_\phi}\phi_m(\widetilde{x},\widetilde{y}).
\end{multline}
Therefore, we must have
\begin{equation}
\phi_j^v(\widetilde{x},\widetilde{y})\propto
\sum_m^{N_\phi}e^{-i2\pi jm/N_\phi}\phi_m(\widetilde{x},\widetilde{y}).
\end{equation}
The proportionality constant cannot depend on $(\widetilde{x},\widetilde{y})$. 
It can be fixed by comparing the two sides at 
$(\widetilde{x},\widetilde{y})=(0,0)$.
We find
\begin{align}
\phi_j^v(0,0)&=\frac{1}{(\sqrt{\pi} L_1 l_B)^{1/2}}\frac{\xi}{\sqrt{N_\phi}}\,
\vartheta_3({\textstyle \frac{j}{N_\phi}|\frac{i\xi^2}{N_\phi}}),\\
\phi_m(0,0)&=\frac{1}{(\sqrt{\pi} L_2 l_B)^{1/2}}\frac{1}{\xi\sqrt{N_\phi}}\,
\vartheta_3({\textstyle \frac{m}{N_\phi}|\frac{i}{\xi^2 N_\phi}}),
\end{align}
where $\xi=\sqrt{L_1/L_2}\,e^{i(\pi/4-\theta/2)}$, and
$\vartheta_3(z|\tau)$ is the third Jacobi-$\vartheta$ function,
\begin{equation}
\vartheta_3(z|\tau)\equiv\sum_{n}^{\mathbb{Z}}
e^{i\pi\tau n^2+i2\pi n z},\quad \mathrm{Im}\,\tau>0.
\end{equation}
Using the discrete Fourier transform formula in 
Ref.~\onlinecite{Ruzzi06:Theta-DFT}, we find
\begin{equation}\nonumber
\vartheta_3({\textstyle \frac{j}{N_\phi}|\frac{i\xi^2}{N_\phi}})=
\frac{1}{\xi\sqrt{N_\phi}}
\sum_m^{N_\phi}e^{-i2\pi jm/N_\phi}
\vartheta_3({\textstyle \frac{m}{N_\phi}|\frac{i}{\xi^2 N_\phi}}).
\end{equation}
The proportionality constant at $(\widetilde{x},\widetilde{y})=(0,0)$ 
is then found to be $e^{i(\pi/4-\theta/2)}/\sqrt{N_\phi}$.
This proves the recombination formula of the LLL orbitals 
Eq.~\eqref{eq:j-jv-wf}.

\section{Proof of Quasi-Isotropy}\label{sec:isotropy-generic}

In this Appendix we generalize the proof in Sec.~\ref{sec:isotropy-proof} to 
cover $C=\pm 1$ with generic values of $\delta_x$ and $\delta_y$.
The only hurdle compared with the solved case with $C=1$ and 
$\delta_x=\delta_y=0$ is that the states $|\Psi;s,r\rangle_\mathrm{lat}$ and 
$|\Psi;s,r\rangle_\mathrm{lat}'$ reside in different momentum sectors 
[Eqs.~\eqref{eq:lattice-K} and~\eqref{eq:lattice-K-prime}].

We discuss the case of $C=1$ and $C=-1$ separately.

\subsection{$C=+1$}
Comparing Eqs.~\eqref{eq:lattice-K} and~\eqref{eq:lattice-K-prime}, we find 
that the state $|\Psi;s',r'\rangle_\mathrm{lat}'$ with $(s',r')$ given by
\begin{align}
s'&=s-\delta_x,&
r'&=r-\delta_y
\end{align}
has the same total momentum as the state $|\Psi;s,r\rangle_\mathrm{lat}$.
The above equation replaces Eq.~\eqref{eq:srprime} in the main text.

We now examine the corresponding FQH states. Substituting in 
Eq.~\eqref{eq:T-psi-s-r} gives
\begin{equation}
|\Psi;s',r'\rangle=e^{i\beta}
(T_\mathrm{cm}^y)^{\delta_x}
(T_\mathrm{cm}^x)^{-\delta_y}
|\Psi;s,r\rangle,
\end{equation}
where $e^{i\beta}=e^{i2\pi \delta_x(\kappa_x+rN_e)/N_\phi}$ is an 
inconsequential phase factor.
We find the action of $T_\mathrm{cm}^x$ and $T_\mathrm{cm}^y$ on 
$|\{j\}\rangle_v$ by gauge transforming Eq.~\eqref{eq:j-vprime-translation}:
\begin{align}
T_\mathrm{cm}^x|\{j\}\rangle_v&=e^{i2\pi\sum j/N_\phi}|\{j\}\rangle_v,&
T_\mathrm{cm}^y|\{j\}\rangle_v&=|\{j\!+\!1\}\rangle_v.
\end{align}
Therefore,
\begin{equation}
{}_v\!\langle\{j\}|\Psi;s',r'\rangle
=e^{i\beta}
e^{-i2\pi\sum (j-\delta_x)\delta_y/N_\phi}
{}_v\!\langle\{j\!-\!\delta_x\}|\Psi;s,r\rangle.
\end{equation}

We now examine the lattice amplitudes $\langle\{k_x,k_y\}|\Psi;s,r\rangle'$ 
obtained by the Wannier mapping.
As explained in Sec.~\ref{sec:isotropy-proof}, we put all the momenta 
$(k_x,k_y)$ in the principal Brillouin zone determined by
\begin{align}
-k_x+\delta_x&\in\range{N_x},&
k_y+\delta_y&\in\range{N_y}.
\end{align}
This enables us to use the Wannier mappings $j^{X,k_y}$ and $j_v^{Y,k_x}$ 
directly.

\begin{widetext}
We find
\begin{align}
{}_v\!\langle\{j_v^{Y,k_x}\}|\Psi;s',r'\rangle=
{}_v\!\langle\{YN_x-k_x+\delta_x\}|\Psi;s',r'\rangle
=e^{i\beta}
e^{-i2\pi\sum (YN_x-k_x)\delta_y/N_\phi}
{}_v\!\langle\{YN_x-k_x\}|\Psi;s,r\rangle.
\end{align}
Plugging in the Fourier transform in Eq.~\eqref{eq:j-jv} gives
\begin{equation}
{}_v\!\langle\{j_v^{Y,k_x}\}|\Psi;s',r'\rangle
=e^{i\beta+i\gamma}
\frac{1}{\sqrt{N_\phi}^{N_e}}
\sum_{\{X\}}
e^{-i2\pi\sum k_xX/N_x}
\sum_{\{l_y\}}
e^{i2\pi\sum l_yY/N_y}
e^{-i2\pi\sum k_xl_y/N_\phi}
\langle\{j^{X,l_y}\}|\Psi;s,r\rangle.
\end{equation}
Here we have rewritten each sum over $m$ in the Fourier transform as a double 
sum over all unit cells $X$ and all $l_y$ points in the principal Brillouin zone,
and $e^{i\gamma}=e^{-iN_e(\frac{\pi}{4}-\frac{\theta}{2})}$ is a constant 
phase factor.
Then, the core part of $\langle\{k_x,k_y\}|\Psi;s',r'\rangle_\mathrm{lat}'$ becomes
\begin{equation}
\frac{1}{\sqrt{N_y}^{N_e}}\sum_{\{Y\}}
e^{-i2\pi\sum k_y Y/N_y}\,
{}_v\!\langle\{j_v^{Y,k_x}\}|\Psi;s',r'\rangle
=e^{i\beta+i\gamma -i2\pi\sum k_x k_y/N_\phi}
\frac{1}{\sqrt{N_x}^{N_e}}
\sum_{\{X\}}e^{-i2\pi \sum k_x X/N_x}
\langle\{j^{X,k_y}\}|\Psi;s,r\rangle.
\end{equation}
\end{widetext}

Plugging the above equation into Eq.~\eqref{eq:bloch-amplitude-prime} gives 
Eq.~\eqref{eq:bloch-amplitude-prime-transform} with an inconsequential extra 
factor $e^{i\beta}$.

\subsection{$C=-1$}
In this case, we need to find $(s',r')$ such that
\begin{equation}
\begin{aligned}
\kappa_x+(s'+\delta_x)N_e&=-\kappa_x-sN_e\text{ mod }N_x,\\
\kappa_y+r'N_e&=-(\kappa_y+(r-\delta_y)N_e)\text{ mod }N_y.
\end{aligned}
\end{equation}
Putting in Eqs.~\eqref{eq:s-sbar} and Eq.~\eqref{eq:r-rbar} gives
\begin{align}
s'=\bar{s}-\delta_x,&
r'=\bar{r}+\delta_y,
\end{align}
where $(\bar{s},\bar{r})$ are related to $(s,r)$ by
\begin{equation}\tag{\ref{eq:FQH-wf-inv}}
\langle \{-j\}|\Psi;s,r\rangle=
\zeta_\Psi\langle\{j\}|\Psi;\bar{s},\bar{r}\rangle.
\end{equation}
Therefore, the FQH amplitudes of the primed states are given by
\begin{equation}
{}_v\!\langle\{j\}|\Psi;s',r'\rangle
=e^{i\beta}
e^{-i2\pi\sum (j-\delta_x)\delta_y/N_\phi}
{}_v\!\langle\{j\!-\!\delta_x\}|\Psi;s,r\rangle.
\end{equation}
We emphasize that the above steps requires the inversion symmetry of the 
\emph{FQH states} only; there is no requirement on the FCI system.

\begin{widetext}
The following development largely parallels the $C=+1$ case. Putting all 
$(k_x,k_y)$ in the principal Brillouin zone, we find
\begin{align}
{}_v\!\langle\{j_v^{Y,k_x}\}|\Psi;s',r'\rangle
=\zeta_\Psi^{-1} e^{i\beta}
e^{+i2\pi\sum (YN_x-Ck_x)\delta_y/N_\phi}
{}_v\!\langle\{-YN_x+Ck_x\}|\Psi;s,r\rangle.
\end{align}
Plugging in the Fourier transform in Eq.~\eqref{eq:j-jv} gives for $C=-1$
\begin{equation}
{}_v\!\langle\{j_v^{Y,k_x}\}|\Psi;s',r'\rangle
=\zeta_\Psi^{-1}e^{i\beta+i\gamma}
\frac{1}{\sqrt{N_\phi}^{N_e}}
\sum_{\{X\}}
e^{i2\pi C\sum k_xX/N_x}
\sum_{\{l_y\}}
e^{-i2\pi C\sum l_yY/N_y}
e^{i2\pi\sum k_xl_y/N_\phi}
\langle\{j^{X,l_y}\}|\Psi;s,r\rangle.
\end{equation}
And the core part of $\langle\{k_x,k_y\}|\Psi;s',r'\rangle_\mathrm{lat}'$ becomes
\begin{equation}
\frac{1}{\sqrt{N_y}^{N_e}}\!\sum_{\{Y\}}
e^{-i2\pi\sum k_y Y/N_y}\,
{}_v\!\langle\{j_v^{Y,k_x}\!\}|\Psi;s',r'\rangle
=\frac{e^{i\beta+i\gamma}}{\zeta_\Psi}
e^{-i2\pi C\sum k_x k_y/N_\phi}
\frac{1}{\sqrt{N_x}^{N_e}}\!
\sum_{\{X\}}e^{-i2\pi\!\sum k_x X/N_x}
\langle\{j^{X,k_y}\!\}|\Psi;s,r\rangle.
\end{equation}
\end{widetext}

Plugging the above equation into Eq.~\eqref{eq:bloch-amplitude-prime} gives 
Eq.~\eqref{eq:bloch-amplitude-prime-transform} with an inconsequential extra 
factor $e^{i\beta}/\zeta_\Psi$.

\section{Continuum Limit}\label{sec:continuum-limit}

We consider the continuum limit of a Chern insulator,
$N_x\rightarrow\infty$, $N_y\rightarrow\infty$ with a fixed aspect ratio $N_x/N_y$.
For simplicity, in the following we focus on the case of a square Bravais 
lattice, with $\mathbf{b}_1\cdot\mathbf{b}_2=0$ and 
$|\mathbf{b}_1|=|\mathbf{b}_2|$. The formulas in the generic 
case can be obtained without much extra effort.
We define the continuous coordinates $\widetilde{x}\in[0,N_x)$, 
$\widetilde{y}\in[0,N_y)$,
and the continuous momenta $\widetilde{k}_x,\widetilde{k}_y\in[0,2\pi)$,
\begin{align}
\widetilde{k}_x&=2\pi k_x/N_x,&
\widetilde{k}_y&=2\pi k_y/N_y.
\end{align}
Normally, the continuum limit involves sending the unit cell size to zero.
To make direct comparison with the formulas in 
Ref.~\onlinecite{Qi11:Wavefunction}, however, we do not take the limit of the 
unit cell size going zero, and work instead in a unit system in which 
$|\mathbf{b}_1|=|\mathbf{b}_2|=1$.

\subsection{Berry Connection}\label{sec:continuum-connection}

We show that the discrete connections defined in 
Eq.~\eqref{eq:discrete-A-nonunitary} 
reduce to the Berry connection in the usual definition in the continuum limit.

Recall that the Bloch state is given by
\begin{align}
|\widetilde{k}_x,\widetilde{k}_y\rangle
&=\sum_\alpha u_\alpha(\widetilde{k}_x,\widetilde{k}_y)
|\widetilde{k}_x,\widetilde{k}_y,\alpha\rangle\\
&=\frac{1}{\sqrt{N_xN_y}}\sum_{x,y}\sum_\alpha
e^{i\widetilde{k}_xx+i\widetilde{k}_yy}
u_\alpha(\widetilde{k}_x,\widetilde{k}_y)
|x,y,\alpha\rangle.
\end{align}
The single-particle orbitals are defined by the real-space wave functions.
\begin{equation}
\langle\widetilde{x},\widetilde{y}|x,y,\alpha\rangle
=\phi_\alpha(\widetilde{x}-x-\epsilon_\alpha^x,\widetilde{y}-y-\epsilon_\alpha^y).
\end{equation}
Here the function $\phi_\alpha(\widetilde{x},\widetilde{y})$ is centered around 
$\widetilde{x}=\widetilde{y}=0$ and we have made explicit the sublattice shift 
$\epsilon_\alpha^{x,y}$.

We assume that that the orbitals are orthonormal in the sense that
\begin{equation}
\langle x,y,\alpha|x',y',\beta\rangle
=\delta_{\alpha\beta}\delta_{xx'}\delta_{yy'}.
\end{equation}
And in the spirit of tight-binding approximation, we further assume that the 
orbitals are localized enough such that
\begin{equation}
\begin{aligned}
\langle x,y,\alpha|\widetilde{x}|x',y',\alpha\rangle
&=(x+\epsilon_\alpha^x)\delta_{\alpha\beta}\delta_{xx'}\delta_{yy'},\\
\langle x,y,\alpha|\widetilde{y}|x',y',\alpha\rangle
&=(y+\epsilon_\alpha^y)\delta_{\alpha\beta}\delta_{xx'}\delta_{yy'}.
\end{aligned}
\end{equation}

We can calculate the Berry connection $\mathbf{a}=(a_x,a_y)$ using the usual 
definition
\begin{equation}\label{eq:berry-connection-continuum}
\mathbf{a}(\widetilde{k}_x,\widetilde{k}_y)=
-i\langle u_{\widetilde{k}_x,\widetilde{k}_y}|\nabla_{\widetilde{k}}|
u_{\widetilde{k}_x,\widetilde{k}_y}\rangle,
\end{equation}
where the periodic part of the Bloch state is given by
\begin{equation}
|u_{\widetilde{k}_x,\widetilde{k}_y}\rangle
=e^{-i\widetilde{k}_x\widetilde{x}-i\widetilde{k}_y\widetilde{y}}
|\widetilde{k}_x,\widetilde{k}_y\rangle.
\end{equation}
The result is
\begin{equation}
\begin{aligned}
a_x(\widetilde{k}_x,\widetilde{k}_y)
&=-\!\sum_\alpha u_\alpha^*(\widetilde{k}_x,\widetilde{k}_y)
(i\partial_{\widetilde{k}_x}+\epsilon_\alpha^x)
u_\alpha(\widetilde{k}_x,\widetilde{k}_y),\\
a_y(\widetilde{k}_x,\widetilde{k}_y)
&=-\!\sum_\alpha u_\alpha^*(\widetilde{k}_x,\widetilde{k}_y)
(i\partial_{\widetilde{k}_y}+\epsilon_\alpha^y)
u_\alpha(\widetilde{k}_x,\widetilde{k}_y).
\end{aligned}
\end{equation}

We assume that the Bloch state $|\widetilde{k}_x,\widetilde{k}_y\rangle$ is 
periodic in the $\widetilde{k}_x$ direction, and satisfies 
\begin{equation}
a_y(\widetilde{k}_x,\widetilde{k}_y)=0
\end{equation}
in the $\widetilde{k}_y$ direction (similar to the Landau gauge). The 
Stokes' theorem is applicable, but the non-zero Chern number does not obstruct 
the construction of a smooth gauge, as we have abandoned the periodic gauge 
condition in the $\widetilde{k}_y$ direction.
This is exactly the same boundary condition adopted in 
Ref.~\onlinecite{Qi11:Wavefunction}.

Since the exponentiated Berry connection restores unitarity in the continuum 
limit, we can relate its phase to $\mathbf{a}(\widetilde{k}_x,\widetilde{k}_y)$.
To order $\mathcal{O}(1/N_x)$, we have
\begin{align}
\nonumber
\mathcal{A}_x(\widetilde{k}_x,\widetilde{k}_y)&=\!\sum_\alpha
e^{-i2\pi\epsilon_\alpha^x/N_x}u_\alpha^*(\widetilde{k}_x,\widetilde{k}_y)
u_\alpha(\widetilde{k}_x+{\textstyle \frac{2\pi}{N_x}},\widetilde{k}_y)\\
\nonumber
&\approx 1\!-i\frac{2\pi}{N_x}\!\sum_\alpha u_\alpha^*(\widetilde{k}_x,\widetilde{k}_y)
(i\partial_{\widetilde{k}_x}+\epsilon_\alpha^x)
u_\alpha(\widetilde{k}_x,\widetilde{k}_y)\\
&\approx e^{i2\pi a_x(\widetilde{k}_x,\widetilde{k}_y)/N_x}.
\end{align}
Similarly, to order $\mathcal{O}(1/N_y)$,
\begin{equation}
\mathcal{A}_y(\widetilde{k}_x,\widetilde{k}_y)\approx
e^{i2\pi a_y(\widetilde{k}_x,\widetilde{k}_y)/N_y}.
\end{equation}
In the thermodynamic limit, the above approximate relations become exact.

\subsection{Wilson Loops}\label{sec:continuum-wilson-loops}

Define the phase angle (polarization)
\begin{equation}
\theta(\widetilde{k}_y)=\int_0^{2\pi}\mathrm{d}\widetilde{p}_x\,
a_x(\widetilde{p}_x,\widetilde{k}_y)
\end{equation}
The Wilson loop along a fixed $\widetilde{k}_y$ is given by 
\begin{equation}
W_x(\widetilde{k}_y)=e^{i\theta(\widetilde{k}_y)}.
\end{equation}
The Bloch state $|\widetilde{k}_x,\widetilde{k}_y+2\pi\rangle$ is connected to 
the state $|\widetilde{k}_x,\widetilde{k}_y\rangle$ by a gauge transform. 
Specifically, since we assume $a_y(\widetilde{k}_x,\widetilde{k}_y)=0$, we have
\begin{align}\nonumber
W_y(\widetilde{k}_x)
&=\exp\!\Big[i\!\int_{\widetilde{k}_y}^{\widetilde{k}_y+2\pi}\!\!\!
\mathrm{d}\widetilde{p}_y\,a_y(\widetilde{k}_x,\widetilde{p}_y)
\Big]\,
\langle\widetilde{k}_x,\widetilde{k}_y+2\pi|
\widetilde{k}_x,\widetilde{k}_y\rangle\\
&=\langle\widetilde{k}_x,\widetilde{k}_y+2\pi|
\widetilde{k}_x,\widetilde{k}_y\rangle.
\end{align}
The gauge transform is thus given explicitly by
\begin{equation}\label{eq:ky-2pi}
|\widetilde{k}_x,\widetilde{k}_y+2\pi\rangle
=[W_y(\widetilde{k}_x)]^*\,
|\widetilde{k}_x,\widetilde{k}_y\rangle.
\end{equation}

The Wilson loop $W_x(\widetilde{k}_y)$ is gauge invariant. Naively, one might 
expect that the phase angle $\theta(\widetilde{k}_y)$ is also periodic. 
However, in the continuum limit, this is \emph{not} true. As explained in 
Sec.~\ref{sec:wannier-j}, 
when $\widetilde{k}_y$ gradually increases by $2\pi$, $W_x(\widetilde{k}_y)$ 
winds around the unit circle at the origin of the complex plane. The winding 
number in the clockwise direction is given by the Chern number $C$.
Since we assume the smooth gauge for the Bloch states, $\theta(\widetilde{k}_y)$ 
must be smooth as well. Therefore, we have the quasi-periodic 
condition~\cite{Qi11:Wavefunction}
\begin{equation}
\theta(\widetilde{k}_y+2\pi)=\theta(\widetilde{k}_y)-2\pi C.
\end{equation}
In the light of Eq.~\eqref{eq:ky-2pi}, this can also be understood as
\begin{equation}\label{eq:theta-transform-C}
\theta(\widetilde{k}_y)\rightarrow \theta(\widetilde{k}_y)-2\pi C
\end{equation}
under the gauge transformation
\begin{equation}\label{eq:ky-2pi-gauge-transform}
|\widetilde{k}_x,\widetilde{k}_y\rangle\rightarrow
[W_y(\widetilde{k}_x)]^*\,|\widetilde{k}_x,\widetilde{k}_y\rangle.
\end{equation}

\subsection{Maximally Localized Wannier States}\label{sec:wannier-continuum}

In the continuum limit (which unfortunately cannot be used to compute 
overlaps with the exact ground states from numerical diagonalization on the 
finite-size lattice), the projected position operator 
$\widehat{\mathcal{X}}=P\widehat{x}P$ becomes unitary. 
Therefore, the eigenstates $|X,k_y\rangle$ of $\widehat{X}$ (the unitary part of 
$\widehat{\mathcal{X}}$) are just the eigenstates of $\widehat{\mathcal{X}}$, the 
1D \emph{maximally} localized Wannier states.
The change in the gauge condition in the $\widetilde{k}_y$ direction does not 
affect the form of the Wannier states, as the construction is performed 
independently at each $\widetilde{k}_y$. 
We can rewrite the definition in Eq.~\eqref{eq:wannier-eigenstate} using the 
continuum variables,
\begin{widetext}
\begin{equation}\label{eq:wannier-eigenstate-continuum}
|X,\widetilde{k}_y\rangle=\frac{1}{\sqrt{N_x}}
\sum_{\widetilde{k}_x}^{N_x}
\exp\Big[
-i\widetilde{k}_x X
+i\widetilde{k}_x\frac{\theta(\widetilde{k}_y)}{2\pi}
-i\int_0^{\widetilde{k}_x}\mathrm{d}\widetilde{p}_x\,
a_x(\widetilde{p}_x,\widetilde{k}_y)
\Big]\,
|\widetilde{k}_x,\widetilde{k}_y\rangle.
\end{equation}
\end{widetext}
This expression is exactly the same as in Ref.~\onlinecite{Qi11:Wavefunction}.
The center position of this state can be related to the phase angle by
$\theta(\widetilde{k}_y)$
\begin{equation}
\langle X,\widetilde{k}_y|\widetilde{x}|X,\widetilde{k}_y\rangle
=X-\frac{\theta(\widetilde{k}_y)}{2\pi}.
\end{equation}

We now move on to discuss the gauge transform of these Wannier states. This is 
where the current discussion diverges from Sec.~\ref{sec:wannier-finite-size} 
and Ref.~\onlinecite{Qi11:Wavefunction}. 
Consider a gauge transformation
\begin{equation}
|\widetilde{k}_x,\widetilde{k}_y\rangle\;\rightarrow\;
e^{i\eta(\widetilde{k}_x,\widetilde{k}_y)}
|\widetilde{k}_x,\widetilde{k}_y\rangle.
\end{equation}
The Berry connection transforms by
\begin{equation}
\mathbf{a}(\widetilde{k}_x,\widetilde{k}_y)\;\rightarrow\;
\mathbf{a}(\widetilde{k}_x,\widetilde{k}_y)
+\nabla_{\widetilde{k}}\,\eta(\widetilde{k}_x,\widetilde{k}_y),
\end{equation}
and thus
\begin{multline}
\int_0^{\widetilde{k}_x}\mathrm{d}\widetilde{p}_x\,
a_x(\widetilde{p}_x,\widetilde{k}_y)\;\rightarrow\;
\int_0^{\widetilde{k}_x}\mathrm{d}\widetilde{p}_x\,
a_x(\widetilde{p}_x,\widetilde{k}_y)\\
+\eta(\widetilde{k}_x,\widetilde{k}_y)-\eta(0,\widetilde{k}_y).
\end{multline}
In particular, for $\widetilde{k}_x=2\pi$,
\begin{equation}
\theta(\widetilde{k}_y)\;\rightarrow\; \theta(\widetilde{k}_y)
+\eta(2\pi,\widetilde{k}_y)-\eta(0,\widetilde{k}_y).
\end{equation}
This is consistent with Eq.~\eqref{eq:theta-transform-C}.
Putting all of these together, we find that
\begin{equation}\label{eq:wannier-gauge-transform-continuum}
|X,\widetilde{k}_y\rangle \rightarrow e^{i\eta(0,\widetilde{k}_y)}
|X\!-{\textstyle\frac{1}{2\pi}}
[\eta(2\pi,\widetilde{k}_y)-\eta(0,\widetilde{k}_y)],\;
\widetilde{k}_y\rangle.
\end{equation}
This is \emph{fundamentally different} from the transformation in the 
periodic, non-smooth gauge on a lattice 
[Eq.~\eqref{eq:wannier-gauge-transform}]: in addition to acquiring a phase, the 
Wannier state may hop to a different unit cell upon a gauge transformation.
The transformation given in Eq.~\eqref{eq:wannier-gauge-transform-continuum} 
differs from that of Ref.~\onlinecite{Qi11:Wavefunction}.

The discrepancy between Eq.~\eqref{eq:wannier-gauge-transform} and 
Eq.~\eqref{eq:wannier-gauge-transform-continuum} comes from the smooth gauge 
condition. 
Due to the winding motion, the argument angle of $W_x(\widetilde{k}_y)$ is 
multi-valued by nature.
At finite size, we require $\lambda_x(k_y)$ to be the $N_x$-th root 
of $W_x(k_y)$ with argument angle in $(-2\pi/N_x,0]$. This corresponds to 
confining the argument angle of $W_x(k_y)$ to the branch $(-2\pi,0]$.
In the continuum limit,
\begin{equation}
[\lambda_x(k_y)]^{k_x}=[W_x(k_y)]^{k_x/N_x}\rightarrow
[e^{i\theta(\widetilde{k}_y)}]^{\widetilde{k}_x/(2\pi)}.
\end{equation}
The smooth gauge condition in the continuum requires $\theta(\widetilde{k}_y)$ 
to be smooth as well, and thus makes it impossible to fix 
$\theta(\widetilde{k}_y)$ to $(-2\pi,0]$.

\subsection{Curvature Fluctuations}\label{sec:curvature-fluctuation-continuum}

We now derive the connection between Wannier states, the continuum analogue of 
Eq.~\eqref{eq:wannier-connection}.
Similar to Eq.~\eqref{eq:berry-connection-continuum}, we use the periodic part 
of $|X,\widetilde{k}_y\rangle$ and calculate
\begin{equation}
\widetilde{a}_y(X,\widetilde{k}_y)\equiv
-i\langle X,\widetilde{k}_y|e^{i\widetilde{k}_y\widetilde{y}}
\,\partial_{\widetilde{k}_y}\,
e^{-i\widetilde{k}_y\widetilde{y}}|X,\widetilde{k}_y\rangle.
\end{equation}
From Eq.~\eqref{eq:wannier-eigenstate-continuum}, we obtain
\begin{widetext}
\begin{equation}
\widetilde{a}_y(X,\widetilde{k}_y)
=\partial_{\widetilde{k}_y}\Phi_y(X,\widetilde{k}_y)
+\frac{1}{2\pi}\int_0^{2\pi}\!\mathrm{d}\widetilde{k}_x
\left[\frac{\widetilde{k}_x}{2\pi}\partial_{\widetilde{k}_y}
\theta(\widetilde{k}_x,\widetilde{k}_y)
-\int_0^{\widetilde{k}_x}\!\mathrm{d}\widetilde{p}_x\,
\partial_{\widetilde{k}_y}a_x(\widetilde{p}_x,\widetilde{k}_y)
+a_y(\widetilde{k}_x,\widetilde{k}_y)
\right].
\end{equation}
\end{widetext}
Here we have used the limit
$\frac{1}{N_x}\sum_{k_x}^{N_x}\rightarrow
\frac{1}{2\pi}\int_0^{2\pi}\mathrm{d}\widetilde{k}_x$ to make the notations coherent.
Recall that $a_y(\widetilde{k}_x,\widetilde{k}_y)=0$, the third term in the 
bracket vanishes, while the first two terms can be related to the Berry curvature
\begin{equation}
f_{xy}(\widetilde{k}_x,\widetilde{k}_y)=
\partial_{\widetilde{k}_x}a_x(\widetilde{k}_x,\widetilde{k}_y)
-\partial_{\widetilde{k}_y}a_y(\widetilde{k}_x,\widetilde{k}_y).
\end{equation}
We end up with
\begin{multline}
\widetilde{a}_y(X,\widetilde{k}_y)
=\partial_{\widetilde{k}_y}\Phi_y(X,\widetilde{k}_y)\\
+\frac{1}{2\pi}\!\int_0^{2\pi}\!\!\mathrm{d}\widetilde{k}_x\!
\int_0^{\widetilde{k}_x}\!\!\mathrm{d}\widetilde{p}_x
\left[
f_{xy}(\widetilde{p}_x,\widetilde{k}_y)
-\bar{f}_{xy}(\widetilde{k}_y)
\right]\!,
\end{multline}
where the average Berry curvature $\bar{f}_{xy}(\widetilde{k}_y)$ is defined by
\begin{equation}
\bar{f}_{xy}(\widetilde{k}_y)=\frac{1}{2\pi}\int_0^{2\pi}\mathrm{d}\widetilde{k}_x
f_{xy}(\widetilde{k}_x,\widetilde{k}_y).
\end{equation}
Similar to the discrete case, the fluctuation of Berry curvature appears in 
the connection between adjacent Wannier states, even more sharply than the 
finite-size formula in terms of $W_{\protect\rule{5pt}{2pt}}(k_x,k_y)$ and 
$\overline{W}_{\protect\rule{5pt}{2pt}}(k_x,k_y)$.

We stop the exposition of the continuum case here, as the rest of the 
construction could be obtained directly by taking the continuum limit of the 
finite-size formulas.

\section{Comments on Qi's Prescription}\label{sec:Qi}

In this Appendix we discuss issues in Ref.~\onlinecite{Qi11:Wavefunction} that 
we have found while trying to extend the prescription to the torus geometry.

The original proposal was to map the eigenstates of the projected 
position operator in the gauge $a_y=0$ to the states in the LLL, and 
use these Wannier states to build the \emph{cylinder} FQH states on a lattice.
One would expect a similar construction would hold for the \emph{torus} FQH states.
As pointed out in the main text, the subtle points are that we need to apply 
the phase fix discussed in Sec.~\ref{sec:orthogonality} to restore orthogonality 
of the Wannier states, and that we have to modify the Wannier-LLL index 
mapping according to Eq.~\eqref{eq:j-Xky}.

Based on the observation that the Wannier center shift by one unit cell 
when $\widetilde{k}_y$ increases by $2\pi$, Eq.~(3) in 
Ref.~\onlinecite{Qi11:Wavefunction} claimed that
$|X,\widetilde{k}_y+2\pi\rangle=|X+C,\widetilde{k}_y\rangle$.
However, as pointed out in Appx.~\ref{sec:continuum-wilson-loops}, we can 
think of $\widetilde{k}_y\rightarrow \widetilde{k}_y+2\pi$ as the gauge 
transform in Eq.~\eqref{eq:ky-2pi-gauge-transform}, and thus the Wannier state 
$|X,\widetilde{k}_y+2\pi\rangle$ is given by 
Eq.~\eqref{eq:wannier-gauge-transform-continuum},
\begin{equation}
|X,\widetilde{k}_y+2\pi\rangle
=[W_y(0)]^*\,|X+C,\widetilde{k}_y\rangle.
\end{equation}
Ref.~\onlinecite{Qi11:Wavefunction} missed the extra factor $[W_y(0)]^*$.

This seemingly innocent problem is actually catastrophic, unless $W_y(0)=1$.
It is directly responsible for the low overlap between the constructed 
many-body state and the ground states for several important FCI models 
examined in Sec.~\ref{sec:overlap}.
Assuming $|X,\widetilde{k}_y+2\pi\rangle=|X+C,\widetilde{k}_y\rangle$ as in 
Ref.~\onlinecite{Qi11:Wavefunction},
we can limit $k_y$ to a \emph{single} Brillouin zone when building the $N_xN_y$ 
Wannier states.

Consider the simplest case with Chern number $C=+1$ and the shift 
parameter $\delta_y=0$. The principal Brillouin zone is given by 
$(k_x,k_y)\in\range{N_x}\times\range{N_y}$. We need to implement of the gauge 
condition $a_y=0$ on the lattice. The lattice analogue of $a_y=0$ is setting 
the exponentiated connection $A_y=1$. As pointed out in 
Sec.~\ref{sec:gauge-freedom}, we only need to enforce this condition along 
$k_x=0$. Namely, we should fix the phase of $|0,k_y+1\rangle$ relative to 
$|0,k_y\rangle$ by requiring $A_y(0,k_y)=1$. Starting from the Bloch state at 
$(k_x,k_y)=(0,0)$, this can be done recursively in the interior of the 
principal Brillouin zone till we reach the boundary $(k_x,k_y)=(0,N_y-1)$.
This introduces a \emph{jump} in the phase of $A_y(0,k_y)$ at the boundary of 
Brillouin zone. The Berry connection along $k_x=0$ is given by 
$A_y(0,k_y)=A_y(0,k_y+N_y)$ and
\begin{equation}\label{eq:Qi-Ay}
A_y(0,k_y)=
\begin{cases}
1 & \text{if }k_y\in\range{N_y-1}, \\
W_y(0) & \text{if }k_y=N_y-1.
\end{cases}
\end{equation}
The value of $A_y(0,N_y-1)$ reflects the relative phase of the Bloch state at 
$(0,N_y)$ as fixed by the periodic boundary and the (gauge-invariant) Wilson 
loop. We emphasize that $A_y(0,N_y-1)$ is the connection between the Bloch 
states at momenta $(0,N_y-1)$ and $(0,0)$, rather than the states at 
$(0,N_y-1)$ and $(0,N_y)$; the value of $A_y(0,N_y-1)$ in Eq.~\eqref{eq:Qi-Ay} 
is not in conflict with $a_y=0$.

In this particular gauge of Bloch states, the original construction amounts to 
Eq.~\eqref{eq:wannier-eigenstate} with $e^{i\Phi_y(X,k_y)}=1$.
The analysis in Sec.~\ref{sec:translational-inv} shows that for this choice of 
$e^{i\Phi_y(X,k_y)}$, the resulting many-body wave 
function is indeed translationally invariant, in agreement with the claim in 
the original paper.~\cite{Qi11:Wavefunction}
However, the connection between adjacent Wannier states 
$\langle X,k_y|\widehat{Y}|X',k_y'\rangle\propto A_y(0,k_y)$ changes abruptly at 
the Brillouin zone boundary $k_y=N_y-1$. Unless $W_y(0)=1$, this strong $k_y$ 
dependence sets the Wannier states apart from the Landau orbitals, as shown in
Sec.~\ref{sec:phase-fixing}, and destroys the FQH character of the 
constructed lattice trial state, in accordance with the numerical results 
presented in Sec.~\ref{sec:overlap}.

We now check the inversion symmetry. Putting $e^{i\Phi_y(X,k_y)}=1$ in 
Eq.~\eqref{eq:Xky-inversion-raw}, we have
\begin{equation}
e^{i\xi_{0,k_y}}\mathcal{P}|X,k_y\rangle=
\begin{cases}
|{-X},-k_y\rangle
& \text{if }W_x(k_y)=1,\\
|{-X-1},-k_y\rangle
& \text{otherwise.}
\end{cases}
\end{equation}
Recall that the sewing matrix elements $e^{i\xi_{0,k_y}}$ are related to the 
Berry connections. Plugging Eq.~\eqref{eq:Qi-Ay} into 
Eq.~\eqref{eq:Ay-inversion}, we find
\begin{equation}
e^{i\xi_{0,k_y}}=
\begin{cases}
e^{i\xi_{0,0}} & \text{if }k_y=0\text{ mod }N_y,\\
W_y(0)e^{i\xi_{0,0}} & \text{otherwise}.
\end{cases}
\end{equation}
Restricted by the inversion symmetry, $W_y(0)$ can be $\pm 1$. If $W_y(0)=1$, 
then all the analysis of \emph{our proposal} in the main text can be reused to 
show that the original prescription in this case preserves the inversion symmetry on 
the many-body level. If $W_y(0)=-1$, however, the constructed many-body wave 
function breaks inversion symmetry. The reason is very simple: since 
$|X,k_y\rangle$ receives a different coefficients upon inversion depending on 
whether $k_y=0$ mod $N_y$, different components of the many-body wave function 
transforms differently under inversion.
We expect the degree of the inversion breaking to decrease as the system size 
increases.

It should be obvious that the above problems in original proposal is not 
limited to the case $C=1$ and $\delta_y=0$; it erupts whenever $W_y(0)=-1$.

Finally, we note that in a related paper,~\cite{Barkeshli11:Nematic}
a slightly modified formula for the 1D localized Wannier states were proposed 
for a generic gauge condition on $a_y$.  In our notations, the updated formula 
reads
\begin{widetext}
\begin{equation}\label{eq:wannier-barkeshli}
|X,\widetilde{k}_y\rangle=
\exp\left[
i\widetilde{k}_y\frac{\theta_y}{2\pi}
-i\int_0^{\widetilde{k}_y}\mathrm{d}\widetilde{p}_y\,a_y(0,\widetilde{p}_y)
\right]
\frac{1}{\sqrt{N_x}}
\sum_{\widetilde{k}_x}^{N_x}
\exp\Big[
-i\widetilde{k}_x X
+i\widetilde{k}_x\frac{\theta(\widetilde{k}_y)}{2\pi}
-i\int_0^{\widetilde{k}_x}\mathrm{d}\widetilde{p}_x\,
a_x(\widetilde{p}_x,\widetilde{k}_y)
\Big]\,
|\widetilde{k}_x,\widetilde{k}_y\rangle.
\end{equation}
\end{widetext}
Here the real number $\theta_y$ is defined by
\begin{equation}
\theta_y=\int_0^{2\pi}\mathrm{d}\widetilde{p}_y\,a_y(0,\widetilde{p}_y).
\end{equation}
The exponential prefactor in Eq.~\eqref{eq:wannier-barkeshli} was introduced 
to fix the gauge choice between the Wannier states at different 
$\widetilde{k}_y$.~\cite{Barkeshli11:Nematic}
Judging by appearance, the formula may look similar to ours 
[Eqs.~\eqref{eq:wannier-eigenstate} and Eq.~\eqref{eq:Phi-recursive}] at first sight.
However, further scrutiny reveals a fundamental difference.
The continuum formalism in Refs.~\onlinecite{Qi11:Wavefunction} 
and~\onlinecite{Barkeshli11:Nematic} requires a smooth gauge that is periodic 
in the $\widetilde{k}_x$ direction. Stokes' theorem then dictates that the 
gauge cannot be periodic in the $k_y$ direction. In particular, 
$a_y(0,\widetilde{p}_y)$ is not periodic. Therefore, the phase $e^{i\theta_y}$ 
is \emph{not} the Wilson loop along $\widetilde{k}_x=0$ and is \emph{not} gauge 
invariant. We can actually remove the added exponential prefactor in 
Eq.~\eqref{eq:wannier-barkeshli} altogether, by going back to the $a_y=0$ gauge.
Therefore, the updated formula for the Wannier state in 
Ref.~\onlinecite{Barkeshli11:Nematic} is actually no different from the original 
one in Ref.~\onlinecite{Qi11:Wavefunction}, and they suffer from the same problem 
detailed earlier.

\section{Overlap between Random Vectors}\label{sec:random-overlap}

Consider two random unit vectors sampled independently and uniformly from the 
vector space $\mathbb{C}^d$. We are interested in the probability distribution 
of the absolute square of the inner product (the overlap) between them.
We seek an expression for the probability $p_d(u)$ for the overlap to 
be larger than $u$.

Since the absolute square does not see the relative phase between the two 
vectors, we only need to study $\mathbb{R}^d$. The uniform sampling of unit 
vectors in $\mathbb{C}^d$ corresponds to the uniform sampling \emph{by area} 
over $S^{d}$, the unit $(d-1)$-dimensional sphere surface embedded in 
$\mathbb{R}^d$. We build a Cartesian 
coordinate system $(x_1,x_2,\ldots,x_d)$ with the $x_1$ axis pointing in the 
direction of one of the two vectors. Then we can see clearly that $p_d(u)$ is just 
two times the surface area of the hyperspherical cap on $S^{d}$ of height 
$(1-\sqrt{u})$, divided by the surface area of $S^{d}$.
Using the hyperspherical polar coordinates, we find that
\begin{equation}
p_d(u)=\frac{2 S_{d-1}}{S_{d}}\int_{\sqrt{u}}^{1}\mathrm{d}x\,
(1-x^2)^{(d-3)/2},
\end{equation}
where $S_{n}$ is the surface area of $S^n$, given by
\begin{equation}
S_{n}=\frac{n\,\pi^{n/2}}{\Gamma(\frac{n}{2}+1)}.
\end{equation}
This can be evaluated using numerical quadrature.
We note that at large $d$, the probability $p_d(u)$ to have overlap higher 
than $u$ is \emph{exponentially} small, bounded from above by
\begin{equation}
\mathcal{O}[(1-u)^{(d-1)/2}].
\end{equation}

\bibliography{cond-mat}
\end{document}